\documentclass[a4paper,11pt]{article}
\pdfoutput=1 
\usepackage[english]{babel}
\usepackage{microtype,xspace}
\usepackage[utf8]{inputenc}	

\usepackage{jheppub} 
\usepackage{bm,amssymb,amsmath,amsfonts,slashed,graphicx,multirow,soul,mathtools,xspace,array,comment,color}
\hypersetup{colorlinks=true, allcolors=blue}  
\usepackage{adjustbox}
\usepackage{siunitx}
\usepackage{bm} 
\usepackage{float}
\usepackage{feynmf}
\allowdisplaybreaks
\usepackage{ bbold }
\usepackage{subfigure}
\usepackage[normalem]{ulem} 
\usepackage{enumerate}
\usepackage{multirow}
\usepackage{xparse}
\usepackage{etoolbox}
\usepackage{physics}
\usepackage{hypcap}

\newcommand{\be}{\begin{equation}}
\newcommand{\ee}{\end{equation}}
\newcommand{\TeV}{\text{TeV}}

\newcommand{\LL}{\mathcal{L}}
\newcommand{\ldoublet}{l}

\NewDocumentCommand{\Op}{ m m O{} o }{
	\O^{\ifblank{#3}{}{#3,}#2 }_{\IfNoValueTF{#4}{#1}{\substack{#1\\#4}}}
}
\NewDocumentCommand{\lwc}{ m m O{} o }{
	L^{\ifblank{#3}{}{#3,}#2 }_{\IfNoValueTF{#4}{#1}{\substack{#1\\#4}}}
}
\NewDocumentCommand{\dlwc}{ m m O{} o }{
	{\dot L}^{\ifblank{#3}{}{#3,}#2 }_{\IfNoValueTF{#4}{#1}{\substack{#1\\#4}}}
}
\NewDocumentCommand{\tlwc}{ m m O{} o }{\widetilde
	L^{\ifblank{#3}{}{#3,}#2 }_{\IfNoValueTF{#4}{#1}{\substack{#1\\#4}}}
}
\NewDocumentCommand{\cwc}{ m m O{} o }{
	C^{\ifblank{#3}{}{#3}#2 }_{\IfNoValueTF{#4}{#1}{\substack{#1\\#4}}}
}

\NewDocumentCommand{\tcwc}{ m m O{} o }{
	\widetilde C^{\ifblank{#3}{}{#3}#2 }_{\IfNoValueTF{#4}{#1}{\substack{#1\\#4}}}
}
\newcommand{\opleft}[3]{\mathcal{O}^{#2,#3}_{#1}}

\title{\boldmath Indirect constraints on top quark operators\\ from a global SMEFT analysis}

\author[1,2]{Francesco Garosi,}
\author[2]{David Marzocca,}
\author[1,2]{Antonio Rodriguez-Sanchez,}
\author[1,2]{Alfredo Stanzione}

\affiliation[1]{SISSA International School for Advanced Studies, Via Bonomea 265, 34136, Trieste, Italy}
\affiliation[2]{INFN, Sezione di Trieste, SISSA, Via Bonomea 265, 34136, Trieste, Italy}

\abstract{We perform a model-independent analysis of top-philic New Physics scenarios, under the assumption that only effective operators involving top quarks are generated at tree level.
Within the SMEFT framework, we derive indirect constraints on Wilson Coefficients by combining a large set of low-energy observables: $B$-meson and kaon decays, meson mixing observables, precision electroweak and Higgs measurements, anomalous magnetic moments, lepton flavour violating processes, lepton flavour universality tests, and measurements of the Cabibbo angle. We consider the renormalization group evolution of the operators and use the one-loop matching of the SMEFT onto the LEFT. 
The global analysis is then used to perform one-parameter, two-parameter, and global fits, as well as applications to explicit ultraviolet models. 
We find that the inclusion of measurements from different physics sectors reveals a strong interplay and complementarity among the observables.
The resulting constraints are also compared to direct bounds provided by top quark productions at the LHC.
}

\begin{document}
\maketitle

\section{Introduction}
The top quark plays a special role both in the Standard Model (SM) and beyond, due to its large mass or, in other terms, to its $\mathcal{O}(1)$ Yukawa coupling to the Higgs.
In the SM, due to the GIM mechanism \cite{Glashow:1970gm}, it generates the dominant loop contribution in many low-energy flavour-changing neutral-current (FCNC) processes. It also induces some of the leading radiative corrections in the electroweak (EW) sector, it is the main driver of the renormalization group evolution of the quartic Higgs coupling, and it is the main culprit of the (meta-)stability of the Higgs potential.
The top quark also generates the largest corrections to the Higgs mass, therefore going beyond the SM it is a crucial actor in all New Physics scenarios that aim at addressing the naturalness problem of the EW scale. For instance, the so-called top partners are expected to be some of the lightest new particles in such scenarios, both in supersymmetric (stops) and in composite Higgs (vectorlike top partners) models. This expectation is reinforced by the fact that experimental constraints from direct searches of these particles are generally weaker than for the partners of the lighter quarks.
Furthermore, due to the large top mass, its couplings to EW gauge bosons are still not so strongly constrained and large new physics (NP) effects could still hide there. Therefore, in many NP scenarios the new states couple most strongly to the top quark. See Ref.~\cite{Franceschini:2023nlp} for a recent review.

For all these reasons it is reasonable to assume that NP might be more strongly coupled to the top quark than to the other fermions. 
If such NP is heavy, then at low energy its effect can be described by effective operators that involve the top quark.
In this work we assume that at a UV scale $\Lambda$ (we fix $\Lambda = 1 \, \TeV$ for concreteness) only effective operators involving the top quark  are generated at the tree level and we study the indirect constraints that can be obtained on the corresponding coefficients by considering the effects they have on a large set of low energy observables, comparing these to the direct constraints that have been obtained by studying top quark processes at the LHC.

We work in the Standard Model Effective Field Theory (SMEFT) and choose the Warsaw basis \cite{Grzadkowski:2010es} of dimension-six operators:\footnote{We assume baryon and lepton number conservation at the scale $\Lambda$.}
\begin{equation}\label{eq:SMEFT}
\mathcal{L}_{\mathrm{SMEFT}}=\mathcal{L}_{\mathrm{SM}}+\sum_{i}\frac{c_i^{(6)}}{\Lambda^2} \mathcal{O}_i^{(6)}+\ldots\, .
\end{equation}
As stated above, we focus on top-quark operators, i.e. operators that among quarks involve only the third generation quark doublet $q^3$ and right-handed top $u^3$, assuming that all other vanish.
The lepton sector is instead left completely general, since we want to consider also tests of lepton flavour universality (LFU) and lepton flavour violation (LFV).
The set of SMEFT operators is then reduced to the list in Table~\ref{table:Operators}. In the following, the superscripts `3' for quark indices in SMEFT coefficients are omitted, as they are in the Table. We  work with the rescaled coefficients $C_i \equiv \frac{c_i^{(6)}}{\Lambda^2}$ in $\TeV^{-2}$ units.

\begin{table}[t]
\renewcommand{\arraystretch}{1.7}
\centering
\begin{tabular}{ |c|c|c|c| } 
 \hline
\multicolumn{2}{|c|}{Semi-leptonic}&\multicolumn{2}{c|}{Four quarks} \\
\hline
  $\mathcal{O}^{(1),\alpha\beta}_{\ldoublet q}$  & $(\bar{\ldoublet}^a\gamma_{\mu}\ldoublet^{\beta})(\bar{q}^3\gamma^{\mu}q^3)$     & $\mathcal{O}^{(1)}_{qq}$ & $(\bar{q}^3\gamma^{\mu}q^3)(\bar{q}^3\gamma_{\mu}q^3)$   \\
  $\mathcal{O}^{(3),\alpha\beta}_{\ldoublet q}$  & $(\bar{\ldoublet}^a\gamma_{\mu} \tau^a \ldoublet^{\beta})(\bar{q}^3\gamma^{\mu}\tau^a q^3)$  & $\mathcal{O}^{(3)}_{qq}$ & $(\bar{q}^3\gamma^{\mu}\tau^{a}q^3)(\bar{q}^3\gamma_{\mu}\tau^{a}q^3)$ \\
  $\mathcal{O}_{\ldoublet u}^{\alpha \beta}$ & $(\bar{\ldoublet}^{\alpha}\gamma^{\mu}\ldoublet^{\beta})(\bar{u}^3\gamma_{\mu}u^3)$ & $\mathcal{O}_{uu}$ & $(\bar{u}^3\gamma^{\mu}u^3)(\bar{u}^3\gamma_{\mu}u^3)$ \\
  $\mathcal{O}_{qe}^{\alpha\beta}$ & $(\bar{q}^3\gamma^{\mu}q^3)(\bar{e}^{\alpha}\gamma_{\mu}e^{\beta})$ & $\mathcal{O}_{qu}^{(1)}$ & $(\bar{q}^3\gamma^{\mu}q^3)(\bar{u}^3\gamma_{\mu}u^3)$ \\
  $\mathcal{O}_{eu}^{\alpha\beta}$ & $(\bar{e}^{\alpha}\gamma^{\mu}e^{\beta})(\bar{u}^3\gamma_{\mu}u^3)$ & $\mathcal{O}_{qu}^{(8)}$ & $(\bar{q}^3\gamma^{\mu}T^Aq^3)(\bar{u}^3\gamma_{\mu}T^Au^3)$ \\
  \cline{3-4}
  \cline{3-4}
  $\mathcal{O}_{\ldoublet equ}^{(1),\alpha \beta}$ & $(\bar{\ldoublet}^{\alpha}e^{\beta})\epsilon(\bar{q}^{3}u^3)$ & \multicolumn{2}{c|}{Higgs-Top}  \\
  \cline{3-4}
  $\mathcal{O}_{\ldoublet equ}^{(3),\alpha \beta}$ & $(\bar{\ldoublet}^{\alpha}\sigma_{\mu\nu}e^{\beta})\epsilon(\bar{q}^{3}\sigma^{\mu\nu}u^3)$ & $\mathcal{O}_{H q}^{(1)}$ & $(H^{\dagger}i\overset{\text{$\leftrightarrow$}}{\mathcal{D}_{\mu}}H)(\bar{q}^3\gamma^{\mu}q^3)$ \\
\cline{1-2}
\multicolumn{2}{|c|}{ Dipoles } & $\mathcal{O}_{H q}^{(3)}$ & $(H^{\dagger}i\overset{\text{$\leftrightarrow$}}{\mathcal{D}^a_{\mu}}H)(\bar{q}^3\gamma^{\mu}\tau^a q^3)$ \\
\cline{1-2}
$\mathcal{O}_{uG}$ & $(\bar{q}^3\sigma^{\mu\nu}T^Au^3)\Tilde{H}G_{\mu\nu}^A$ & $\mathcal{O}_{H u}$ & $(H^{\dagger}i\overset{\text{$\leftrightarrow$}}{\mathcal{D}_{\mu}}H)(\bar{u}^3\gamma^{\mu}u^3)$ \\ 
$\mathcal{O}_{uW}$ & $(\bar{q}^3\sigma^{\mu\nu}u^3)\tau^a\Tilde{H}W_{\mu\nu}^a$ & $\mathcal{O}_{uH}$ & $(H^{\dagger}H)(\bar{q}^3u^3\Tilde{H})$ \\ 
$ \mathcal{O}_{uB}$ & $(\bar{q}^3\sigma^{\mu\nu}u^3)\Tilde{H}B_{\mu\nu}$ & & \\ 
\hline
 \end{tabular}
\caption{The 19 dimension-six operators considered in this work. They can be split in four classes, depending on the fields coupled to the Top quark. We keep the lepton flavour structure arbitrary.}
\label{table:Operators}
 \renewcommand{\arraystretch}{1}
 \end{table}

While in the limit where only the top quark mass is considered the gauge and mass bases are the same, we want to describe also the mixing with light generations via the Cabibbo-Kobayashi-Maskawa (CKM) matrix and therefore we must consider all quark masses.
In this case, the flavour basis at the scale $\Lambda$ is in general different than the quark mass bases, forcing us to make a flavour assumption at the high scale.
Two common choices in the literature are the up or down quark mass bases. If only the operators listed above are considered, then the results of the fit will of course depend on this choice.\footnote{One alternative could be to introduce a consistent flavour symmetry, for instance $U(2)_q \times U(2)_u \times U(2)_d$ \cite{Barbieri:2011ci,Barbieri:2012uh,Barbieri:2012tu}, and symmetry-breaking spurions, and consider then also the operators suppressed by the spurions, see e.g. Ref.~\cite{Faroughy:2020ina}. This would however increase greatly the number of coefficients to fit and goes beyond the purpose of our work.}
Since we are assuming that new physics is mostly coupled to the top quark, it is logical to work in the up-quark and charged-lepton mass basis, where $q^i = (u_L^i, \, V_{ij} d_L^j)$, $\ldoublet^\alpha = (\nu_L^\alpha, \, \ell_L^\alpha)$, and $V$ is the CKM matrix. This will induce CKM-suppressed operators involving first and second generation left-handed down quarks.

We aim at deriving indirect constraints on all these operators by considering a large set of low-energy observables in a global analysis. We include rare $B$-meson and kaon decays, meson mixing observables, all the processes used to measure the Cabibbo angle, anomalous magnetic moments of the electron and muon, lepton flavour universality tests in charged-current lepton decays, charged lepton flavour-violating processes, and precision electroweak and Higgs measurements.
To obtain the dependence of these observables on our SMEFT coefficients, we evolve them from the scale $\Lambda$ down to the low-energy scale relevant for each observable using the renormalization group evolution in the SMEFT and in the low-energy effective field theory (LEFT), including the one-loop matching between the two.
Finally, we build a global likelihood and perform several fits on the coefficients of top quark operators.

Several groups studied indirect constraints on top anomalous interactions or top quark operators. Refs.~\cite{Fox:2007in,Grzadkowski:2008mf,Drobnak:2011aa,Brod:2014hsa,Altmannshofer:2023bfk} analyzed anomalous top couplings to $W$ and $Z$ bosons, and top quark flavour violation, considering rare meson decays and electroweak precision data. Dipole and scalar SMEFT operators with top quarks have been considered in Refs.~\cite{Cirigliano:2016nyn,Bissmann:2019gfc}, where indirect bounds from $b\to s \gamma$ and electric dipole moments have been derived after the RG evolution down to the low scale. More recently, Ref.~\cite{Bissmann:2020mfi} performed a more global analysis, where several top quark operators have been considered and the indirect constraints from $B$-meson observables and $Z\to b\bar{b}$ have been derived and compared with direct limits from LHC.
In our work we go beyond these previous analyses by considering a much larger set of top quark operators (all the ones that involve top quarks) and by substantially enlarging the scope of the observables considered.

The paper is structured as follows. In Section~\ref{sec:Methodology} we present the setup of our analysis, describing in general terms how we build the global likelihood. Section~\ref{sec:observables} contains an overview of all the observables we take into account, divided into different classes. In Section~\ref{sec:globalanalysis} we employ the global likelihood to derive fits in some simplified scenarios and discuss the results: one coefficient at a time, interesting pairs of coefficients, and a Gaussian fit with all coefficients except those of semileptonic operators. To showcase some other applications of our analysis, in Section~\ref{sec:UVmodels} we study two specific UV models that we match to the SMEFT. The first contains a scalar leptoquark coupled only to the third generation quark and lepton doublets. The second simplified model, inspired by the Cabibbo anomaly, contains a scalar and a vector leptoquark.
Finally, we conclude in Section~\ref{sec:conclusions}.
Appendix~\ref{App:Observables} contains details on all the observables included in our analysis.

\section{Methodology}
\label{sec:Methodology}

Our goal is to constrain TeV-scale top-philic scenarios deriving bounds from a large set of low-energy observables. The EFT approach represents a suitable framework for this task, as it allows us to consistently study and keep track of the scale dependence of the operators in a multi-step procedure. In practice, RGEs connect different energy scales within the range of validity of the EFT, while matching procedures allow us to integrate out heavy degrees of freedom, linking EFTs valid above or below the threshold.

The description of low energy observables below the electroweak scale relies on the LEFT Lagrangian defined in Ref.~\cite{Jenkins:2017jig}:
\be
\mathcal{L}_{\mathrm{LEFT}}=\mathcal{L}_{\mathrm{QED}+\mathrm{QCD}} + \sum_{i} L_i^{(5)} O_i^{(5)}+ \sum_{i} L_i^{(6)} O_i^{(6)}\,,
\label{eq:LEFT}\ee
where the Higgs, $W$ and $Z$ bosons and the top quark have been integrated out, leaving the QCD sector with only two up-type quarks. Higher-dimension non-renormalizable operators are generated both by the SM heavy particles and by SMEFT contributions, again in a model-independent EFT framework. The complete list of LEFT operators is provided in Table~\ref{tab:oplist1}.

\begin{figure}[t]
    \centering
    \includegraphics[height=0.55\textwidth]{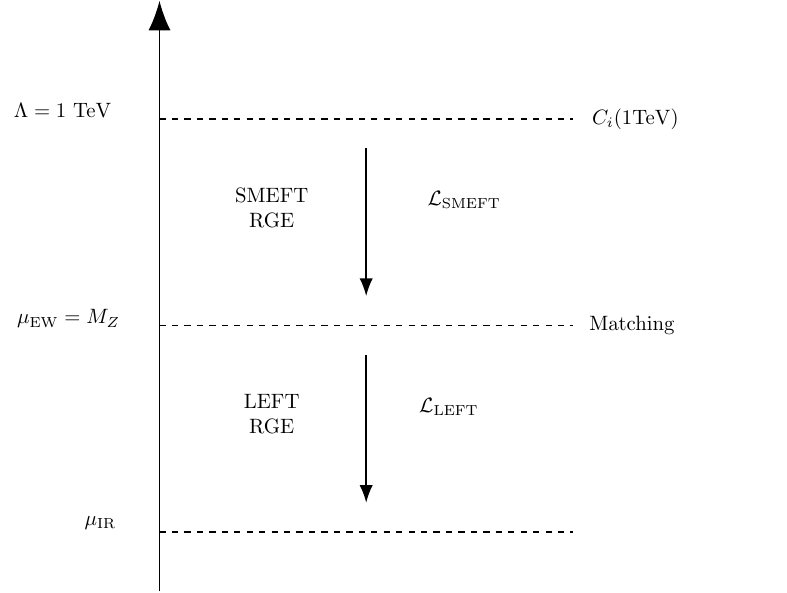}
    \caption{\label{fig:EFTprocedure} Sketch of the EFT analysis procedure adopted in this work. $\mu_{\mathrm{IR}}$ is the scale relevant to the low energy observable under consideration.}
\end{figure}

After having defined our set of top-quark operators at the UV scale $\Lambda = 1$~TeV, as in Eq.~\eqref{eq:SMEFT}, the matching to the whole set of observables is done through the following steps.
\begin{itemize}
\item We evolve the SMEFT coefficients $C_i$ from $\Lambda$ down to the electroweak matching scale, that we fix to $\mu_{\mathrm{EW}} = M_{Z}\simeq 91.2$~GeV. We perform this running procedure numerically using the DSixTools package  \cite{Celis:2017hod,Fuentes-Martin:2020zaz}, which implements the RG equations provided in \cite{Jenkins:2013zja,Jenkins:2013wua,Alonso:2013hga}.

\item As discussed in Ref.~\cite{Braathen:2017jvs}, we combine the one-loop matching with one-loop RG evolution. For the matching between the SMEFT and the LEFT we use the results computed in Ref.~\cite{Dekens:2019ept} and refer to its ancillary files for the complete expressions (see also 
Ref.~\cite{Aebischer:2015fzz}).

\item The LEFT coefficients \cite{Jenkins:2017dyc} are then evolved from the electroweak scale down to the energy scales $\mu_{\mathrm{IR}}$ relevant to the specific experiment, e.g. $\mu_{\mathrm{IR}} = 4.2$~GeV for $B$ decays, again using the DSixTools numerical routines. 

\item We express the low-energy observables in terms of the LEFT Wilson coefficients generated by the previous steps.
\end{itemize}
The whole process is sketched in Figure~\ref{fig:EFTprocedure}, where the main steps and the corresponding energy scales are outlined. QCD effects are known to be relevant, especially at low energies, and a resummation beyond the leading order is required. So, as an exception, we use the five-loop QCD correction for quark Yukawas and the gauge coupling from Refs.~\cite{vanRitbergen:1997va,Vermaseren:1997fq,Baikov:2017ujl} and the four-loop strong coupling beta function and quark mass anomalous dimension from Ref.~\cite{Chetyrkin:2000yt}. These QCD corrections are implemented in DSixTools as well as the three-loop SM RGEs from Refs.~\cite{Bednyakov:2012rb,Bednyakov:2012en,Bednyakov:2013eba,Bednyakov:2014pia}.

The procedure depicted in Fig. \ref{fig:EFTprocedure} does not apply to the observables defined above/at the EW scale, when running the SMEFT coefficients up to $\Lambda_{\mathrm{UV}}$ is the only necessary step. In some cases, previous studies and results in the literature allowed us to partially or completely skip some steps. These cases will be mentioned and discussed in the following.

Once we have expressed the low energy observables and pseudo-observables in terms of SMEFT Wilson coefficients, we can build the log-likelihood:
\begin{equation}
-2\,\mathrm{log}\mathcal{L}(C_{k})\equiv\chi^2(C_k)=\sum_{i,j}(\mathcal{O}_i(C_k)-\mu_i) \, (\sigma^2)^{-1}_{ij} \, (\mathcal{O}_j(C_k)-\mu_j) \,,
\label{eq:likelihood}
\end{equation}
where $\mathcal{O}_i$ are the observables, $C_k$ are the Wilson coefficients defined at the TeV scale, $\mu_i$ are the experimental values and $(\sigma^2)^{-1}_{ij}$ the corresponding covariance matrix. The likelihood is maximized by looking at the minimum of the chi-square, $\chi^2_{0}$, so that the quantity $\Delta \chi^2=\chi^2-\chi^2_{0}$ allows to define the 68\% or 95\% CL regions.
We use our global likelihood to perform individual, pairwise and global fits on Wilson coefficients. In practice, we allow only one, two or a subset of parameters in the global $\chi^2(C_i)$ to vary and set to zero all the remaining ones. Results and applications are discussed in Sections \ref{sec:globalanalysis} and \ref{sec:UVmodels}.

\section{Observables}
\label{sec:observables}

The up quark basis we employ for the left-handed quark doublet makes several top quark operators, such as the ones from Table \ref{table:Operators}, contribute to FCNC $d^i_L\to d^j_L$ transitions even at tree-level, only suppressed by $V_{tj}^{*}V_{ti}$ factors. Very strong bounds to such BSM contributions can be obtained from rare $B$ and $K$ decays as well as meson mixing observables.
Operators with top quarks are also expected to give relevant contributions, via loop effects, to anomalous magnetic moments of leptons, tau decays, electroweak precision data, Higgs measurements and precision measurements entering in Cabibbo-angle analyses.
Finally, operators that violate lepton flavour are constrained by LFV tests in lepton and meson decays.
In the following subsections we provide some more details on these observables, postponing more detailed discussions to Appendix~\ref{App:Observables}.

\subsection{$B$ physics}\label{sec:Bphys}

SM-suppressed FCNC $b \rightarrow s$ transitions are powerful channels to probe new physics, in particular in those observables for which the SM prediction is free from poorly-known long-distance QCD effects.
Among these, after the recent update by LHCb \cite{LHCb:2022qnv}, the lepton flavour universality (LFU) ratios $R_K$ and $R_{K^{*}}$ provide stringent constraints on NP coupled to light leptons:
\begin{equation}
\begin{split}
& R_K[1.1,6]=0.949\pm0.047\,,\\
& R_{K^*}[1.1,6]=1.027\pm0.077\,.
\end{split}\end{equation}
Other powerful decay channels for testing new physics are the so-called \emph{golden-channel} decays, for which it can be useful to define the ratios with the SM predictions as
\be
    R^\nu_{K^{(*)}} \equiv \frac{\mathcal{B}(B\to K^{(*)} \nu \bar{\nu})}{\mathcal{B}(B\to K^{(*)} \nu \bar{\nu})_{\rm SM}}~,
\ee
using the updated SM predictions from Ref.~\cite{Becirevic:2023aov}. Notably, the Belle-II experiment recently presented the first evidence for a signal in the $B^+ \to K^+ \nu \bar{\nu}$ channel:
$\mathcal{B}(B^+\to K^+ \nu \bar{\nu})_{\rm Belle-II} = (2.3 \pm 0.5^{+ 0.5}_{- 0.4}) \times 10^{-5}$ \cite{Belle-II:2023esi}. Once combined with previous upper limits, this becomes $\mathcal{B}(B^+\to K^+ \nu \bar{\nu})_{\rm Comb.} = (1.3 \pm 0.4) \times 10^{-5}$, to be compared with the SM prediction $\mathcal{B}(B^+\to K^+ \nu \bar{\nu})_{\rm SM} = (0.444 \pm 0.030) \times 10^{-5}$.\footnote{This is the SM prediction with the long-distance $B^+ \to \bar{\nu}_\tau \tau^+ \to K^+ \bar{\nu}_\tau \nu_\tau$ removed, since the experimental collaboration treats this as background.}
The corresponding value for $R^\nu_K$ is reported in Table~\ref{table:expBphys}.
On the other hand, the strongest upper limit on the $K^*$ mode comes from Belle, with $R^\nu_{K^{*}} < 3.21$ at 95\%CL \cite{Belle:2017oht} after combining the different modes.
Several NP models affect these decay modes as a rescaling of the SM short-distance amplitude, that involves only $V-A$ currents, in such a way that $R^\nu_{K} = R^\nu_{K^{*}}$. In this case it is interesting to combine the two modes, obtaining:
\be
    R^\nu_{K^{(*)}} = 2.16 \pm 0.70~,
    \label{eq:RnuKKstComb}
\ee
compatible with the SM value of 1 at $1.7\sigma$. We emphasize that, since we are not assuming lepton flavour number conservation, the inclusive sum on neutrino final states takes into account LFV cases: 
\begin{equation}
\mathcal{B}(B\to K^{(*)}\nu\bar{\nu})=\sum_{\alpha,\beta=1}^{3}\mathcal{B}(B\to K^{(*)}\nu_{\alpha}\bar{\nu}_{\beta})\,.
\end{equation}
We conclude the list of $B$ decays by including $B_s \to X_s \gamma$, the leptonic $B_s \to \ell_\alpha \ell_\beta$ decays as well as semileptonic and leptonic LFV modes. The complete list of $B$-physics observables we consider is displayed Table~\ref{table:expBphys}, together with measurements provided by the LHCb, Belle(-II) and BaBar experiments.

\begin{table}[t]
\renewcommand{\arraystretch}{1.2}
\begin{minipage}{.5\linewidth}
\centering
\begin{tabular}{ |c|c| } 
 \hline
 Observable & Experimental value \\ \hline
$B \to X_s\gamma$ & $(3.49\pm0.19)\times10^{-4}$ \cite{Workman:2022ynf} \\ \hline \hline
$R^{\nu}_K$ & $2.93 \pm 0.90$ \cite{Belle-II:2023esi,Becirevic:2023aov} \\ \hline 
$R^{\nu}_{K^*}$ & $ < 3.21$ \cite{Belle:2017oht,Becirevic:2023aov} \\ \hline\hline
$R_K[1.1,6]$&$0.949\pm0.047$ \cite{LHCb:2022qnv} \\ \hline
$R_{K^*}[1.1,6]$&$1.027\pm0.077$ \cite{LHCb:2022qnv} \\ \hline
$\mathcal{B}(B\to K e \mu)$ & $<4.5\times10^{-8}$ \cite{BELLE:2019xld} \\ \hline
$\mathcal{B}(B\to K e \tau)$&$<3.6\times10^{-5}$ \cite{BaBar:2012azg} \\ \hline
$\mathcal{B}(B\to K \mu \tau)$& $<4.5\times10^{-5}$ \cite{LHCb:2020khb} \\ \hline 
\end{tabular} \vspace{1.1cm}
\end{minipage}%
\begin{minipage}{.5\linewidth}
\centering
\begin{tabular}{ |c|c| } 
 \hline
 Observable & Experimental value \\ \hline
$\mathcal{B}(B_s \to ee)$&$<11.2\times10^{-9}$ \cite{LHCb:2020pcv} \\ \hline
$\mathcal{B}(B_s \to \mu\mu)$&$(3.01\pm0.35)\times10^{-9}$ \cite{LHCb:2021awg} \\ \hline
$\mathcal{B}(B_s \to \tau\tau)$&$<6.8\times10^{-3}$ \cite{LHCb:2017myy} \\ \hline
$\mathcal{B}(B_s \to e\mu)$& $<6.3\times10^{-9}$ \cite{LHCb:2017hag} \\ \hline
$\mathcal{B}(B_s \to \mu\tau)$&  $<4.2\times10^{-5}$ \cite{LHCb:2019ujz} \\ \hline
$\mathcal{B}(B_d \to ee)$&$<3.0\times 10^{-9}$ \cite{LHCb:2020pcv} \\ \hline
$\mathcal{B}(B_d \to \mu\mu)$&$<2.6\times10^{-10}$ \cite{LHCb:2021awg} \\ \hline
$\mathcal{B}(B_d \to \tau\tau)$&$<2.1\times10^{-3}$ \cite{LHCb:2017myy} \\ \hline
$\mathcal{B}(B_d \to e\mu)$&$<1.3\times10^{-9}$ \cite{LHCb:2017hag} \\ \hline
$\mathcal{B}(B_d \to \mu\tau)$&$<1.4\times10^{-5}$ \cite{LHCb:2019ujz} \\ \hline
\end{tabular}
\end{minipage}
\caption{Experimental values for $B$-physics observables included in this work. Bounds are given at 95\% CL. In the last column, prospects of future precision are included.}
\label{table:expBphys}
\end{table}

Theoretical predictions are extensively discussed in Appendix \ref{App:Observables}, where branching ratios expressed in terms of low energy EFT coefficients are given. Here, we limit ourselves to show, in Table \ref{table:BphysWC}, a comprehensive sketch of SMEFT contributions to the observables considered in the work. We briefly discuss WC's relevance and interplay in the rest of this Section.

\begin{table}[t]
\centering
\begin{tabular}{ |c|c|c|  }
 \hline
 &  Tree level matching & RG and 1-loop matching\\
 \hline
 $B \to X_s \gamma$     &  & $C_{Hq}^{(1,3)},~ C_{uB}$,~ $C_{uW},~ C_{uG}$ \\ \hline
                                            
  {$R_{K^{(*)}}^{\nu}$}   & 
  \multirow{2}{*}{$C_{Hq}^{(1,3)},~C_{\ldoublet q}^{(1,3),\alpha\beta}$} & $C_{Hu},~C_{qq}^{(1,3)},~C_{\ldoublet u}^{\alpha\beta},~C_{qe}^{\alpha\beta} $  \\[1mm]
 {$K\to \pi \nu\bar{\nu}$} &                         &$C_{qu}^{(1,8)},~C_{uu},~C_{uW}$  \\ \hline
                                            
  {$B \to K^{(*)} \ell _{\alpha} \ell _{\beta} $}     & \multirow{4}{*}{$C_{\ldoublet q}^{(1,3),\alpha\beta} ,~C_{qe}^{\alpha\beta},~C_{Hq}^{(1,3)}$}&\multirow{4}{*}{$C_{\ldoublet u}^{\alpha\beta},~C_{eu}^{\alpha\beta},~C_{qq}^{(1,3)}$}\\[1mm]
    {$B_{s,d} \to \ell _{\alpha} \ell _{\beta} $}     &                         & \\[1mm]
    {$K \to \pi\ell _{\alpha} \ell _{\beta} $}     &                         & \\[1mm]
    {$K \to \ell _{\alpha} \ell _{\beta} $}     &                         & \\ \hline
                                            
 $R_{K^{(*)}}$        & $C_{\ldoublet q}^{(1,3),\ell\ell },~C_{qe}^{\ell\ell}$&$C_{\ldoublet u}^{\ell\ell}$  \\[1mm]
 \hline
\end{tabular}
\caption{Most relevant WC to B-physics observables. The operators that contribute at tree level are displayed in the left column, while operators generated by radiative corrections are listed in the right column. The indices take values $\alpha,\beta$=1,2,3 and $\ell=e,\mu$. Transpose conjugate operators are not listed since related to the already mentioned ones, e.g. $C_{\ldoublet q}^{(1),\beta\alpha}=C_{\ldoublet q}^{(1),\alpha\beta \, *}$.
}
\label{table:BphysWC}
\end{table}

The semileptonic $C_{qe}$ and the combination $(C_{\ldoublet q}^{(1)}+C_{\ldoublet q}^{(3)})$ contribute at tree level to $b(d)\to s \ell_\alpha \ell_\beta$ processes, getting constrained by both rare meson $B_{s,d}\to \ell_\alpha \ell_\beta$ and semileptonic $B\to K \ell_\alpha \ell_\beta$ decays. Within these cases, the recent analysis of $R_{K^{(*)}}$ provides strong constraints on light lepton operators (see discussion in Section \ref{sec:twoparsanalysis} and figures therein) up to the $|C_i| \le 10^{-2}\,\text{TeV}^{-2}$ level.
These bounds on semileptonic current-current operators are completed and complemented by dineutrino modes $B \to K \nu\bar{\nu}$, which are sensitive at tree level to the combination $(C_{\ldoublet q}^{(1)}-C_{\ldoublet q}^{(3)})$ and to the Higgs-top operators $C_{Hq}^{(1)}$ and $C_{Hq}^{(3)}$. 

In general, contributions from 4-quark operators arise in radiative corrections, from both the one-loop matching and the RG evolution. Similarly, the up-type dipole operators $C_{uG},C_{uB}$ and $C_{uW}$ enter the one-loop matching expression for down-type dipoles, contributing then to the $B \to X_s \gamma$ decays. Inclusive radiative decays and rare $B\to\ell_\alpha \ell_\beta$ decays also constrain the $C_{Hq}^{(1/3)}$ coefficients, making $B$ physics bounds almost comparable to the EW precision tests (see Figure (\ref{fig:CHq13})). 

Remarkably, the lack of direct limits on semileptonic dimension-six operators from top-quark measurements, such as inclusive $t\bar{t}$ and single top productions, makes indirect bounds from $B$ mesons crucial in interpreting top-philic NP scenarios. At the same time, these flavour observables provide competitive or stronger constraints on 4-quark and Higgs-top operators, testing the robustness of global fits.

\subsection{Kaon physics}

Analogously to $B$ physics, several operators in Table \ref{table:Operators} induce $s\to d$ transitions, driving FCNC decays of kaons. The list of observables we considered in this work and the corresponding experimental measurements are reported in Table \ref{table:expK}, while more details can be found in Appendix~\ref{App:Observables}.
The discussion for kaon physics follows the same lines as the one above for $B$ decays: the relevant coefficients constrained by each observables are reported in Table~\ref{table:BphysWC}.
Bounds on WC from rare kaon decays involving leptons are less relevant then the corresponding ones of $B$ mesons, mainly due to the strong constraints from $R_{K^{(*)}}$ and $B_s \to \mu \mu$, while $K\to \pi \nu\bar{\nu}$ gives competitive bounds on $C_{\ldoublet q}^{(1/3),33}$, see the relevant discussions in Section~\ref{sec:globalanalysis}. 

\begin{table}[t]\renewcommand{\arraystretch}{1.2}
\centering
\begin{tabular}{ |c|c| }
\hline
Observable & Experimental value \\ \hline
$\mathcal{B}(K^+ \to \pi^+\nu\bar{\nu})$ & $(1.14^{+0.4}_{-0.33})\times10^{-10}$ \cite{NA62:2021zjw} \cite{E949:2008btt}\\ 
\hline
$\mathcal{B}(K_L \to \pi^0\nu\bar{\nu})$ & $<3.6 \times 10^{-9}$ \cite{KOTO:2018dsc}\\
\hline \hline
$\mathcal{B}(K_S\to \mu^+\mu^-)$ &  $<2.5\times 10^{-10}$ \cite{LHCb:2020ycd}\\
\hline
$\mathcal{B}(K_L\to \mu^+\mu^-)_{SD}$ &  $<2.5\times 10^{-9}$ \cite{Isidori:2003ts}\\
\hline
$\mathcal{B}(K_L\to \mu^\pm e^\mp)$ &  $<5.6\times 10^{-12}$ \cite{BNL:1998apv}\\
\hline \hline
$\mathcal{B}(K_L\to \pi^0 \mu^+\mu^-)$ &  $<4.5\times 10^{-10}$ \cite{KTEV:2000ngj}\\
\hline
$\mathcal{B}(K_L\to \pi^0 e^+e^-)$ &  $<3.3\times 10^{-10}$ \cite{KTeV:2003sls}\\
\hline
$\mathcal{B}(K_L\to \pi^0 e^+\mu^-)$ &  $<9.1\times 10^{-11}$ \cite{KTeV:2007cvy}\\
\hline
$\mathcal{B}(K^+\to \pi^+ e^+\mu^-)$ &  $<7.9\times 10^{-11}$ \cite{NA62:2021zxl}\\
\hline
\end{tabular}
\caption{Experimental values for kaon physics observables included in this work. Bounds are given at 95\% CL. }
\label{table:expK}
\end{table}

\subsection{$\Delta F = 2$}

\begin{table}[t]
\centering
\begin{tabular}{ |c|c|c| }
\hline
Observable & Experimental value & SM prediction \\ \hline
$\epsilon_K$ & $(2.228 \pm 0.011)\times 10^{-3}$ & $(2.14 \pm 0.12)\times 10^{-3}$ \\
$\Delta M_s$ & $(17.765 \pm 0.006) \, \text{ps}^{-1}$ & $(17.35 \pm 0.94) \, \text{ps}^{-1}$ \\
$\Delta M_d$ & $(0.5065 \pm 0.0019) \, \text{ps}^{-1}$ & $(0.502 \pm 0.031) \, \text{ps}^{-1}$ \\\hline
\end{tabular}
\caption{Experimental values \cite{Workman:2022ynf} and SM predictions \cite{Buras:2022wpw} for meson mixing observables.}
\label{table:DF2}
\end{table}

Meson mixing observables offer some of the most stringent constraints for several scenarios of beyond the SM physics. Among the top quark operators we consider, $\mathcal{O}_{qq}^{(1,3)}$ induce tree-level contributions to these processes, while many more contribute at the loop level.

Model-independent expressions of new physics contributions to $\Delta F=2$ amplitudes in terms of SMEFT coefficients have been derived in Ref.~\cite{Aebischer:2020dsw}. The authors considered  hadronic matrix elements of the various LEFT operators at the low scale, the RG evolution to the electroweak scale, the matching to SMEFT coefficients in the Warsaw basis (and the up quark mass basis), and finally the SMEFT RG evolution up to a scale $\Lambda = 5 \, \TeV$. We neglect the small mismatch in the RG evolution between this scale and the scale  at which we define our coefficients, 1 TeV, as this can be well included among the theory uncertainties of working with the LL RG evolution.
For the predictions of the SM contribution to meson mixing we use the results of Ref.~\cite{Buras:2022wpw}, specifically the values obtained with the exclusive $V_{cb}$ measurements and the inclusive $V_{ub}$ one (so-called hybrid scenario).
Finally, we take the experimental values for $\epsilon_K$, $\Delta M_s$, and $\Delta M_d$ from the PDG combination \cite{Workman:2022ynf}. In Table~\ref{table:DF2} we collect the values employed in our analysis, where experimental and theory uncertainties are added in quadrature.

\subsection{Cabibbo angle decays}\label{sec:cabibbo}

Nuclear beta, baryon, pion, kaon decay and semileptonic tau decay data are also precise new physics probes. Since the studied decay modes in the SM are suppressed by $\mathcal{O}\left( \frac{1}{v^2} \right)$, new physics contributions to the observables are only suppressed with respect to the SM by $\mathcal{O} \left(\frac{v^2}{\Lambda^2}\right)$. Taking into account the per-mil level precision reached in some of the observables, potential new physics beyond the $\mathrm{TeV}$ scale is probed by these decays. Within the SM picture, this sector is known to lead to some tensions, known as Cabibbo anomalies. It is then interesting to incorporate it to the analysis and to check whether, within our BSM set-up, constraints from other observables leave room for potential new physics explanations of these anomalies.

We make use the global analysis to those observables made in Ref.~\cite{Cirigliano:2021yto} in terms of low-energy EFT coefficients $\epsilon_i$, which updates the EFT analyses made in Refs.~\cite{Gonzalez-Alonso:2016etj,Falkowski:2020pma} and incorporates hadronic tau decays. Consistently with the assumptions of that analysis, we match the $\epsilon_i$ to the LEFT and then to the SMEFT at tree-level (see~\ref{subapp:cabibbomatch}). The leading contributions induced by the studied top operators appear through leading logs in the SMEFT running proportionally to the top Yukawa squared.

In our BSM set-up, the combined fit to this set of observables translates into $\mathcal{O}(\mathrm{TeV})$ sensitivity to $C_{\ldoublet q}^{(3),\ell\ell}$.  At least part of the Cabibbo tension can in principle be alleviated by a nonzero $C_{\ldoublet q}^{(3),22}$ value, which can play an important role in the unitarity relation. In Section~\ref{sec:Cabibbo} we discuss a UV model inspired by this anomaly.
Currently, efforts in the area are focused in understanding whether the so-called Cabibbo anomalies are genuine new physics hints or due to underestimated uncertainties. Overall one may not expect any major improvement in the sensitivity to new physics with respect to the quoted precision from this sector in the short term.

\subsection{Magnetic moments and LFU in $\tau$ decays}

\begin{table}[t]\renewcommand{\arraystretch}{1.2}
\centering
\begin{tabular}{ |c|c|c| }
\hline
\multirow{2}{*}{Observable} & \multicolumn{2}{c|}{Experimental value} \\ \cline{2-3}
                            & \(\ell = e\) & \(\ell = \mu\) \\ \hline \hline
  $\Delta a_{\ell}$ & $(2.8 \pm 7.4) \, \times 10^{-13}$ &$(20.0 \pm 8.4) \, \times 10^{-10}$ \\ \hline
$g_\tau/g_\ell-1$ & $(2.7 \pm 1.4)\, \times 10^{-3}$ & $(0.9 \pm 1.4)\, \times 10^{-3}$ \\\hline
\end{tabular}
\caption{BSM contributions to anomalous magnetic moment of the leptons, $\Delta a_{\ell}\equiv a_{\ell}^{\mathrm{exp}}-a_{\ell}^{\mathrm{SM}}$, and LFU in $\tau$ decays~\cite{HFLAV:2022pwe}, $g_\tau/g_\ell-1$. The correlation between the $g_\tau/g_e$ and the $g_\tau/g_\mu$ values is of a $51\,\%$.}
\label{table:gandLFU}
\end{table}

The anomalous magnetic moments of electrons and muons, $a_{\ell}=(g_{\ell} -2)/2$, are among the most precisely measured quantities in experimental physics,
\be\begin{split}
a_{e}^{\mathrm{exp}}&=(11596521807.3 \pm 2.8)\, \times 10^{-13} \, ,\\
a_{\mu}^{\mathrm{exp}}&= (11659205.9 \pm 2.2)\, \times 10^{-10}\, .
\end{split}\ee
Remarkably, the theoretical precision of the corresponding SM predictions is similar, making them stringent SM tests. In practice, some tensions in the associated evaluations using different inputs slightly limit the current precision. Namely, as pointed out in Ref.~\cite{Aebischer:2021uvt}, the value of $a_{e}^{\mathrm{SM}}$ is sensitive to the input value of the fine-structure constant, $\alpha_{\mathrm{QED}}$, and the two most precise determinations, based on Cesium and Rubidium atomic recoils \cite{Morel:2020dww,Parker:2018vye}, differ by more than $5\, \sigma$. Similarly, the SM leading-order hadronic vacuum polarization (HVP) prediction quoted in the muon $g-2$ Theory White Paper (WP) \cite{Aoyama:2020ynm,Aoyama:2012wk,Aoyama:2019ryr,Czarnecki:2002nt,Gnendiger:2013pva,Davier:2017zfy,Keshavarzi:2018mgv,Colangelo:2018mtw,Hoferichter:2019mqg,Davier:2019can,Keshavarzi:2019abf,Kurz:2014wya,Melnikov:2003xd,Masjuan:2017tvw,Colangelo:2017fiz,Hoferichter:2018kwz,Gerardin:2019vio,Bijnens:2019ghy,Colangelo:2019uex,Blum:2019ugy,Colangelo:2014qya}, which constitutes the dominant source of uncertainty for $a_\mu^{\mathrm{SM}}$, is $2 \sigma$ below the lattice BMW value~\cite{Borsanyi:2020mff} and new results on related observables \cite{Borsanyi:2020mff,Lehner:2020crt,Ce:2022eix,Ce:2022kxy,Wang:2022lkq,Alexandrou:2022amy,Colangelo:2022vok,Aubin:2022hgm,Blum:2023qou,Bazavov:2023has,Davier:2023hhn,CMD-3:2023alj,Benton:2023dci,Davier:2023cyp} suggest that the source of the difference may go beyond a mere statistical fluctuation. We then take the weighted average of the $a_{e}^{\mathrm{SM}}$ result obtained from $\alpha_{\mathrm{QED}}^{\mathrm{Cs}}$ and $\alpha_{\mathrm{QED}}^{\mathrm{Rb}}$ and the weighted average of the $a_{\mu}^{\mathrm{SM}}$ one using $a_{\mu, \mathrm{BMW}}^{\mathrm{HVP,LO}}$ and $a_{\mu,\mathrm{WP}}^{\mathrm{HVP,LO}}$ as inputs, but adding half their differences as additional sources of systematic uncertainties. The corresponding values of $\Delta a_{\ell}= a_{\ell}^{\mathrm{exp}}-a_{e}^{\mathrm{SM}}$ are compiled in Table~\ref{table:gandLFU}.

The main way to generate an extra $a_{\ell}$ contribution in our BSM set-up is through the $\mathcal{O}_{\ldoublet equ}^{(3),\ell \ell}$ operator, since top-antitop annihilation will generate, through mixing, the dipole operator associated to the anomalous magnetic moment.
The process has been recently studied within the LEFT-SMEFT framework in Ref.~\cite{Aebischer:2021uvt}. Using the results from that reference, running up to $\Lambda = 1\, \mathrm{TeV}$ and keeping only the studied operators we find, in $\mathrm{TeV}$ units,
\begin{equation}\begin{split}
\Delta a_{e}&=-4.8 \times 10^{-8}\, \cwc{\ldoublet equ}{}[(3),11][]+ \, 7.1 \times 10^{-11}\, \cwc{\ldoublet equ}{}[(1),11][]  ,\\
\Delta a_{\mu}&=-1.0 \times 10^{-5}\, \cwc{\ldoublet equ}{}[(3),22][]+ \, 1.5 \times 10^{-8}\, \cwc{\ldoublet equ}{}[(1),22][]     \, .
\end{split}\end{equation}
Thus, barring some bizarre cancellation mechanism, $|\cwc{\ldoublet equ}{}[(3),11][]| \gtrsim 10^{-4} \, \mathrm{TeV}^{-2}$ and  $|\cwc{\ldoublet equ}{}[(3),22][]| \gtrsim 10^{-3} \, \mathrm{TeV}^{-2}$ at $\Lambda=1\, \mathrm{TeV}$ can already be excluded by current g-2 measurements.

Ratios of leptonic decays of $\tau$ and $\mu$ provide very clean tests of lepton flavor universality~\cite{Pich:2013lsa,HFLAV:2022pwe}. Deviation from the SM predictions are often parameterized by ratios of effective charges, $g_{\ell}/g_{\ell'}$, which in the SM limit are equal to $1$, and whose experimental values are given in Table \ref{table:gandLFU}. In the LEFT one has
\begin{equation}\begin{split}
\frac{g_\tau}{g_e}-1 &= \frac{v^2}{2} \left( \lwc{\nu e }{LL}[V][\mu e e \mu] - \lwc{\nu e }{LL}[V][\tau \mu\mu \tau] \right) \, , \\
\frac{g_\tau}{g_\mu}-1 &= \frac{v^2}{2} \left( \lwc{\nu e }{LL}[V][\mu e e \mu] - \lwc{\nu e }{LL}[V][\tau e e \tau] \right) \, ,
\end{split}\end{equation}
where $v\approx 246$~GeV. In our set-up this translates, for $\Lambda=1\, \mathrm{TeV}$, into
\begin{equation}\begin{split}
\frac{g_{\tau}}{g_{e}}-1 &= 0.0038\, (C_{\ldoublet q}^{(3),33}-C_{\ldoublet q}^{(3),11}) \, ,\\
\frac{g_{\tau}}{g_{\mu}}-1 &=0.0038\, (C_{\ldoublet q}^{(3),33}-C_{\ldoublet q}^{(3),22}) \, ,
\end{split}\end{equation}
where the $C_{\ldoublet q}$ coefficients are in $\mathrm{TeV}^{-2}$ units.

\subsection{Charged Lepton Flavor-Violating decay modes}
\label{sec:cLFV}

\begin{table}[t]
\renewcommand{\arraystretch}{1.2}
\begin{minipage}{.5\linewidth}
\centering
\begin{tabular}{ |c|c| } 
 \hline
 Observable & Experimental limit \\ \hline
$\mathcal{B}(\mu\to  e \gamma)$ & $5.0\times 10^{-13}$ \cite{MEG:2016leq} \\ \hline
$\mathcal{B}(\mu\to  3e)$ &$1.2 \times 10^{-12}$ \cite{SINDRUM:1987nra}  \\ \hline 
$\mathcal{B}(\mu\,\mathrm{Au}\to e\, \mathrm{Au})$ &$8.3 \times 10^{-13}$ \cite{SINDRUMII:2006dvw} \\ \hline 
$\mathcal{B}(\tau\to e\gamma)$& $3.9 \times 10^{-8}$ \cite{BaBar:2009hkt} \\ \hline
$\mathcal{B}(\tau\to  3e)$ &$3.2 \times 10^{-8}$ \cite{Hayasaka:2010np} \\ \hline
$\mathcal{B}(\tau \to e\, \bar{\mu}\mu)$& $3.2 \times 10^{-8}$ \cite{Hayasaka:2010np} \\ \hline
$\mathcal{B}(\tau \to e\pi^0)$&  $9.5 \times 10^{-8}$ \cite{Belle:2007cio} \\ \hline
$\mathcal{B}(\tau \to e\eta)$&  $1.1 \times 10^{-7}$ \cite{Belle:2007cio} \\ \hline
$\mathcal{B}(\tau \to e\eta')$&  $1.9 \times 10^{-7}$ \cite{Belle:2007cio} \\ \hline
\end{tabular} \vspace{0.62cm}
\end{minipage}%
\begin{minipage}{.5\linewidth}
\centering
\begin{tabular}{ |c|c| } 
 \hline
 Observable & Experimental limit \\ \hline
$\mathcal{B}(\tau \to e\pi^+\pi^-)$&  $2.7 \times 10^{-8}$ \cite{Belle:2012unr} \\ \hline
$\mathcal{B}(\tau \to e K^+ K^-)$&  $4.1 \times 10^{-8}$ \cite{Belle:2012unr} \\ \hline
$\mathcal{B}(\tau \to \mu\gamma)$ & $5.0 \times 10^{-8}$ \cite{Belle:2021ysv} \\ \hline
$\mathcal{B}(\tau \to  3\mu)$ & $2.5 \times 10^{-8}$ \cite{Hayasaka:2010np} \\ \hline
$\mathcal{B}(\tau \to \mu\, \bar{e}e)$& $2.1 \times 10^{-8}$ \cite{Hayasaka:2010np} \\ \hline
$\mathcal{B}(\tau \to \mu\pi^0)$&  $1.3 \times 10^{-7}$ \cite{BaBar:2006jhm} \\ \hline
$\mathcal{B}(\tau \to \mu\eta)$&  $7.7 \times 10^{-8}$ \cite{Belle:2007cio} \\ \hline
$\mathcal{B}(\tau \to \mu\eta')$&  $1.5 \times 10^{-7}$ \cite{Belle:2007cio} \\ \hline
$\mathcal{B}(\tau \to \mu\pi^+\pi^-)$&  $2.5 \times 10^{-8}$ \cite{Belle:2012unr} \\ \hline
$\mathcal{B}(\tau \to \mu K^+ K^-)$&  $5.2 \times 10^{-8}$ \cite{Belle:2012unr} \\ \hline
\end{tabular}
\end{minipage}
\caption{Current $95 \%$ CL limits on studied LFV branching ratios.}
\label{tab:BRlfv}
\end{table}

Experimental searches of neutrinoless lepton flavor-violating decay modes of leptons are suitable to test potential BSM scenarios at energy scales beyond the reach of searches at high-energy colliders.

The very stringent limits on $\mu \to e\gamma$, $\mu \to 3e$ transitions and $\mu \to e$ conversion in nuclei, currently coming respectively from MEG at PSI and SINDRUM \cite{MEG:2016leq,SINDRUM:1987nra,SINDRUMII:2006dvw}, constitute a BSM probe of scales of up to $\Lambda_{\mathrm{BSM}}\sim 10^{3}-10^{4} \, \mathrm{TeV}$ \cite{Davidson:2022jai}. They were studied within the LEFT in Ref.~\cite{Crivellin:2017rmk}. For our specific SMEFT set-up, the corresponding limits translate into limits on linear combinations of Wilson coefficients involving semileptonic operators that violate lepton flavour. 
They are the only operators that at the same time contain the needed BSM LFV insertion and satisfy the assumed top-philic condition. If only one (but any) parameter is switched on, one is able to constrain it at the $|C_{i}|\lesssim 10^{-4}\, \mathrm{TeV}^{-2}$ level. Studies at PSI, MEG II ~\cite{MEGII:2018kmf}, Mu2e~\cite{Mu2e:2014fns} and Mu3e~\cite{Mu3e:2020gyw}, are expected to significantly improve these limits in the near future. We compile the present bounds in Table~\ref{tab:BRlfv}.

\begin{figure}[t]
    \centering
    \includegraphics[height=0.4\textwidth]{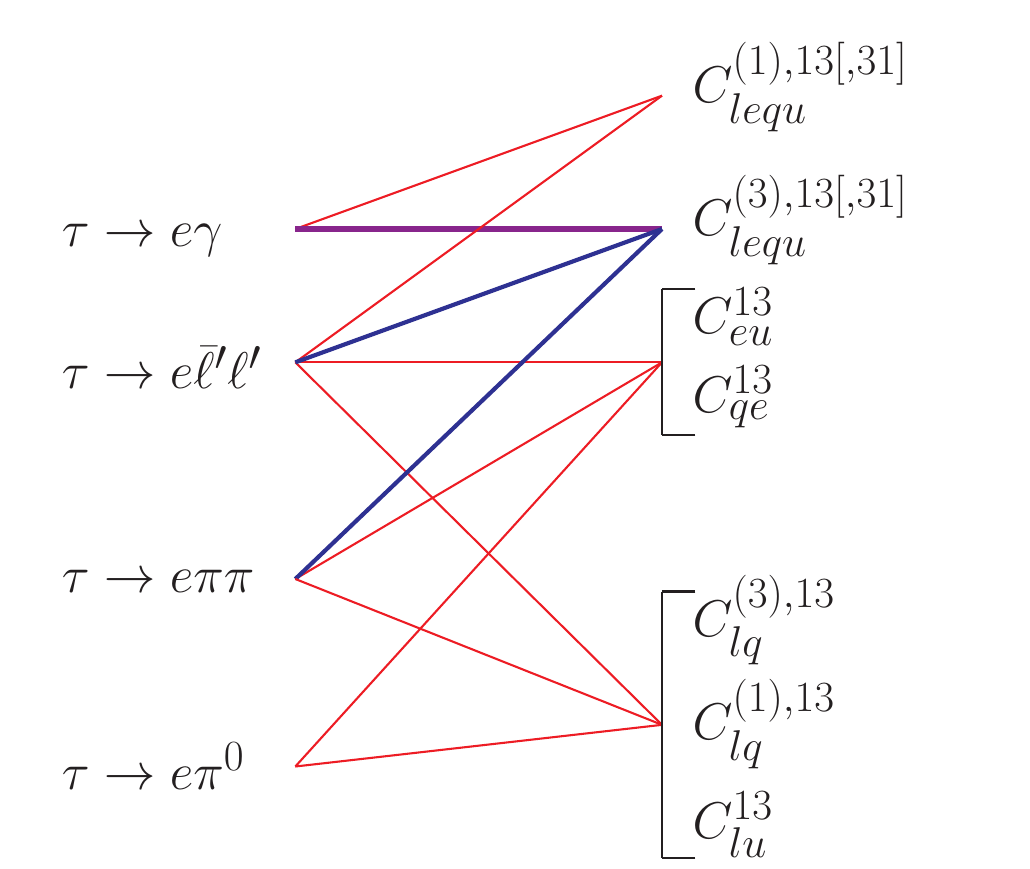}
    \caption{\label{fig:taulfv}Sensitivity of the different $\tau \to e$ decay modes (similar hierarchy is observed for the $\tau \to \mu$ ones) to the different studied Wilson coefficients using the current experimental bounds. Red, blue and purple lines correspond, respectively, to upper bounds to the coefficient below $0.4$, $4\times 10^{-3}$ and $4\times 10^{-4}$ ($\mathrm{TeV}$ units), assuming one parameter is present at a time.}
\end{figure}

While the EFT description at the SMEFT-LEFT level of the tau decays is qualitatively equivalent, leading to bounds on Wilson coefficients to the same kind of semileptonic operators for our case, the underlying LFV studies are different \cite{Celis:2014asa,Cirigliano:2021img,Husek:2020fru,Banerjee:2022xuw}. With the existing and future probes one does not expect to test new physics much higher than $\Lambda_{\mathrm{BSM}}\sim 10 \, \mathrm{TeV}$ \cite{Davidson:2022jai}. However, many more potential decay channels can be experimentally accessed, as a consequence of a tau mass that is large enough to produce hadrons. In this sense, if charged LFV were discovered in the tau sector, the information from the different decay channels would unlock the full power of the EFT approach to stringently discriminate among different BSM scenarios, by having direct experimental access to the values of the different Wilson Coefficients. In the meantime we can use the existing limits to set bounds on different combinations of them. Current limits, coming from Belle and BaBar~\cite{Workman:2022ynf}, are expected to be improved by Belle-II~\cite{Belle-II:2022cgf}. 
In Figure~\ref{fig:taulfv} the sensitivity of the different decay modes to the different studied operators is shown.

\subsection{Electroweak and Higgs data}\label{sec:EWHiggs}

Electroweak precision data ($Z$ and $W$ decays) and Higgs measurements provide strong indirect constraints on new physics involving the top quark, via loop effects.
The authors of Ref.~\cite{Falkowski:2019hvp} performed a global SMEFT analysis of $Z$ and $W$ pole data, $WW$ pair production data at LEP2, and Higgs signal strength measurements from ATLAS and CMS. SMEFT contribution in the Warsaw basis are included at the tree-level except for observables that are loop-generated in the SM (such as $h \to \gamma\gamma, Z\gamma$ and $gg \to h$), in which case one-loop contributions are included as well. For our numerical analysis we use updated results for the global fit (including correlations), kindly provided by the authors of Ref.~\cite{Falkowski:2019hvp}.
The constraints on LFV $Z$ couplings to charged leptons are taken from the fit in Ref.~\cite{Efrati:2015eaa}, while those the LFV couplings to neutrinos are constrained via the measurement of the effective number of neutrinos $N_\nu$ updated in Ref.~\cite{Janot:2019oyi}. 
The fit is performed in terms of specific combinations of Wilson coefficients of the Warsaw basis evaluated at the weak scale shown in Table~\ref{table:EPWTcoeff}. Their definition is reported in App.~\ref{app:HiggsBasis}. 

\begin{table}[t]
\centering
\begin{tabular}{ c c c c c c c c }
\hline
 $C_W$ & $\delta c_z$ & $c_{z\Box}$ & $c_{gg}$ & $c_{\gamma\gamma}$ & $c_{zz}$ & $c_{z\gamma}$ \vspace{0.2cm}\\ 
  $[\delta g_L^{W\ell}]_{\alpha\beta}$ & $[\delta g_L^{Z\ell}]_{\alpha\beta}$ & $[\delta g_R^{Z\ell}]_{\alpha\beta}$ & $[\delta g_L^{Zd}]_{\alpha\alpha}$ & $[\delta g_R^{Zd}]_{\alpha\alpha}$ & $[\delta g_L^{Zu}]_{\alpha\alpha}$ & $[\delta g_L^{Zu}]_{\alpha\alpha}$ \vspace{0.2cm} \\
  $[C_{uH}]_{33}$ & $[C_{dH}]_{33}$ & $[C_{eH}]_{22}$ & $[C_{eH}]_{33}$ & $[C_{\ldoublet \ldoublet}]_{1221}$ \\ \hline
\end{tabular}
\caption{List of Wilson coefficients used in the fit of Ref.~\cite{Falkowski:2019hvp}. The ones with capital $C$ are already in the Warsaw basis, while the definition of the others is reported in App.~\ref{app:HiggsBasis}.}
\label{table:EPWTcoeff}
\end{table}

We interpret the resulting likelihood for SMEFT coefficients as evaluated at the EW scale (we fix it at $M_Z$ for concreteness), and then include the RG evolution up to 1 TeV to obtain the final likelihood for our analysis, in terms of the coefficients of the operators listed in Table~\ref{table:Operators}. 
The coefficients that provide relevant constraints to our analysis are listed here, together with the TeV-scale coefficients that contribute the most to each one:
\begin{equation}
    \begin{split}
    \text{EW/Higgs coeff. } & \qquad \qquad \text{TeV-scale coefficient} \\ \hline
        \delta g_{L}^{Z\ell} \quad &\longleftarrow \quad C_{uB}, ~ C_{uW}, C_{Hu}, ~ C_{Hq}^{(1,3)}, ~ C_{\ldoublet q}^{(1,3), \ell\ell}, ~ C_{\ldoublet u}^{\ell\ell}, ~ \ldots \\
        \delta g_{L}^{W\ell} \quad &\longleftarrow \quad C_{uB}, ~ C_{uW}, ~ C_{Hu}, ~ C_{Hq}^{(1,3)}, ~ C_{\ldoublet q}^{(3), \ell\ell}, ~ \ldots \\
        \delta g_{R}^{Z\ell} \quad &\longleftarrow \quad C_{uB}, ~ C_{uW}, ~ C_{Hu}, ~ C_{Hq}^{(1,3)}, ~ C_{eu}^{\ell\ell}, ~ C_{qe}^{\ell\ell}, ~ \ldots \\
        \delta g_{L}^{Zb} \quad &\longleftarrow \quad C_{Hq}^{(1,3)}, ~ C_{Hu}, ~ C_{qq}^{(1,3)}, ~ \ldots \\
         \delta g_{R}^{Zb} \quad &\longleftarrow \quad C_{Hq}^{(1)}, ~ C_{Hu}, ~ C_{qq}^{(1,3)}, ~ C_{uB}, ~ C_{uW}, ~ \ldots \\
        c_{\gamma \gamma} \quad &\longleftarrow \quad C_{uB}, ~ C_{uW}, ~ C_{uG} \\
        c_{g g} \quad &\longleftarrow \quad C_{uG} \\
        [C_{eH}]_{\alpha \alpha} \quad &\longleftarrow \quad C_{\ldoublet equ}^{(1), \alpha \alpha} \\
        [C_{uH}]_{33} \quad &\longleftarrow \quad C_{uH}, ~ C_{uG}, ~ C_{Hq}^{(1,3)}, ~ C_{qu}^{(1,8)}, ~ \ldots
    \end{split}
\end{equation}

\subsection{Direct bounds from LHC} 
\label{sec:LHCconstr}

SMEFT interpretations of top quark production and decay measurements at LHC have been discussed in several works \cite{Aguilar-Saavedra:2018ksv,Maltoni:2019aot,Bissmann:2019gfc,Brivio:2019ius,Durieux:2019rbz,Hartland:2019bjb,Bruggisser:2021duo,Ethier:2021bye,Miralles:2021dyw,Durieux:2022cvf,deBlas:2022ofj,Bruggisser:2022rhb,Giani:2023gfq,Kassabov:2023hbm,Grunwald:2023nli,Altmannshofer:2023bfk}, providing direct bounds on 4-quark, dipole and Higgs-top operators. 

The analysis by the SMEFiT collaboration \cite{Ethier:2021bye} includes observables from the LHC Run-II dataset and studies connections with Higgs and diboson data. 
The SMEFiT fitting framework has been released as a Python open source package \cite{Giani:2023gfq}. We exploit the flexibility of this toolbox to perform a SMEFT analysis of Higgs, top quark and electroweak production data (see Section 3 of Ref.~\cite{Ethier:2021bye} for details) including our operators of Table~\ref{table:Operators}, with exception of the semi-leptonic ones. The resulting constraints are reported in Appendix~\ref{app:LHCconstraints}.
In the next section we compare these with the indirect constraints derived from our global analysis.

\section{Global analysis}\label{sec:globalanalysis}

The observables described in the previous section, except for the direct constraints from LHC of Section~\ref{sec:LHCconstr}, are used to build the log-likelihood as in Eq.~\eqref{eq:likelihood}.
This provides global indirect constraints on top quark operators, that we can compare with the direct ones from measurements at LHC.
In the following we present examples of fits derived using our global likelihood.
For simplicity we assume all Wilson coefficients to be real in the following.

\subsection{One-parameter fits}

First, we perform one-parameter fits setting all the Wilson coefficients to zero except for one.
While such one-parameter set-up is not a realistic low-energy description of typical UV scenarios,  it can provide meaningful information about the new physics scale that can be probed if that operator is generated. 
It also provides a way to compare the sensitivity of different observables or, alternatively, which observables/sectors one should look at first if a specific operator is induced by the studied UV model.

The results associated to the different operators, respectively nonleptonic, semileptonic lepton-flavor conserving and semileptonic lepton-flavor violating, are shown in Tables~\ref{tab:indivnosemi}, \ref{tab:indivsemilfc} and~\ref{tab:indivsemilfv}, together with which (isolated) single observable currently gives the most precise determination of the associated Wilson coefficient. We have defined $C_{iq}^{(\pm)}\equiv C_{iq}^{(1)}\pm C_{iq}^{(3)}$ for $i=H,q,\ldoublet$. As shown in Section \ref{sec:observables}, these linear combinations are typically the ones appearing at leading order in the most constraining observables.

\begin{table}[t]
\renewcommand{\arraystretch}{1.4}
\centering
\begin{tabular}{|c|c|c|}
\hline
Wilson & Global fit [TeV$^{-2}$] & Dominant \\
\hline \hline
$C_{qq}^{(+)}$ & $(-1.9 \pm 2.3) \times 10^{-3}$ & $\Delta M_s$ \\ 
\hline
$C_{qq}^{(-)}$ & $(-2.0 \pm 1.0) \times 10^{-1}$ & $B_{s} \to \mu\mu$ \\ 
\hline
$C_{qu}^{(1)}$ & $\phantom{-}(1.3 \pm 1.0) \times 10^{-1}$ & $\Delta M_s$ \\ 
\hline
$C_{qu}^{(8)}$ & $(-1.7 \pm 4.4) \times 10^{-1}$ & $\Delta M_s$ \\ 
\hline
$C_{uu}$ & $(-3.0 \pm 1.7) \times 10^{-1}$ & $\delta g^{Ze}_{L,11}$ \\ 
\hline
$C_{Hq}^{(+)}$ & $(18.7 \pm 8.8)
\times 10^{-3}$ & $B_s \to\mu\mu$ \\ 
\hline
$C_{Hq}^{(-)}$ & $\phantom{-}(5.8 \pm 4.5) \times 10^{-2}$
& $\delta g^{Ze}_{L,11}$ \\ 
\hline
$C_{Hu}$ & $(-4.3 \pm 2.3) \times 10^{-2}$ & $\delta g^{Ze}_{L,11}$ \\ 
\hline
$C_{uB}$ & $(-0.6 \pm 2.0) \times 10^{-2}$ & $c_{\gamma\gamma}$ \\ 
\hline
$C_{uG}$ & $(-0.1 \pm 2.0) \times 10^{-2}$ & $c_{gg}$ \\ 
\hline
$C_{uH}$ & $(-0.3 \pm 5.2) \times 10^{-1}$ & $C_{uH,33}$ \\ 
\hline
$C_{uW}$ & $(-0.1 \pm 3.1) \times 10^{-2}$ & $c_{\gamma\gamma}$ \\ 
\hline
\end{tabular}
\caption{Combined fit for one at a time Wilson coefficients for nonleptonic operators. The dominant observable giving the most precise bound is also displayed.}
\label{tab:indivnosemi}
\end{table}

\begin{table}[t]
\renewcommand{\arraystretch}{1.4}
\begin{minipage}{.5\linewidth}
\small
\centering
\begin{tabular}{|c|c|c|}
\hline
Wilson & Global fit [TeV$^{-2}$] & Dominant \\
\hline \hline
$C_{\ldoublet q}^{(+),11}$ & $\phantom{-}(2.4 \pm 3.5) \times 10^{-3}$  & $R_K$ \\ 
\hline
$C_{\ldoublet q}^{(+),22}$ & $(-4.0 \pm 3.4) \times 10^{-3}$ & $R_K$ \\ 
\hline
$C_{\ldoublet q}^{(+),33}$ & $\phantom{-}(7.2 \pm 4.4) \times 10^{-1}$ & $g_{\tau}/g_i$ \\ 
\hline
$C_{\ldoublet q}^{(-),11}$ & $(10.9 \pm 7.6) \times 10^{-2}$ & 
$R_{K^{(*)}}^{\nu}$ \\ 
\hline
$C_{\ldoublet q}^{(-),22}$ & $(-6.0 \pm 7.0) \times 10^{-2}$ & $R_{K^{(*)}}^{\nu}$ \\ 
\hline
$C_{\ldoublet q}^{(-),33}$ & $(-1.8 \pm 1.0) \times 10^{-1}$ & $R_{K^{(*)}}^{\nu}$ 
\\ 
\hline
$C_{\ldoublet u}^{11}$ & $(-1.7 \pm 7.0) \times 10^{-2}$ & $\delta g^{Ze}_{L,11}$ \\ 
\hline
$C_{\ldoublet u}^{22}$ & $(-4.3 \pm 1.8) \times 10^{-1}$ & $\delta g^{Ze}_{L,22}, \, R_K$ \\ 
\hline
$C_{\ldoublet u}^{33}$ & $\phantom{-}(0.5 \pm 2.4) \times 10^{-1}$ & $\Delta g^{Ze}_{L,33}$ \\ 
\hline
$C_{qe}^{11}$ & $(-0.7 \pm 3.9) \times 10^{-2}$ & $R_{K^*}$ \\ 
\hline
$C_{qe}^{22}$ & $(12.1 \pm 9.2) \times 10^{-3}$ & $B_s\to\mu\mu$ \\ 
\hline
$C_{qe}^{33}$ & $\phantom{-}(2.2 \pm 2.4) \times 10^{-1}$ & $\delta g^{Ze}_{R,33}$ \\ 
\hline
\end{tabular}
\end{minipage}
\begin{minipage}{.5\linewidth}
\centering
\small
\begin{tabular}{|c|c|c|}
\hline
Wilson & Global fit [TeV$^{-2}$] & Dominant \\
\hline \hline
$C_{eu}^{11}$ & $\phantom{-}(5.0 \pm 8.1) \times 10^{-2}$ & $\Delta g^{Ze}_{R}{}_{11}$ \\ 
\hline
$C_{eu}^{22}$ & $\phantom{-}(4.8 \pm 2.1) \times 10^{-1}$ & $\Delta g^{Ze}_{R}{}_{22}$ \\ 
\hline
$C_{eu}^{33}$ & $(-2.3 \pm 2.5) \times 10^{-1}$ & $\Delta g^{Ze}_{R}{}_{33}$ \\ 
\hline
$C_{lequ}^{(1),11}$ & $\phantom{-}(0.4 \pm 1.0) \times 10^{-2}$ & $(g-2)_e$ \\ 
\hline
$C_{lequ}^{(1),22}$ & $\phantom{-}(1.8 \pm 1.6) \times 10^{-2}$ & $C_{eH}{}_{22}$ \\ 
\hline
$C_{lequ}^{(1),33}$ & $\phantom{-}(8.0 \pm 9.1) \times 10^{-2}$ & $C_{eH}{}_{33}$ \\ 
\hline
$C_{lequ}^{(3),11}$ & $(-0.6 \pm 1.5) \times 10^{-5}$ & $(g-2)_e$ \\ 
\hline
$C_{lequ}^{(3),22}$ & $(-19.3 \pm 8.1) \times 10^{-5}$ & $(g-2)_\mu$ \\ 
\hline
$C_{lequ}^{(3),33}$ & $(-7.0 \pm 7.8) \times 10^{-1}$ & $C_{eH}{}_{33}$ \\ 
\hline 
\end{tabular} \vspace{2cm}
\end{minipage}
\caption{Combined fit for one at a time Wilson coefficients for semileptonic lepton-flavor-conserving operators. For reference, we also display the single observable giving the most precise bound.}
\label{tab:indivsemilfc}
\end{table}

\begin{table}[t]
\centering\renewcommand{\arraystretch}{1.4}
\begin{tabular}{|c||c|c||c|c||c|c|}
\hline
\multicolumn{1}{|c||}{} & \multicolumn{2}{c||}{$\mu \rightarrow e$ } & \multicolumn{2}{c||}{$\tau \rightarrow \mu$ } & \multicolumn{2}{c|}{$\tau \rightarrow e$ } \\
\cline{2-7}
\multicolumn{1}{|c||}{\multirow{-2}{*}{Wilson}} & Limit & Dominant & Limit & Dominant & Limit & Dominant \\
\hline
$C_{lequ}^{(3)}$ & $3.9 \times 10^{-9}$ & $\mu \rightarrow e \gamma$ & $5.0 \times 10^{-5}$ & $\tau \rightarrow \mu \gamma$ & $4.4 \times 10^{-5}$ & $\tau \rightarrow e \gamma$ \\
$C_{lequ}^{(1)}$ & $3.6 \times 10^{-5}$ & $\mu \rightarrow 3e,e\gamma$ & $2.7 \times 10^{-2}$ & $\tau \rightarrow \mu \gamma$ & $2.4 \times 10^{-2}$ & $\tau \rightarrow e \gamma$ \\
$C_{\ldoublet q}^{(3)}$ & $6.7 \times 10^{-5}$ & $\mu \mathrm{Au} \rightarrow e \mathrm{Au}$ & $7.1 \times 10^{-2}$ & $\tau \rightarrow \mu \pi \pi$ & $7.4 \times 10^{-2}$ & $\tau \rightarrow e \pi \pi$ \\
$C_{\ldoublet q}^{(1)}$ & $4.0 \times 10^{-5}$ & $\mu \mathrm{Au} \rightarrow e \mathrm{Au}$ & $1.1 \times 10^{-1}$ & $\tau \rightarrow \mu \pi \pi$ & $1.1 \times 10^{-1}$ & $\tau \rightarrow e \pi \pi$ \\
$C_{\ldoublet u}$ & $4.0 \times 10^{-5}$ & $\mu \mathrm{Au} \rightarrow e \mathrm{Au}$ & $1.0 \times 10^{-1}$ & $\tau \rightarrow \mu \pi \pi$ & $1.1 \times 10^{-1}$ & $\tau \rightarrow e \pi \pi$ \\
$C_{eu}$ & $3.6 \times 10^{-5}$ & $\mu \mathrm{Au} \rightarrow e \mathrm{Au}$ & $
1.0\times 10^{-1}$ & $\tau \rightarrow \mu \pi\pi$ & $
1.1\times 10^{-1}$ & $\tau \rightarrow e\pi\pi$ \\
$C_{qe}$ & $3.6 \times 10^{-5}$ & $\mu \mathrm{Au} \rightarrow e \mathrm{Au}$ & $%
1.0  \times 10^{-1}$ & $\tau \rightarrow \mu \pi\pi$ & $%
1.0 \times 10^{-1}$ & $\tau \rightarrow e\pi\pi$ \\
\hline
\end{tabular}
\caption{Upper limits ($68 \% \, \mathrm{CL}$) in  TeV$^{-2}$ for the different LFV Wilson coefficients. Family superscripts are omitted (in the working approximation bounds on $C_{lequ}^{(1,3),ij}$ and $C_{lequ}^{(1,3),ji}$ are found to be the same). For reference we also display the single observable giving the most precise bound.}
\label{tab:indivsemilfv}
\end{table}

\begin{figure}[t]
    \centering
    \includegraphics[width=0.95\textwidth]{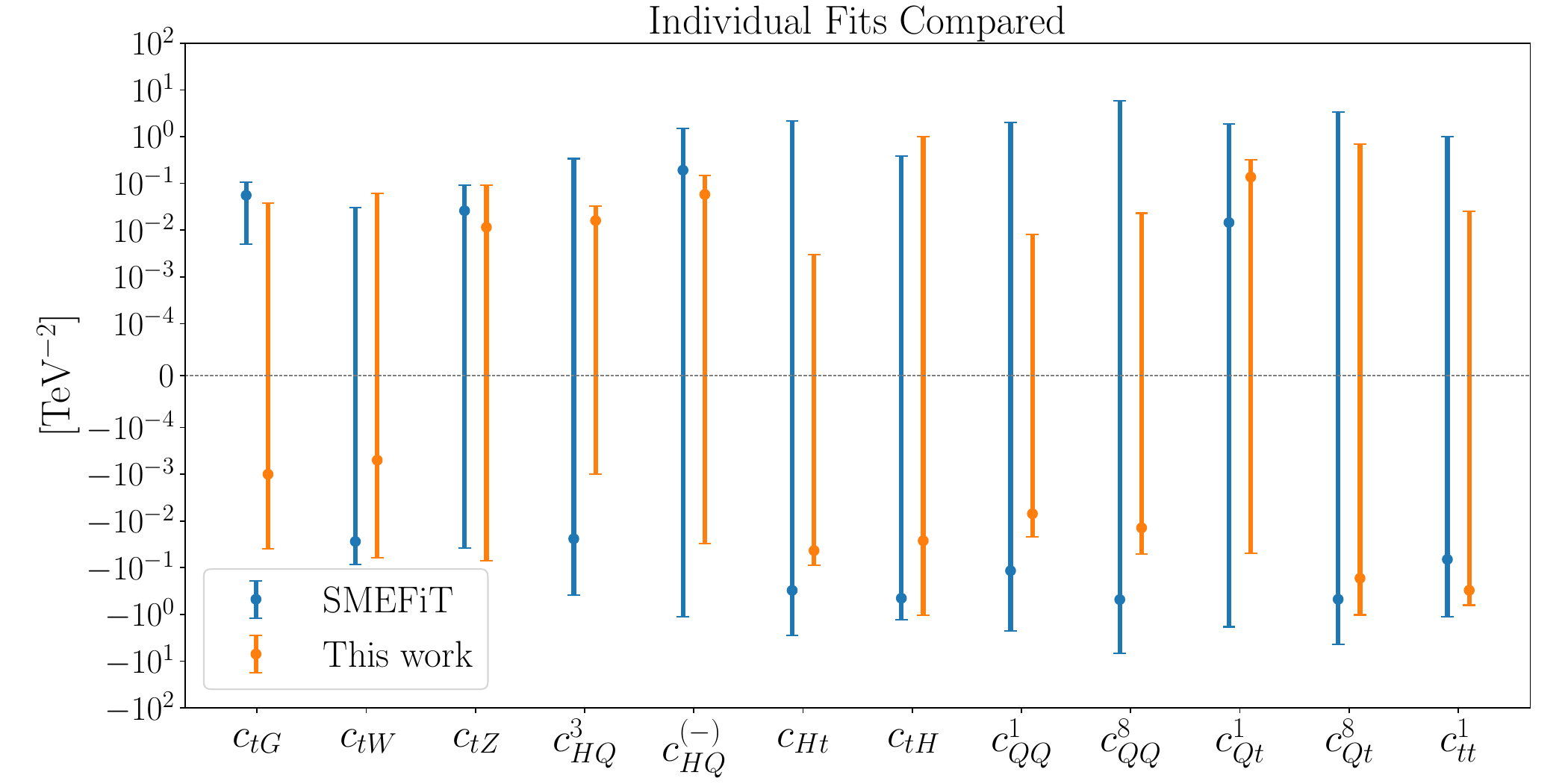}
    \caption{Comparison of the indirect constraints derived in this work and the direct bounds obtained through the SMEFiT toolbox. We display the 95\% CL bounds for Wilson coefficients in the SMEFiT basis and notation (see Table \ref{tab:SMEFIT} for their expression in terms of the Warsaw basis).}
    \label{Fig:CompareIndividual}
\end{figure}

We study next the complementarity of the indirect bounds with the direct LHC ones. 
In Fig.~\ref{Fig:CompareIndividual} we show the results of the individual fits for Higgs-Top, Dipoles and Four Quarks operators, together with corresponding bounds derived through the SMEFiT fitting framework.
We stress that, exclusively in this plot, Wilson coefficients are displayed according to the SMEFiT basis (lower case $c_{AB}$), whose expression in terms of the Warsaw basis can be found in Table~\ref{tab:SMEFIT}.

A few comments are in order concerning the comparison of these bounds. The combinations involving the coefficients $C_{qq}^{(1/3)}$ are constrained up to the $10^{-2}$ level, due to the $\Delta F=2$ meson mixing observables that strongly constrain their sum. This result remarkably improves the bounds derived by the ATLAS and CMS direct measurements of top quark cross-sections by 2-3 orders of magnitude. An improvement is also clear for the combinations of the Higgs-Top operators $C_{Hq}^{(1/3)}$ and $C_{Hu}$ ($c_{Ht}$), mostly constrained in this analysis by the electroweak and Higgs observables discussed in Section~\ref{sec:EWHiggs}.
The $C_{uG}$ ($c_{tG}$) coefficient represents the only exception to the general improvement trend observed in almost all the cases. However, as discussed in Ref.~\cite{Ethier:2021bye}, the numerical fit of $C_{uG}$ performed through the SMEFiT routine seems to be unstable, 
so that we do not consider this direct bound reliable (see Section 5.3 of Ref.~\cite{Ethier:2021bye} for an extended discussion).

\subsection{Two-parameters fits}\label{sec:twoparsanalysis}

In this Section we carry out two-parameter fits, i.e. two Wilson coefficients at a time are allowed to vary under the assumption that they are generated at the same scale, while all the others are set to zero. These analyses can provide useful information on the interplay between pairs of coefficients, highlighting what are the most constrained combinations and giving thus intuition on their correlations. This is also the first step towards a UV interpretation, as WC's allowed regions can be contrasted with the relations predicted by specific UV scenarios.
Among all the possible pairs we only show some interesting cases, e.g. when the importance of combining different sectors is highlighted and/or when the pair can be interesting from a NP perspective.

\begin{figure}[t]
    \centering
    \includegraphics[height=6.5cm]{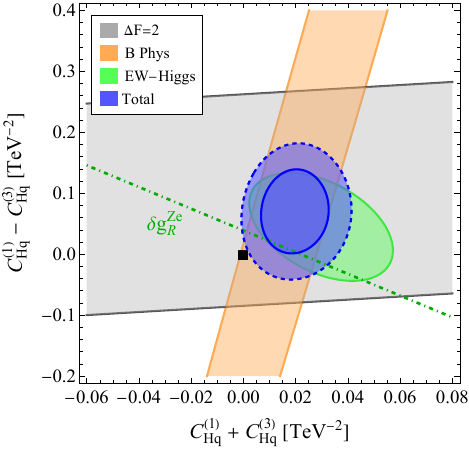}
    \hfill
    \includegraphics[height=6.5cm]{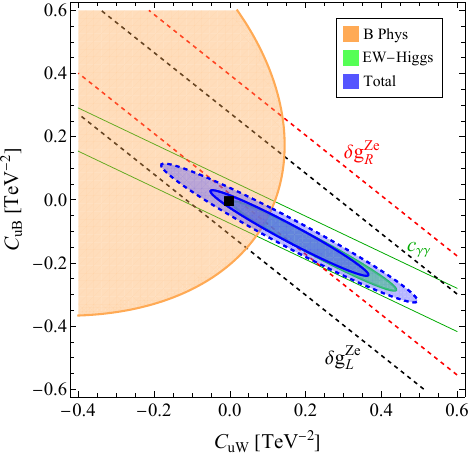}
    \caption{Left panel: Higgs-Top coefficients $C_{Hq}^{(+)}$ vs $C_{Hq}^{(-)}$. Right panel: dipole coefficients $C_{uW}$ vs $C_{uB}$.}
    \label{fig:CHq13}
\end{figure}

\begin{figure}[t]
    \centering
    \includegraphics[height=6.5cm]{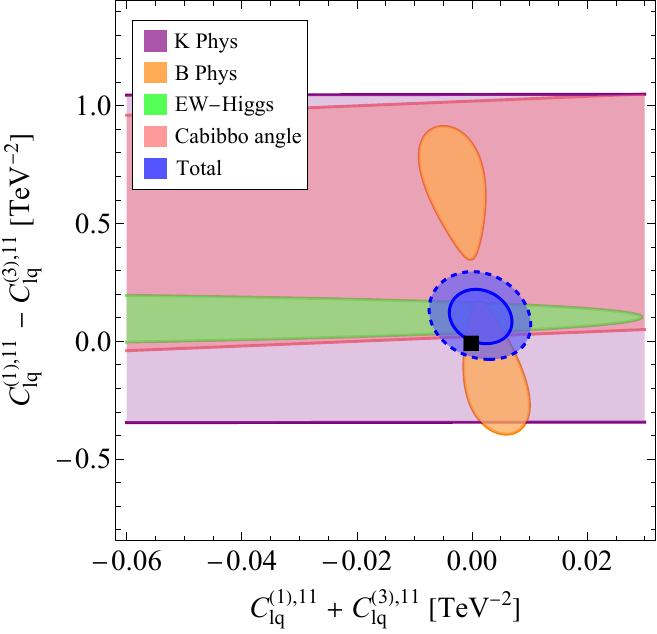}
    \hfill
    \includegraphics[height=6.5cm]{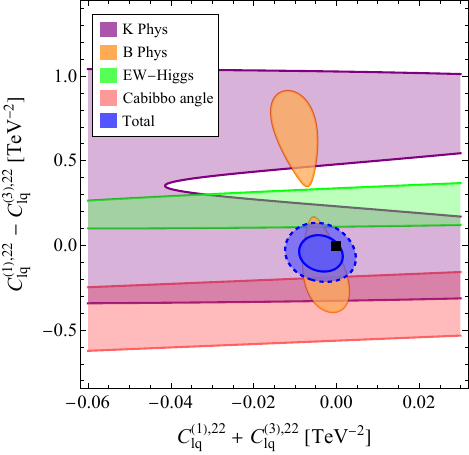}
    \caption{Semileptonic $C_{\ldoublet q}^{(+)}$ vs $C_{\ldoublet q}^{(-)}$ coefficients involving electrons and muons.}
    \label{fig:Clq1122}
\end{figure}

\begin{figure}[t]
    \centering
    \includegraphics[height=6.5cm]{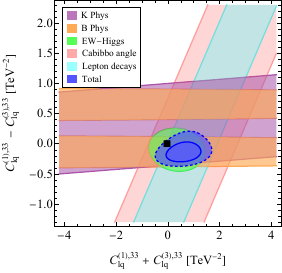}
    \hfill
    \includegraphics[height=6.5cm]{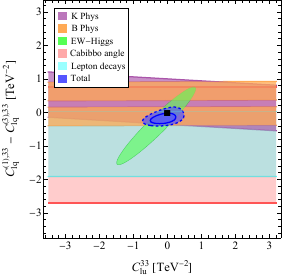}
    \caption{Left panel: semileptonic $C_{\ldoublet q}^{(+)}$ vs $C_{\ldoublet q}^{(-)}$ coefficients involving tau leptons. Bounds from the EW sector are improved by the synergy with the other sectors. Right panel: Semileptonic $C_{\ldoublet q,33}^{(-)}$ vs 
$C_{\ldoublet u,33}$ coefficients.}
    \label{fig:Clq1m3vsClu33}
\end{figure}

\begin{figure}[t]
    \centering
    \includegraphics[width=0.46\textwidth]{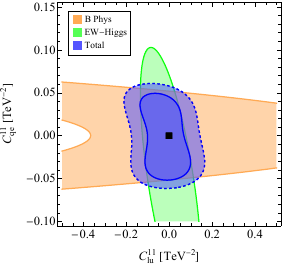}
    \hfill
    \includegraphics[width=0.46\textwidth]{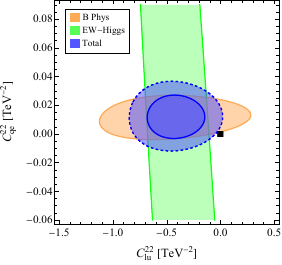}
    \caption{Here we show the semileptonic coefficients $C_{\ldoublet u}$ vs. $C_{q e}$ for electrons (left) and muons (right).}
    \label{fig:CluCqe}
\end{figure}

\begin{figure}[t]
    \centering
    \includegraphics[height=6.5cm]{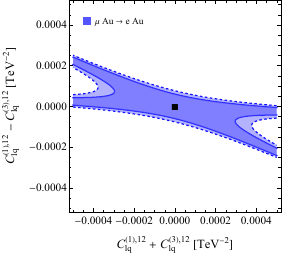}
    \hfill
    \includegraphics[height=6.5cm]{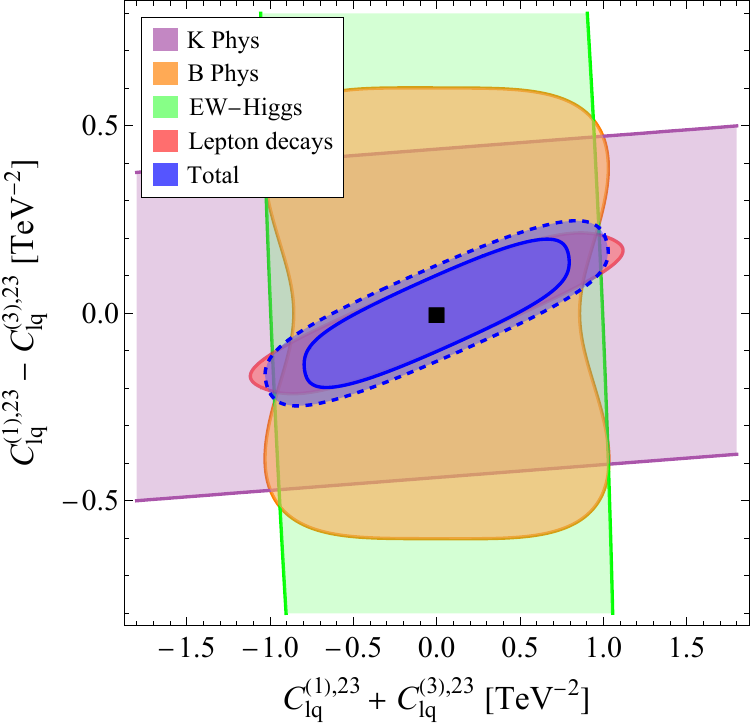}
    \caption{Two-parameters fit for the LFV semileptonic coefficients $C_{\ldoublet q,12}^{(+/-)}$ and $C_{\ldoublet q,23}^{(+/-)}$. The stronger constraints come from the LFV decay modes of tau lepton.}
    \label{fig:LFVcoeff}
\end{figure}

\begin{figure}[t]
\centering
    \includegraphics[width=0.46\textwidth]{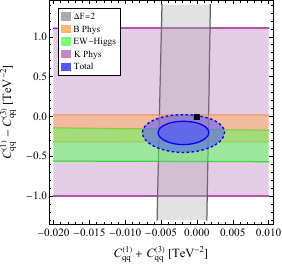}
    \hfill
    \includegraphics[width=0.46\textwidth]{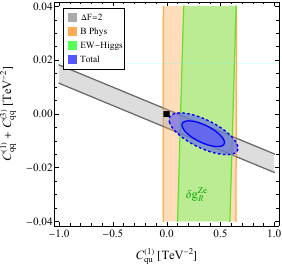}
    \caption{Left panel: four-quarks $C_{qq}^{(1)+(3)}$ vs $C_{qq}^{(1)-(3)}$ coefficients. $\Delta F=2$ processes constraint the $C_{qq}^{(+)}$ combination up to the $10^{-3}$ level. The difference is constrained by the other sectors. Right panel: $C_{qu}^{(1)}$ and $C_{qq}^{(+)}$.}
    \label{fig:4Q}
\end{figure}

The results of two-parameters fits are illustrated in Figs.~\ref{fig:CHq13} - \ref{fig:4Q}. We display the $68\%$CL regions obtained when different subsets of data are taken as input and the $68\%$ and $95\%$CL regions resulting from the whole set of observables. 
In order to better understand the underlying phenomenology, in some plots we also show single-observable $1\sigma$ contours.
In this way, we can investigate the constraining power of each sector/observable on coefficients pairs.

To start with, in Fig.~{\ref{fig:CHq13}} we show the allowed regions for the $C_{Hq}^{(+)}$ and $C_{Hq}^{(-)}$ coefficients. Noteworthy, constraints from the $B$ sector, mostly due to leptonic $B\to \ell \ell$ and radiative decays, are competitive to the EW and Higgs bounds, resulting in a (slightly more than) 2$\sigma$ deviation from the SM prediction. This pull is mostly due to the $Z$ boson coupling to right-handed electrons, $\delta g_R^{Ze} = (-7.3 \pm 4.4) \times 10^{-3}$  \cite{Falkowski:2019hvp}, as we show in the plot with the dot-dashed green line marking the $1\sigma$ bound. In the right panel, as an example of dipole operators, we show the $C_{uW}$ vs $C_{uB}$ plane. As expected, a strong constraint is set on the photon direction, whose main responsible is the $c_{\gamma \gamma}$ coefficient discussed in Section \ref{sec:EWHiggs}. The dominant constraints on the orthogonal direction are set again by the electroweak sector, as a combined effect of the deviations in the Z boson coupling to leptons and other (pseudo-)observables. 

In Fig.~\ref{fig:Clq1122} and the left panel of Fig.~\ref{fig:Clq1m3vsClu33}, we show the two-parameters fits for the $C_{\ldoublet q}^{(+/-),\alpha\alpha}$ coefficients. The existence of flat directions points out the importance of performing SMEFT analysis using various datasets, exploiting the complementarity between different physics sectors to constraint the EFT space. Focussing on $B$-physics, the strong bound on the positive combinations $C_{\ldoublet q}^{(+), \alpha \alpha}$, with $\alpha=1,2$, can be ascribed to the $R_{K^{(*)}}$ measurements. In addition, the recently announced measurement of $R_{K}^{\nu}$, with a pull from the SM, drives the allowed region for the negative combination, splitting the allowed band in two regions depending on the interference with the SM. Interestingly, in case of electrons and muons (Fig.~\ref{fig:Clq1122}) the deviation in $R_{K}^{\nu}$ could be compatible with the Cabibbo angle anomaly but is in some tension with the bounds from EW precision data, which drives the global fit towards the SM.
In case of tau leptons (Fig.~\ref{fig:Clq1m3vsClu33}), instead, the looser EW bounds and the vanishing contribution to $R_K$ allow the global analysis to overlap with the preferred region from $R_{K}^{\nu}$.
This feature is visible also in the right panel of Fig.~\ref{fig:Clq1m3vsClu33}, where we display the allowed regions in the plane $C_{\ldoublet q}^{(-),33}$ vs $C_{\ldoublet u}^{33}$. 
The semileptonic coefficients $C_{\ldoublet q}^{(1,3),33}$ will be further discussed in Section~\ref{sec:UVmodels}, where the scenario of a single scalar leptoquark $S_1$ coupling tau leptons and top quarks is explored.
In Fig.~\ref{fig:CluCqe} we show the fit in the plane of $C_{\ldoublet u}$ vs. $C_{q e}$ for both electrons (left panel) and muons (right panel). In both cases the global fit shows an interesting interplay of $B$ and EW physics.


In Fig.~\ref{fig:LFVcoeff}, we show the pairwise fits on the LFV coefficients $C_{\ldoublet q,12}^{(+/-)}$ and $C_{\ldoublet q,23}^{(+/-)}$. In the first case, the global fit is controlled entirely by the $\mu \textrm{Au} \to e \textrm{Au}$ measurement, as expected from Table~\ref{tab:indivsemilfv}. In the second case, the sensitivity of the coefficients is mostly driven by the tau LFV decays discussed in Section~\ref{sec:cLFV}. However, LFV decay modes of $B$ and $K$ mesons, as well bounds on LFV $Z$ decays, can improve the constraints. This discussion also applies to the $C_{\ldoublet q,13}^{(+/-)}$ coefficients, whose bounds are very similar to the ones derived for $C_{\ldoublet q,23}^{(+/-)}$.

Regarding four-quark operators, it is clear from the left panel of Fig.~\ref{fig:4Q} that the combinations $C_{qq}^{(+)}$ and $C_{qq}^{(-)}$ are the relevant degrees of freedom in the EFT space. In particular, the sum is severely constrained by meson oscillations, while the orthogonal combination is mainly constrained by $B_s \to \mu^+ \mu^-$, kaon decays and EW and Higgs observables. In the right panel we display the $C_{qu}^{(1)}$ vs $C_{qq}^{(+)}$ pair. We illustrate the effect of the $\delta g_{R}^{Ze}$ measurement, responsible again for the main discrepancy in the EW sector.  The resulting allowed region shows a tension from the SM value within of slightly more than two sigmas.

\subsection{Gaussian fit with no semileptonic operators}\label{sec:gaussianfits}

Plots pose a limitation in only allowing fits to pairs of coefficients. There are two main possibilities to go beyond that: providing the complete likelihood or employing the Gaussian approximation around the global maximum of the likelihood to derive a multi-dimensional fit. While the former solution is more general, it is not a fit and the complexity of our analysis make the resulting numerical function unwieldy to publish in a paper.\footnote{However, we can provide it in electronic form upon request.} Therefore, in this work we opt to perform a multi-dimensional Gaussian fit.

We consider all the operators in Table~\ref{table:Operators} except the semileptonic ones:\footnote{An analogous Gaussian fit including also semileptonic operators does not provide a physically meaningful result. This is due to the non-Gaussianities, that become very important in light of the mild deviations from the SM expectation present in some observables. In this case, performing a Gaussian expansion around the global minimum provides a bad approximation to the full likelihood.}
\be
    \vec{C} = (C_{qq}^{(+)}, C_{qq}^{(-)}, C_{uu}, C_{qu}^{(1)}, C_{qu}^{(8)}, C_{Hq}^{(+)}, C_{Hq}^{(-)}, C_{Hu}, C_{uH}, C_{uG}, C_{uW}, C_{uB})~.
\ee
The best-fit point improves the $\chi^2$ from the SM value by $\Delta \chi^2 \equiv \chi^2_{\rm SM} - \chi^2_{\rm best-fit} \approx 10.9$. This comes mostly from the EW-Higgs sector, however it is a result in mild improvements in several observables rather than a resolution of a specific large anomaly.
The fit presents some almost flat directions, which imply correlations among some coefficients very close to $\pm 1$. For this reason, we report the result in terms of the eigenvectors of the Hessian matrix around the best-fit minimum:
\be
    \chi^2 = \chi^2_{\rm best-fit} + (C_i - \mu_{C_i}) (\sigma^2)^{-1}_{ij} (C_j - \mu_{C_j}) =
        \chi^2_{\rm best-fit} + \frac{(K_i - \mu_{K_i})^2}{\sigma_{K_i}^2}.
\ee
\begin{table}[t]
\centering
\begin{tabular}{|c|c||c|c|}
\hline
    Coefficient & Gaussian fit [$\TeV^{-2}$] & Coefficient & Gaussian fit  [$\TeV^{-2}$] \\ \hline
        $K_1$ & $0.0019 \pm 0.0023$     & $K_7$ & $0.54 \pm 0.79$ \\
        $K_2$ & $0.0179 \pm 0.0083$     & $K_8$ & $0.74 \pm 0.88$ \\
        $K_3$ & $-0.002 \pm 0.015$      & $K_9$ & $-0.8 \pm 1.3$ \\
        $K_4$ & $-0.016 \pm 0.021$      & $K_{10}$ & $-0.7 \pm 1.8$ \\
        $K_5$ & $0.044 \pm 0.029$       & $K_{11}$ & $12 \pm 13$ \\
        $K_6$ & $-0.30 \pm 0.38$        & $K_{12}$ & $-11 \pm 16$ \\ \hline
\end{tabular}
\caption{Result of our multidimensional gaussian fit of non-leptonic coefficients, in terms of the eigenvectors of the Hessian matrix.}
\label{tab:GaussianFitK}
\end{table}
The results for the best-fit values $\mu_{K_i}$ and uncertainties $\sigma_{K_i}$ are reported in Table~\ref{tab:GaussianFitK},
while the rotation matrix from these coefficients to our $C_i$ is given by $\vec{K} = U_{KC} \vec{C}$, with
\be
{\tiny 
U_{KC} = 
\left(
\begin{array}{cccccccccccc}
 -1.00 & 0.000 & 0.000 & -0.016 & -0.004 & -0.004 & 0.021 & -0.001 & 0.000 & 0.000 & 0.000 & 0.000 \\
 -0.005 & -0.089 & -0.015 & 0.058 & 0.000 & 0.984 & -0.004 & -0.117 & 0.000 & -0.009 & 0.044 & 0.063 \\
 0.004 & 0.011 & -0.039 & 0.018 & -0.001 & -0.1 & 0.145 & -0.28 & 0.015 & -0.494 & 0.447 & 0.667 \\
 -0.007 & -0.013 & 0.09 & -0.053 & -0.003 & 0.081 & -0.316 & 0.64 & 0.024 & -0.673 & -0.126 & -0.059 \\
 0.005 & 0.007 & -0.074 & 0.042 & -0.002 & -0.025 & 0.259 & -0.525 & 0.025 & -0.548 & -0.213 & -0.55 \\
 -0.004 & -0.041 & 0.025 & 0.067 & 0.006 & -0.004 & -0.128 & 0.084 & 0.006 & 0.022 & 0.853 & -0.492 \\
 -0.006 & -0.137 & 0.078 & 0.196 & 0.96 & -0.017 & 0.09 & 0.047 & -0.065 & -0.007 & -0.017 & 0.008 \\
 0.002 & -0.349 & -0.006 & 0.646 & -0.248 & -0.029 & 0.545 & 0.318 & 0.014 & 0.001 & -0.012 & 0.006 \\
 0.005 & 0.007 & 0.028 & -0.138 & 0.077 & 0.017 & 0.145 & 0.06 & 0.973 & 0.037 & 0.017 & -0.003 \\
 0.023 & 0.221 & 0.074 & -0.569 & 0.053 & 0.092 & 0.684 & 0.292 & -0.212 & -0.002 & 0.095 & -0.057 \\
 0.006 & -0.798 & 0.451 & -0.364 & -0.071 & -0.059 & -0.038 & -0.122 & -0.039 & 0.000 & -0.012 & 0.007 \\
 -0.004 & 0.404 & 0.876 & 0.235 & -0.058 & 0.025 & 0.017 & -0.093 & 0.013 & 0.000 & -0.01 & 0.006 \\
\end{array}
\right) .
}
\ee
The last two coefficients, $K_{11}$ and $K_{12}$, correspond to two flat directions with only very weak constraints from our observables, given approximately by:
\be\begin{split}
    K_{11} &\approx - 0.80 C_{qq}^{(-)} + 0.45 C_{uu} - 0.36 C_{qu}^{(1)} - 0.12 C_{Hu} + \ldots ~, \\
    K_{12} &\approx + 0.40 C_{qq}^{(-)} + 0.88 C_{uu} + 0.24 C_{qu}^{(1)} - 0.09 C_{Hu} + \ldots ~.
\end{split}\ee
\sloppy Using the naive power counting of Ref.~\cite{Gavela:2016bzc}, one may estimate the allowed EFT hyper-volume spanned by the studied non-leptonic Wilson coefficients as $V_{\mathrm{EFT}}\sim \frac{\pi^6}{720}\left(\frac{(4\pi)}{\mathrm{TeV}^2}\right)^3\left(\frac{(4\pi)^2}{\mathrm{TeV}^2}\right)^{8}\left(\frac{(4\pi)^3}{\mathrm{TeV}^2}\right)$, assuming $\Lambda = 1\,\TeV$. 
From Table~\ref{tab:GaussianFitK}, experimental constraints restrict the potential SMEFT hyperspace to a very tiny fraction of its volume, $\sim 10^{-31}$.

\section{Applications for UV models}
\label{sec:UVmodels}

In this Section we apply our global analysis to two UV scenarios of New Physics at the TeV scale, coupled mainly to the top quark.
The first is a simple minimal extension of the SM by one scalar leptoquark $S_1 \sim ({\bf \bar{3}}, {\bf 1})_{+1/3}$ coupled only to the third generation quark and lepton doublets.
The second scenario is an application guided by the Cabibbo anomaly, it includes the scalar leptoquark $S_3 \sim ({\bf \bar{3}}, {\bf 3})_{+1/3}$ as well as the vector leptoquark $U_1 \sim ({\bf 3}, {\bf 1})_{+2/3}$.

\subsection{Single leptoquark $S_1$}

\begin{figure}
    \centering
    \includegraphics[width=0.5\textwidth]{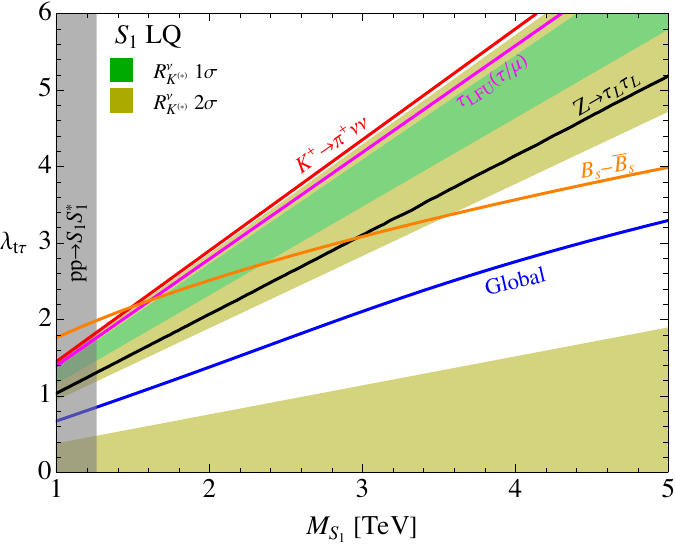}
    \caption{Present constraints on the $S_1$ leptoquark model, coupled only to third generation quark and lepton doublets. The green (yellow) region is preferred at $1\sigma$ ($2\sigma$) by $R^\nu_{K^{(*)}}$, while the gray region is excluded at 95\%CL from direct searches at LHC. The region above each line is excluded at 95\%CL by the corresponding observable, or by our global analysis (blue line).}
    \label{fig:S1_top_tau}
\end{figure}

Let us consider the scalar leptoquark $S_1 \sim ({\bf \bar{3}}, {\bf 1})_{+1/3}$, coupled only to the third generation of quark and lepton $SU(2)_L$ doublets:
\be
    \LL \supset \lambda_{t \tau} \, \bar{q}_3^c i \sigma_2 \ldoublet_3 \, S_1 + \text{h.c.}~,
\ee
where $q_3 = (t_L, \, V_{tj} d_L^j)$, $\ldoublet_3 = (\nu_{\tau}, \, \tau_L)$ and $\sigma_i$ are the Pauli matrices.
While the recent interest in this leptoquark stems from its ability to address the $R(D^{(*)})$ and $(g-2)_\mu$ anomalies, our setup will not address either of those since in our case $S_1$ doesn't couple to the charm nor the muon.
Indeed, our goal is only to showcase a simple application of the global analysis.

Matching at tree level this Lagrangian to the SMEFT gives \cite{Dorsner:2016wpm,deBlas:2017xtg}
\be
    C_{\ldoublet q}^{(1),33} = - C_{\ldoublet q}^{(3),33} = \frac{|\lambda_{t \tau}|^2}{4 M_{S_1}^2}~.
\ee
For precision studies of such scenarios, however, the tree level matching is not sufficient since one-loop contributions can be important for some observables, as discussed in Refs.~\cite{Arnan:2019olv,Crivellin:2019dwb,Saad:2020ihm,Crivellin:2020ukd,Gherardi:2020qhc,Marzocca:2021miv}.
For our goal it is sufficient to add the leading contributions to four-quark operators, which induce contributions to meson-mixing observables. From the complete one-loop matching of this scalar leptoquark to SMEFT, done in Ref.~\cite{Gherardi:2020det}, we extract the relevant contribution:
\be
    C_{qq}^{(1)} = C_{qq}^{(3)} = - \frac{|\lambda_{t \tau}|^4}{256 \pi^2 M_{S_1}^2}~.
\ee
These coefficients are generated at the $M_{S_1}$ scale, that we assume to be near 1~TeV, so that we neglect the RG evolution between $M_{S_1}$ and 1~TeV.

The constraints from the global analysis, as well as from the most relevant observables, in the plane of the $\lambda_{t \tau}$ coupling vs. the leptoquark mass, are shown in Fig.~\ref{fig:S1_top_tau}. The shaded gray region is excluded by ATLAS from leptoquark pair-production searches \cite{ATLAS:2021jyv}.
The regions preferred at 1$\sigma$ and 2$\sigma$ by the $R^\nu_{K^{(*)}}$ combination, Eq.~\eqref{eq:RnuKKstComb}, are show in green and yellow, respectively. Interestingly, the intermediate white region is disfavored at the 95\%CL by $R^\nu_{K^{(*)}}$ due to the negative interference with the SM, which suppresses the branching ratio below the $2\sigma$ level.
We observe an interesting interplay of constraints from different classes of observables: electroweak precision data and $\tau$ physics, meson mixing, $B$ and kaon rare decays, as well as direct searches from LHC.

\subsection{Two leptoquarks for the Cabibbo anomaly?}
\label{sec:Cabibbo}

In this section we illustrate how our combined likelihood may also be used to check whether experimental results in tension with the SM predictions can be partially accommodated in top-philic extensions (and which ones), taking into account the restrictions imposed by experimental results in other sectors. Here we focus on the longstanding tensions within the SM involving the determination of the inputs of the first row of the CKM matrix, known as Cabibbo anomalies.

Regardless of whether nonzero values are allowed or not for the remaining Wilson coefficients, the likelihood of the Cabibbo angle observables displays a $~3\, \sigma$ preference for a nonzero value of $C_{\ldoublet q}^{(3),22}$ at $\Lambda=1\, \mathrm{TeV}$,
\begin{equation}\label{eq:cabanom}
[C_{\ldoublet q}^{(3),22}]^{\mathrm{Cabibbo}}=(0.19 \pm 0.06) \, \mathrm{TeV}^{-2} \, .
\end{equation}
Let us first develop on where this preferred nonzero value comes from. $C_{\ldoublet q}^{(3),\alpha\alpha}$ induces an unusually large $C_{H\ldoublet}^{(3),\alpha \alpha}$ at the EW scale, due to a mixing that involves a double top Yukawa insertion. The leading log approximation, which one may expect to give a first approximation for $\Lambda \gtrsim  \mathrm{TeV}$, is displayed in Eq.~(\ref{eq:leadlogcab}). After EWSB $C_{H\ldoublet}^{(3),\alpha \alpha}$ induces, at tree level, a re-scaling of the corresponding lepton charged current vertices, modifying the ratios of muon, beta and kaon decays. As a consequence, the corresponding apparent unitarity relation is\footnote{$V_{ub}$ has a completely negligible numerical role in this relation.}
\begin{align}\nonumber
\Delta_{\mathrm{CKM}}&\equiv |V_{ud}^\beta|^2+ |V_{us}^{K_{\ell 3}}|^2-1\approx -2 v^2\, (|V_{ud}|^2 C_{H\ldoublet}^{(3),22} - |V_{us}|^2(C_{H\ldoublet}^{(3),\ell\ell} -C_{H\ldoublet}^{(3),11} -C_{H\ldoublet}^{(3),22}) )
\\&\approx-2 v^2\,  C_{H\ldoublet}^{(3),22} \sim - \left[\frac{N_c}{2\pi^2}\frac{m_t^2}{\Lambda^2_{\mathrm{UV}}}\log\frac{\Lambda_{\mathrm{UV}}}{M_Z} \,  \right]\Lambda_{\mathrm{UV}}^2 C_{\ldoublet q}^{(3),22}(\Lambda^2_{\mathrm{UV}}) \, .
\end{align}
Experimentally $\Delta_{\mathrm{CKM}}$ is known beyond the per-mil level and gives the main pull behind the result of Eq.~(\ref{eq:cabanom}).

If we set all the parameters to zero except for $C_{\ldoublet q}^{(3),22}$ for the EW/Higgs likelihood, we also obtain a slight preference for nonzero values, but with the opposite sign,
\begin{equation}
[C_{\ldoublet q}^{(3),22}]^{\mathrm{EW/Higgs}}=(-0.11\pm 0.06) \, \mathrm{TeV}^{-2} \, .
\end{equation}
This is to be expected, since it is well known that explaining $\Delta_{\mathrm{CKM}}$ through increasing the muon decay matrix element is in tension with fits in other EWPOs, e.g. see Refs.~\cite{Belfatto:2019swo,Cirigliano:2022qdm,Crivellin:2022ctt}. We may yet respect CKM unitarity and decrease the muon decay rate with respect to the SM expectation by breaking LFU, with $0<C_{H\ldoublet}^{(3),22}<-C_{H\ldoublet}^{(3),11}$. Let us then allow for nonzero values at the $\mathrm{TeV}$ scale for $C_{\ldoublet q}^{(3),\alpha\alpha}$ for the first two lepton families and also for the corresponding $C_{\ldoublet q}^{(1),\alpha\alpha}$, typically induced in UV models by the same couplings. The minimum $\chi^2$ follows the expected pattern, plus a preference for $C_{\ldoublet q}^{(1),11}\approx C_{\ldoublet q}^{(3),11}$ and $C_{\ldoublet q}^{(1),22}\approx 3 C_{\ldoublet q}^{(3),22}$. Both these relations and the obtained signs happen to match the couplings induced by top-philic $U_{1}$ and $S_3$ leptoquarks coupled, respectively, to the first and the second lepton family,
\begin{equation}
\mathcal{L}\supset \lambda_{U_1} \; \bar{q}_{3} \gamma_{\mu} \ldoublet_1 \, U_1^{\mu} 
+ \lambda_{S_3} \;  \bar{q}_3^c i \sigma_2 \sigma_a \ldoublet_2 \, S_3^a
 + \mathrm{h.c.} \, ,
\end{equation}
giving \cite{Dorsner:2016wpm,deBlas:2017xtg}
\begin{equation}
C_{\ldoublet q}^{(3),11}=-\frac{|\lambda_{U_1}|^2}{2 M_{U_1}^2} \, , \quad C_{\ldoublet q}^{(3),22}=\frac{|\lambda_{S_3}|^2}{4 M_{S_3}^2} \, ,
\end{equation}
plus the relations with $C_{\ldoublet q}^{(1)}$ above. Imposing them as strict equalities one finds, in this two-parameter scenario,
\begin{equation}
C_{\ldoublet q}^{(3),11}=(-0.19\pm 0.06) \, \mathrm{TeV}^{-2} \, , \, C_{\ldoublet q}^{(3),22}=(0.14\pm 0.04) \, \mathrm{TeV}^{-2} \, .
\end{equation}
With respect to the SM, this minimum has a $\Delta \chi^2=8.0$ preference for the Cabibbo sector and a $\Delta \chi^2=5.7$ for the EW/Higgs one.

\begin{figure}[t]
    \centering
    \includegraphics[width=0.46\textwidth]{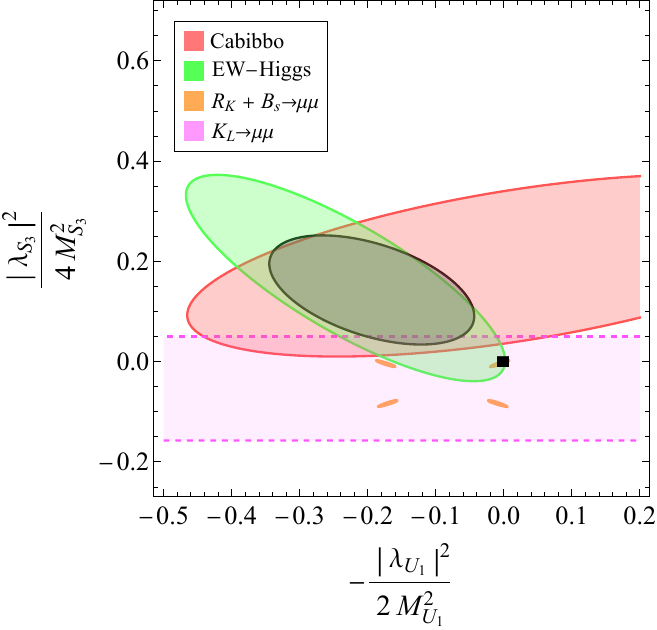}
    \caption{Allowed regions ($95 \%\, \mathrm{CL}$) for the studied leptoquark couplings (see text) from the Cabibbo and the EW Higgs sector and combined, in black. Masses are in TeV units. The preferred nonzero values can be ruled out from $B$-physics and $K$-physics observables, unless one defines the top-philic set up in the down-quark basis.}
    \label{fig:leptocabibbo}
\end{figure}

It is only when adding constraints from $B$ and $K$ physics observables when this scenario becomes strongly disfavored. In a generic top-philic set-up defined in any specific flavor basis in which $C_{\ldoublet q}^{(1,3),ii}$ are induced, FCNCs are generated when rotating to the mass basis. In the down sector FCNCs are very strongly constrained by the processes studied in this work, as we have explicitly shown when defining the top-philic condition in the up quark basis. Indeed we show in Figure~\ref{fig:leptocabibbo} how including those constraints we can rule out this, and practically any, top-philic explanation to the Cabibbo anomaly.

Let us finish this section by remarking that there is, however, a possible way out to the strong constraints on FCNCs coming from $K$ and $B$ physics. Assuming that there exists some, admittedly bizarre from our infra-red perspective, mechanism in the UV theory to select a top-philic set-up in the only basis where this operator does not induce tree-level FCNCs in the down sector, i.e. the down-quark basis, the previous constraints from $K$ and $B$ physics do not hold, while the constraints on Cabibbo angle and EW/Higgs would give, in a first approximation, the same results, since the leading mechanism to modify the SM, through modifying lepton vertices, is largely independent on quark-basis rotation. In that scenario, this leptoquark model may still be a feasible (partial) solution to the Cabibbo anomaly. Complementary constraints on the space of parameters may come from $D-\bar{D}$ mixing (since FCNCs are yet induced in the up sector) and direct leptoquark searches, bounding the possible masses from below. Further studying this scenario is, however, well beyond the goals and the scope of this work.

\section{Conclusions}
\label{sec:conclusions}

This paper provides a global analysis of indirect constraints on SMEFT operators involving the top quark. In fact, several motivated UV scenarios predict new physics coupled mostly with the top. 
Our global analysis combines a large number of observables that do not involve directly the top quark. These include: $B$, $K$ and $\tau$ decays, meson mixing observables, magnetic moments of leptons, measurements of the Cabibbo angle, EW precision measurements, Higgs physics, and LFV tests.
Top quark operators contribute to these observables either at the tree-level via the $SU(2)_L$ connection with left-handed down quarks or via loop effects. Assuming that top quark operators are generated at the 1TeV scale, we evolve them down to the scale relevant for each observable using the relevant RG and matching equations.
The result of this process is a global likelihood expressed in terms of the high-scale Wilson coefficients.

Using this likelihood we derive global indirect constraints on single coefficients, comparing the result with the bounds obtained from LHC analysis of processes involving top quarks. In all cases we find that the indirect constraints are stronger than the direct ones, often by several orders of magnitude.
We then perform several 2D fits to study interesting correlations among coefficients and the complementarity between different observables, that showcase the importance of performing such global analyses.
We also perform a multi-dimensional Gaussian fit of all the coefficients of non-leptonic operators, that can be useful to identify the directions in parameter space that have strongest or weakest constraints. Indeed, among these coefficients we identify two almost-flat directions.

Finally, as examples of possible applications of our analysis we study the EFT coefficients generated by two simple UV models. The first includes a scalar singlet leptoquark coupled only to the third generation of quark and lepton doublets. We show how the different indirect constraints one can derive from different sectors combine in providing strong bounds on the model, and how these are complementary to direct searches at the LHC.
As a second example we study a two-leptoquark scenario inspired by the Cabibbo anomaly, with only two free couplings. This model accommodates the anomaly and improves the fit quality for the rest of EW-Higgs precision observables. However, $B$ and kaon physics constraints are able to completely rule out this scenario unless a very specific flavor alignment is imposed.

Our results show how global indirect constraints on top quark operators can be powerful in constraining new physics scenarios coupled mainly to the top quark. Such constraints are often much stronger, or in any case complementary, to those derived from high-energy top quark physics at colliders.
Furthermore, several of the observables providing the strongest indirect constraints are expected to be measured with a substantially better precision in the future: $R^\nu_{K^{(*)}}$ and $\tau$ decays at Belle-II, $R_{K^{(*)}}$ and $B_s \to \mu \mu$ by LHCb, ATLAS and CMS, $K\to\pi\nu\nu$ by NA62, and several experiments are expected to improve the sensitivity on $\mu \to e$ LFV by several orders of magnitude.
All this will further increase the relevance of such global analysis of indirect bounds in the future.

\section*{Acknowledgments}

We thank Adam Falkowski for providing updated versions of the likelihoods in the EW-Higgs and Cabibbo angle sectors and Benedetta Belfatto for useful discussions. DM and ARS acknowledge partial support by MIUR grant PRIN 2017L5W2PT.

\appendix

\section{Observables}
\label{App:Observables}

\subsection{$B \to K^{(*)}\nu\bar{\nu}$ and $K \to \pi\nu\bar{\nu}$}

The Standard Model prediction and SMEFT parametrization of dineutrino modes $B \to K^{(*)}\nu\bar{\nu}$ are deeply discussed in Ref.~\cite{Buras:2014fpa}. The effective Hamiltonian for these processes reads 
\begin{equation}
\begin{split}
\mathcal{H}_{eff}&=-\frac{4G_F}{\sqrt{2}}\frac{\alpha}{4\pi}\sum_{i,j}V_{ti}^{*}V_{tj}\left(C_L^{ij\alpha\beta}O_L^{ij\alpha\beta}+C_R^{ij\alpha\beta}O_R^{ij\alpha\beta}  \right)\\
&=-\sum_{i,j}\left(\frac{1}{2} \lwc{\nu d}{LL}[V][\alpha\beta ij] O_L^{ij\alpha\beta} + \frac{1}{2}  \lwc{\nu d}{LR}[V][\alpha\beta ij] O_R^{ij\alpha\beta} \right)~,
\end{split}
\end{equation}
where
\begin{equation}
O_L^{ij\alpha\beta}=(\bar{d_i}\gamma_{\mu}P_{L}d_j)(\bar{\nu}_\alpha\gamma^{\mu}(1-\gamma_5)\nu_\beta)\,,\quad O_R^{ij\alpha\beta}=(\bar{d_i}\gamma_{\mu}P_{R}d_j)(\bar{\nu}_\alpha\gamma^{\mu}(1-\gamma_5)\nu_\beta)\,.
\end{equation}
and $L_{\nu d}^{V,LL(R)}$ are LEFT coefficients given in Table~\ref{tab:oplist1}. The ratios
\begin{equation}
    R_{K}^{\nu}=\frac{\mathcal{B}(B\to K\nu\bar{\nu})}{\mathcal{B}(B\to K\nu\bar{\nu})_{\mathrm{SM}}}\,, \quad R_{K^*}^{\nu}=\frac{\mathcal{B}(B\to K^*\nu\bar{\nu})}{\mathcal{B}(B\to K^*\nu\bar{\nu})_{\mathrm{SM}}}\,,
\end{equation}
can be expressed in terms of two parameters $\epsilon>0$ and $\eta \in [-1/2,1/2]$ as:
\begin{equation}
\begin{split}
R_{K}^{\nu}=\sum_{\alpha\beta}&\frac{1}{3}(1-2\eta_{\alpha,\beta})\epsilon^2_{\alpha\beta}\,, \quad R^{\nu}_{K^*}=\sum_{\alpha\beta}\frac{1}{3}(1+\kappa_{\eta}\eta_{\alpha,\beta})\epsilon_{\alpha\beta}^2,\\
    &\epsilon_{\alpha\beta}=\frac{\sqrt{|C_{L,\mathrm{SM}}^{sb}\delta_{\alpha\beta}+C_L^{sb\alpha\beta}|^2 + |C_R^{sb\alpha\beta}|^2}}{|C_{L,\mathrm{SM}}^{sb}|}  ,\\
    & \eta_{\alpha\beta}=-\frac{\text{Re}\left[\left(C_{L,\mathrm{SM}}^{sb}\delta_{\alpha\beta}+C_L^{sb\alpha\beta} \right)\left(C_R^{sb\alpha\beta}\right)^* \right]}{|C_{L,\mathrm{SM}}^{sb}\delta_{\alpha\beta}+C_L^{sb\alpha\beta}|^2+|C_R^{sb\alpha\beta}|^2}\,,
\end{split}
\end{equation}
where the parameter $\kappa_{\eta}$ depends on form factors. There are not isospin asymmetries between the charged and neutral meson decays to neutrinos, so that the only difference between $\mathcal{B}(B^{\pm} \to K^{\pm (*)}\nu\bar{\nu})$ and $\mathcal{B}(B^0 \to K^{0(*)}\nu\bar{\nu})$ is in the lifetime $\tau_{B^{\pm}}$ and $\tau_{B^0}$, that cancel in the ratio. Remarkably, in our setup the coefficient $C_R$ is zero at the level of dimension-six SMEFT contributions, so that $\eta_{\alpha\beta}=0$ and then the theoretical predictions for $R_{K}^{\nu}$ and $R^{\nu}_{K^*}$ coincide.\newline
For kaons, we use directly the branching ratios. Since $C_R=0$ in our setup, we can just write them in terms of the SM values rescaling the $C_L$ coefficients
\begin{equation}
\begin{split}
\mathcal{B}(K^+\to\pi^+\nu\bar{\nu})_{th} &= \mathcal{B}(K^+\to\pi^+\nu_e\bar{\nu}_e)_{\textrm{SM}}\sum_{\alpha,\beta=1,2}\left|\delta_{\alpha\beta}+\frac{C_L^{ds\alpha\beta}}{C_{L,\textrm{SM}}^{ds11}}\right|^2\\
&+\mathcal{B}(K^+\to\pi^+\nu_\tau\bar{\nu}_\tau)_{\textrm{SM}}\left[\left|1+\frac{C_L^{ds33}}{C_{L,\textrm{SM}}^{ds33}}\right|^2+\sum_{\alpha=1,2}\left(\left|\frac{C_L^{ds\alpha 3}}{C_{L,\textrm{SM}}^{ds33}}\right|^2+\left|\frac{C_L^{ds3\alpha}}{C_{L,\textrm{SM}}^{ds33}}\right|^2\right)\right]\,,
\end{split}
\end{equation}

\begin{equation}
\begin{split}
\mathcal{B}(K_L\to\pi^0\nu\bar{\nu})_{th} &= \frac{1}{3}\mathcal{B}(K_L\to\pi^0\nu\bar{\nu})_{\textrm{SM}}\Bigg[\sum_{\alpha,\beta=1,2}\left(\delta_{\alpha\beta}+\frac{\textrm{Im}[N_{ds}^{-1}C_L^{ds\alpha\beta}]}{\textrm{Im}[N_{ds}^{-1}C_{L,\textrm{SM}}^{ds11}]}\right)^2\\
&+\left(1+\frac{\textrm{Im}[N_{ds}^{-1}C_L^{ds33}]}{\textrm{Im}[N_{ds}^{-1}C_{L,\textrm{SM}}^{ds33}]}\right)+\sum_{\alpha=1,2}\left(\left(\frac{\textrm{Im}[N_{ds}^{-1}C_L^{ds\alpha 3}]}{\textrm{Im}[N_{ds}^{-1}C_{L,\textrm{SM}}^{ds33}]}\right)+\left(\frac{\textrm{Im}[N_{ds}^{-1}C_L^{ds3\alpha}]}{\textrm{Im}[N_{ds}^{-1}C_{L,\textrm{SM}}^{ds33}]}\right)\right)\Bigg]\,,
\end{split}
\end{equation}
where we defined $N_{ds}=(\sqrt{2}G_F\alpha V_{td}^*V_{ts}/\pi)^{-1}$ and the SM values for the branching ratios and Wilson coefficients are 
\begin{equation}
\begin{split}
\mathcal{B}(K^+\to\pi^+\nu_e\bar{\nu}_e)_{\textrm{SM}} &= 3.06\times 10^{-11}\,,\\
\mathcal{B}(K^+\to\pi^+\nu_\tau\bar{\nu}_\tau)_{\textrm{SM}}&= 2.52\times 10^{-11}\,,\\
\mathcal{B}(K_L\to\pi^0\nu\bar{\nu})_{\textrm{SM}}&= 3.4\times 10^{-11}\,,
\end{split}
\end{equation}
\begin{equation}
C_{L,\textrm{SM}}^{ds\alpha\beta} = -\frac{1}{s_W^2}\left(X_t+\frac{V_{cd}^*V_{cs}}{V_{td}^*V_{ts}}X^\alpha_c\right)\delta_{\alpha\beta}\,,
\end{equation}
with $X_\tau =1.481$, $X^e_c=X^\mu_c = 1.053\times 10^{-3}$ and $X^t_c = 0.711 \times 10^{-3}$.
%

\subsection{$B_s\to \ell_\alpha^- \ell_\beta^+$ and $K_{L,S}\to \ell_\alpha^- \ell_\beta^+$}
The branching ratio for rare leptonic $B$ decays is discussed, for example, in Ref.~\cite{Becirevic:2016zri} and the same description holds also for kaons. In our framework, these processes are induced by the operators 
\begin{equation}\label{eq:C9C10}
O_{9}^{ij\alpha\beta}=(\bar{d}_i\gamma_{\mu}P_{L}d_j)(\bar{\ell}_\alpha\gamma^{\mu}\ell_\beta)\,,\quad   O_{10}^{ij\alpha\beta}=(\bar{d}_i\gamma_{\mu}P_{L}d_j)(\bar{\ell}_\alpha\gamma^{\mu}\gamma_5\ell_\beta)\,,
\end{equation}
appearing in the effective Hamiltonian
\begin{equation}\label{eq:C9C10WETLEFT}
\mathcal{H}_{eff}=-\frac{4G_F}{\sqrt{2}}\frac{\alpha}{4\pi}\sum_{k=9,10}V_{ti}^{*}V_{tj}\mathcal{C}_k^{ij\alpha\beta} \mathcal{O}_k^{ij\alpha\beta}.
\end{equation}
We do not take into account the other possible contributions: the operators $O_{9',10'}$, with $P_L\to P_R$, are identically zero in our setup and we do not consider the usual scalar and pseudoscalar operators $O_S$ and $O_P$ (see e.g. in Ref.~\cite{Becirevic:2016zri}), since they are generated with negligible coefficients with respect to $O_9$ and $O_{10}$.
The relation between these coefficients and the LEFT ones is
\begin{equation}
    C_{9(10)}^{ij\alpha\beta} = \frac{\sqrt{2}\pi}{G_F\alpha V_{ti}^*V_{tj}}\left( 
    \lwc{de}{LR}[V][ij\alpha\beta]  \pm 
    \lwc{ed}{LL}[V][\alpha\beta ij]
    \right)
    \,.
\end{equation}
The branching ratio for the leptonic $B_s$ decays is given by
\begin{equation}
    \begin{split}
         \mathcal{B}(B_s\to \ell_{\alpha}^- & \ell_{\beta}^+)_{th}  =\frac{\tau_{B_s}}{64\pi^3}\frac{\alpha ^2G_F^{2}}{m_{B_s}^3}f_{B_s}^2|V_{tb}V_{ts}^*|^2 \lambda^{1/2}(m_{B_s},m_{\alpha},m_{\beta})\times\\
         & \times \Bigg[ \left[m_{B_s}^2-(m_{\alpha}+m_{\beta})^2\right]\times \left|(C_{9,\mathrm{SM}}^{sb\alpha\beta}+C_{9}^{sb\alpha\beta})(m_{\alpha}-m_{\beta}) \right|^2 +\\
         & + \left[m_{B_s}^2-(m_{\alpha}-m_{\beta})^2\right]\times \left|(C_{10,\mathrm{SM}}^{sb\alpha\beta}+C_{10}^{sb\alpha\beta})(m_{\alpha}+m_{\beta})\right|^2 \Bigg]\,,
    \end{split}
    \label{eq:BrBll}
\end{equation}
where $\lambda(a,b,c)=[a^2-(b-c)^2][a^2-(b+c)^2]$. 
When comparing the theoretical prediction of $B$ decays to untagged experimental data, the sizeable decay width differences in the $B_{s}^0-\bar{B_{s}}^0$ system must be taken into account. This is done by using an effective lifetime. To a good approximation one has \cite{DeBruyn:2012wj,DeBruyn:2012wk} 
\be
    \mathcal{B}(B_s\to \ell_{\alpha}^{-}\ell_{\beta}^{+})_{\rm eff}\simeq\frac{1}{1-y_s}\mathcal{B}(B_s \to \ell_{\alpha}^{-}\ell_{\beta}^{+})_{\rm th} \, ,
\ee
with $y_s=\Delta\Gamma_{B_s}/(2\Gamma_{B_s})=0.064(4)$, according to the current PDG and HFLAV average \cite{Workman:2022ynf}. Numerical values for couplings and parameters appearing in the formula can be found in Refs.~\cite{Aoki:2016frl,Workman:2022ynf}.
The expression in Eq.~\eqref{eq:BrBll} can be applied to $B_{d} \to \ell_{\alpha}^{-}\ell_{\beta}^{+}$ with the obvious $s\to d$ replacement.

The leptonic kaon decays are instead given by
\begin{equation}
    \begin{split}
         \mathcal{B}(K_{L(S)}\to \ell_\alpha^- & \ell_\beta^+)_{th}  =\frac{\tau_{K_L}}{128\pi^3}\frac{\alpha^2G_F^{2}}{m_{K_0}^3}f_{K}^2|V_{td}V_{ts}^*|^2 \lambda^{1/2}(m_{K},m_\alpha,m_\beta)\times\\
         & \times \Bigg[ \left[m_{K_0}^2-(m_\alpha+m_\beta)^2\right]\times \left|\left(C_{9,\mathrm{SM}}^{ds\alpha\beta}\pm C_{9,\mathrm{SM}}^{sd\alpha\beta}+C_9^{ds\alpha\beta}\pm C_9^{sd\alpha\beta}\right)\right|^2(m_\alpha-m_\beta)^2 +\\
         & + \left[m_{K_0}^2-(m_\alpha-m_\beta)^2\right]\times \left|\left(C_{10,\mathrm{SM}}^{ds\alpha\beta}\pm C_{10,\mathrm{SM}}^{sd\alpha\beta}+C_{10}^{ds\alpha\beta}\pm C_{10}^{sd\alpha\beta}\right)\right|^2(m_\alpha+m_\beta)^2 \Bigg]\,.
    \end{split}
    \label{eq:BrKLll}
\end{equation}
Notice that for kaons both $ds$ and $sd$ indices appear, since $K_{L,S}$ are linear combinations of $K^0$ and $\bar{K}^0$
\begin{equation}
    \ket{K_{L(S)}} = \frac{\ket{K^0}\pm\ket{\bar{K}^0}}{\sqrt{2}}\,.
\end{equation}
In all cases, as one can expect, for $\ell_\alpha=\ell_\beta=\ell$ the coefficient $C_9$ does not contribute.

\subsection{$B \to K^{(*)}\ell^+ \ell^-$}
 In Ref.~\cite{Hiller:2014ula}, the leading contributions to the decay width of $B\to K^{(*)}\ell^{+}\ell^{-}$ are computed. Under some justified assumptions, the $R_{K}$ and $R_{K^{*}}$ ratios read
\begin{equation*}
\begin{split}
R_{K}([1.1,6]) \approx & 1.00+0.24\text{Re}[C_9 - C_{10}]+ 0.24\text{Re}[C'_{9} - C'_{10}]+0.058\text{Re}[C_9^{*} C'_9+ \\ & + C_{10}^{*} C'_{10}]+0.029 \left(| C_9|^2+| C'_9|^2+| C_{10}|^2+| C'_{10}|^2 \right)\,,
\end{split}
\end{equation*}
\begin{equation*}
\begin{split}
R_{K^{*}}([1.1,6]) \approx & 1.00+0.24\text{Re}[C_9 - C_{10}] -0.18\text{Re}[C'_9]+0.17\text{Re}[ C'_{10}]-0.042\text{Re}[ C_9^{*}C'_9+ \\ &+ C_{10}^{*} C'_{10}]+ 
+0.029 \left(| C_9|^2+| C'_9|^2+| C_{10}|^2+| C'_{10}|^2 \right)\,,
\end{split}
\end{equation*}
where the operators and the Hamiltonian are defined as in App.~\ref{eq:C9C10}-\ref{eq:C9C10WETLEFT}. The numerical coefficients are compatible with the ones computed, for example, in Refs.~\cite{Ciuchini:2019usw,Hiller:2017bzc,Geng:2017svp} and with our calculation of $R_{K^{*}}$, derived using expressions and form factors provided in Refs.~\cite{Altmannshofer:2008dz,Bharucha:2015bzk}. The primed coefficients $C_{9}^{\prime}$ and $C_{10}^{\prime}$ are displayed for completeness, but they are anyway set to zero as they are not generated by the operators in Table \ref{table:Operators}.\\

\subsection{$K_L \to \pi^0\ell^+ \ell^-$}
We consider again the Hamiltonian in Eq.~\eqref{eq:C9C10WETLEFT} and we define $\tilde{C}_k^{ij\alpha\beta} = V_{ti}^*V_{tj}C_k^{ij\alpha\beta}$. Rearranging the formula given in Ref.~\cite{Aebischer:2022vky} to fit our notation, we obtain for the branching ratio
\begin{equation}
\mathcal{B}(K_L\to\pi^0\ell^+\ell^-)_{th} =a_\ell\left[(\omega_{7V}^\ell)^2+(\omega_{7A}^\ell)^2\right]+b_\ell\omega_{7V}^\ell+c_\ell,
\end{equation}
where
\begin{equation}
\begin{split}
   \omega_{7V}^\ell &= \frac{1}{2\pi}\frac{1}{1.407\times 10^{-4}}\left[P_0\textrm{Im}(V_{ts}^*V_{td})+\textrm{Im}(C_{9,\mathrm{SM}}^{sd\ell\ell}+C_9^{sd\ell\ell})\right],\\
   \omega_{7A}^\ell &= \frac{1}{2\pi}\frac{1}{1.407\times 10^{-4}}\textrm{Im}(C_{10,\mathrm{SM}}^{sd\ell\ell}+C_{10}^{sd\ell\ell}).\\
\end{split}
\end{equation}
The numerical coefficients $a$, $b$ and $c$ are given by
\begin{equation}
    \begin{split}
        a_e &= ~4.62\times 10^{-12},\qquad a_\mu = 1.09\times 10^{-12},\\
        b_e &= 13.56\times 10^{-12},\qquad b_\mu = 3.156\times 10^{-12},\\
        c_e &= 20.88\times 10^{-12},\qquad c_\mu = 10.0384\times 10^{-12}.
    \end{split}
\end{equation}
%

\subsection{$P\to M \ell_\alpha^- \ell_\beta^+$ }
We consider here the semileptonic decay of a meson $P=\{B,K_L,K^+\}$ into $M=\{K^{(*)},\pi\}$ and two different leptons.
The branching ratios for the decays are \cite{Becirevic:2016zri,Angelescu:2020uug}
\begin{equation}\label{eq:BKlldecays}
\mathcal{B}(P\to M\ell_\alpha^-\ell_\beta^+)_{th} = \frac{1}{8 G_F^2}\left(\alpha_V^{+}\left| \lwc{de}{LR}[V][ij\alpha\beta]  + 
    \lwc{ed}{LL}[V][\alpha\beta ij] \right|^2 
    + \alpha_V^{-}\left| \lwc{de}{LR}[V][ij\alpha\beta]  - 
    \lwc{ed}{LL}[V][\alpha\beta ij] \right|^2\right)\,.
\end{equation}
Numerical values for the multiplicative coefficients are given in Table \ref{table:BKcoeff}. The indices $ij$ correspond to the transition $q_j\to q_i \ell_\alpha^-\ell_\beta^+$ and for $K_L$ it is understood the average $(ds+sd)/2$.
\begin{table}[ht!]\renewcommand{\arraystretch}{1.2}
\centering
\begin{tabular}{ |c|c c| }
\hline
Process & $\alpha_V^{+}$ & $\alpha_V^{-}$\\ \hline
$B\to K e^+ \mu^-$ & 8.2(6) & 8.2(6) \\ 
$B\to K e^+ \tau^-$ & 5.3(2) & 5.3(2) \\ 
$B\to K \mu^+ \tau^-$ & 5.2(2) & 5.2(2) \\ \hline \hline
$B\to K^* e^+ \mu^-$ & 2.8(5) & 2.8(5) \\ 
$B\to K^* e^+ \tau^-$ & 1.4(2) & 1.4(2) \\ 
$B\to K^* \mu^+ \tau^-$ & 1.5(2) & 1.3(2) \\ \hline \hline
$K^+\to \pi^+ e^+ \mu^-$ & 0.596(4) & 0.598(4) \\ 
$K_L\to \pi^0 e^+ \mu^-$ & 2.75(2) & 2.76(2) \\ \hline
\end{tabular}
\caption{Values for the factors defined in Eq.~\ref{eq:BKlldecays}, for all the possible final states. Details on their computation can be found in Ref.~\cite{Angelescu:2020uug}. }
\label{table:BKcoeff}
\end{table}

\subsection{$B \to X_s\gamma$} 

The inclusive radiative B decays can be parametrized in terms of the operators \begin{equation} O_{7}=\frac{e}{16\pi^2}m_b(\bar{s}\sigma_{\alpha\beta}P_{R(L)}b)F^{\alpha\beta} \,,\quad O_{8}=\frac{g}{16\pi^2}m_b(\bar{s}\sigma_{\alpha\beta}P_{R(L)}T^{a}b)G^{\alpha\beta}_a\,,
\end{equation}
and the effective Hamiltonian
\begin{equation}
\mathcal{H}_{eff}=-\frac{4G_F}{\sqrt{2}}V_{ts}^{*}V_{tb}\sum_{i}\mathcal{C}_i \mathcal{O}_i\,.
\end{equation}
The branching ratio is given then by \cite{Misiak:2015xwa,Czakon:2015exa} 
\begin{equation}
    \mathcal{B}(\bar{B}\to X_s\gamma)_{E_{\gamma}>1.6\,\mathrm{GeV}}=10^{-4}\times \left(3.36 \pm 0.23 -8.22 C_7 -1.99  C_8 \right)\,,
\end{equation}
where the coefficients are defined at the matching scale $\mu_0=160$ GeV \cite{Misiak:2015xwa}. In terms of LEFT coefficients, we have:
\begin{equation}
\begin{split}
C_7=\frac{N_{sb}^{-1}e}{m_b}&\times \lwc{d \gamma}{}[][s b] \,,\quad\quad C_8=\frac{N_{sb}^{-1}e^2}{g m_b} \lwc{dG}{}[][s b]\,,\\
&N_{sb}=\frac{4G_F}{\sqrt{2}}\frac{\alpha_{em}}{4\pi}V_{ts}^{*}V_{tb}~.
\end{split}
\end{equation}

\subsection{Charged lepton flavor violation}
\label{sec:LFV}

For charged lepton flavour violation processes we employ the expressions from Refs.~\cite{Crivellin:2017rmk,Cirigliano:2021img}, after translating them to the LEFT basis.
In the following we give the explicit expressions used in this work.

\begin{itemize}
\item $\ell_H \rightarrow \ell_L \gamma$

We may take a reference renormalization scale $\mu= 2\, \mathrm{GeV}$, where the LEFT is yet well defined. One has \cite{Crivellin:2017rmk,Dekens:2018pbu,Cirigliano:2021img}
\begin{equation}
\Gamma_{\ell \rightarrow \ell' \gamma}=\frac{m_{\ell}^3}{4\pi}[|\lwc{ e \gamma}{}[][\ell' \ell]|^2+|\lwc{ e \gamma}{}[][\ell \ell']|^2] \, .
\end{equation}
The (nonperturbative) effect of four-fermion operators involving tensor light-quark currents can be taken into account by making the following transformations
\begin{align}
\lwc{ e \gamma}{}[][\ell' \ell]&\rightarrow \lwc{ e \gamma}{}[][\ell' \ell]- e \left(c_{3}^{\ell'\ell}+\frac{c_{8}^{\ell'\ell}}{\sqrt{3}} \right) i \Pi_{VT}(0) \, ,\\
\lwc{ e \gamma}{}[][\ell \ell']&\rightarrow \lwc{ e \gamma}{}[][\ell \ell']+ e \left(c_{3}^{\ell\ell'}+\frac{c_{8}^{\ell\ell'}}{\sqrt{3}} \right) i \Pi_{VT}(0) \, ,
\end{align}
with
\begin{align}
c_{3}^{ij}&=\lwc{ eu }{RR}[T][ij 11]-\lwc{ ed }{RR}[T][ij 11]  \, ,\\
c_{8}^{ij}&=\frac{1}{\sqrt{3}}\left[ \lwc{ eu }{RR}[T][ij 11]+\lwc{ ed }{RR}[T][ij 11] -2\lwc{ ed }{RR}[T][ij 22]\right]  \, .
\end{align}
As in Ref.~\cite{Cirigliano:2021img} we take $i\Pi_{VT}(0)\approx 0.04\, \mathrm{GeV}$.

\item $\tau \to 3e$, $\tau \to 3\mu$, $\mu \to 3e$

We use the analytic expressions given in Ref.~\cite{Cirigliano:2021img}, adapting them to the LEFT normalization of this work. Numerically we find, in $\mathrm{GeV}$ units,
\begin{align}\nonumber
\mathcal{B}(\tau \to 3e) &\approx 1.6 \times 10^8 (|\lwc{ ee }{RR}[S][1113]|^2+|\lwc{ ee }{LR}[V][1113]|^2+|\lwc{ ee }{LR}[V][1311]|^2)\\ \nonumber
&+1.3 \times 10^{9}(|\lwc{ ee }{LL}[V][1311]|^2+|\lwc{ ee }{RR}[V][1311]|^2)+2.1 \times 10^{9} (|\lwc{ e \gamma}{}[][31]|^2 + |\lwc{ e \gamma}{}[][13]|^2)\\ \nonumber
&+[\lwc{ e \gamma}{}[][31](-4.5 \times 10^8 \lwc{ ee }{LL}[V][1113]-1.1 \times 10^{8}  \lwc{ ee }{LR}[V][1311]) +c.c.  ]\\
&+[\lwc{ e \gamma}{*}[][13](-4.5 \times 10^8 \lwc{ ee }{RR}[V][1113]-1.1 \times 10^{8}  \lwc{ ee }{LR}[V][1113]) +c.c.  ] \, .
\end{align}

The analytic expressions trivially generalize for $\tau \to 3\mu$, $\mu \to 3e$. Numerically one finds
\begin{align}\nonumber
\mathcal{B}(\tau \to 3\mu) &\approx 1.6 \times 10^8 (|\lwc{ ee }{RR}[S][1123]|^2+|\lwc{ ee }{LR}[V][1123]|^2+|\lwc{ ee }{LR}[V][2311]|^2)\\ \nonumber
&+1.3 \times 10^{9}(|\lwc{ ee }{LL}[V][2311]|^2+|\lwc{ ee }{RR}[V][2311]|^2)+4.4 \times 10^{8} (|\lwc{ e \gamma}{}[][32]|^2 + |\lwc{ e \gamma}{}[][23]|^2)\\ \nonumber
&+[\lwc{ e \gamma}{}[][32](-4.5 \times 10^8 \lwc{ ee }{LL}[V][1123]-1.1 \times 10^{8}  \lwc{ ee }{LR}[V][2311]) +c.c.  ]\\
&+[\lwc{ e \gamma}{*}[][23](-4.5 \times 10^8 \lwc{ ee }{RR}[V][1123]-1.1 \times 10^{8}  \lwc{ ee }{LR}[V][1123]) +c.c.  ] \, ,
\end{align}
and
\begin{align}\nonumber
\mathcal{B}(\mu \to 3e) &\approx 9.2 \times 10^8 (|\lwc{ ee }{RR}[S][1112]|^2+|\lwc{ ee }{LR}[V][1112]|^2+|\lwc{ ee }{LR}[V][1211]|^2)\\ \nonumber
&+7.4 \times 10^{9}(|\lwc{ ee }{LL}[V][1211]|^2+|\lwc{ ee }{RR}[V][1211]|^2)+1.9 \times 10^{12} (|\lwc{ e \gamma}{}[][21]|^2 + |\lwc{ e \gamma}{}[][12]|^2)\\ \nonumber
&+[\lwc{ e \gamma}{}[][21](-4.2 \times 10^{10} \lwc{ ee }{LL}[V][1112]-1.1 \times 10^{10}  \lwc{ ee }{LR}[V][1211]) +c.c.  ]\\
&+[\lwc{ e \gamma}{*}[][12](-4.2 \times 10^{10} \lwc{ ee }{RR}[V][1112]-1.1 \times 10^{10}  \lwc{ ee }{LR}[V][1112]) +c.c.  ] \, ,
\end{align}

\item $\tau \to e \bar{\mu}\mu$, $\tau \to \mu \bar{e}e$.

Once again we adapt the analytic results as given in Ref.~\cite{Cirigliano:2021img}, finding
\begin{align}\nonumber
\mathcal{B}(\tau \to e \bar{\mu}\mu) &\approx 2.5 \times 10^9 (|\lwc{ ee }{LL}[V][1223]|^2+|\lwc{ ee }{RR}[V][1223]|^2)\\ \nonumber
&+1.6 \times 10^{8}(|\lwc{ ee }{LR}[V][2213]|^2+|\lwc{ ee }{LR}[V][1322]|^2)+1.0 \times 10^{8} (|\lwc{ e \gamma}{}[][31]|^2 + |\lwc{ e \gamma}{}[][13]|^2)\\ \nonumber
&-2.1 \times 10^{8} [\lwc{ e \gamma}{}[][31] \lwc{ ee }{LL}[V][1223]+c.c.  ]\\
&+[\lwc{ e \gamma}{*}[][13](-5.2 \times 10^{7} \lwc{ee }{LR}[*V][1322]-2.6 \times 10^{8}  \lwc{ ee }{RR}[V][1223]) +c.c.  ] \, ,
\end{align}
and
\begin{align}\nonumber
\mathcal{B}(\tau \to \mu \bar{e}e) &\approx 2.6 \times 10^9 (|\lwc{ ee }{LL}[V][1123]|^2+|\lwc{ ee }{RR}[V][1123]|^2)\\ \nonumber
&+1.6 \times 10^{8}(|\lwc{ ee }{LR}[V][1123]|^2+|\lwc{ ee }{LR}[V][2311]|^2)+5.0 \times 10^{8} (|\lwc{ e \gamma}{}[][32]|^2 + |\lwc{ e \gamma}{}[][23]|^2)\\ \nonumber
&-2.2 \times 10^{8} [\lwc{ e \gamma}{}[][32] \lwc{ ee }{LL}[V][1123]+c.c.  ]\\
&+[\lwc{ e \gamma}{*}[][23](-5.5 \times 10^{7} \lwc{ee }{LR}[*V][2311]-2.8 \times 10^{8}  \lwc{ ee }{RR}[V][1123]) +c.c.  ] \, .
\end{align}

\item $\mu \mathrm{Au} \to e \mathrm{Au}$.

We adapt the results of Ref.~\cite{Crivellin:2017rmk}, neglecting the power-suppressed contributions from $c$ and $b$ quarks. In principle they are given at a $\mu=1\, \mathrm{GeV}$, but within the theoretical uncertainties the same expressions can also be used at $\mu=2\, \mathrm{GeV}$.\footnote{The large QCD running of the scalar quark current is fully taken into account by simply taking as input $m_{q}(\mathrm{\mu=2\, \mathrm{GeV}})$ instead of $m_{q}(\mathrm{\mu=1\, \mathrm{GeV}})$.} We have also checked that we agree with the adaptation given in Ref.~\cite{Dekens:2018pbu}. The numerical result is rather large. For reference, the first quadratic coefficients are
\begin{align}
\mathcal{B}(\mu \mathrm{Au} \to e \mathrm{Au}) &\approx 1.1  \times 10^{13} [(\lwc{ ed }{RL}[S][1211])^2+(\lwc{ ed }{RL}[S][2111])^2+(\lwc{ ed }{RR}[S][2111])^2+(\lwc{ ed }{RR}[S][1211])^2 +(d\rightarrow u)] 
\end{align}

\item $\tau \to \ell \pi^{0}$

We take the numerical result of Ref.~\cite{Cirigliano:2021img}. Adapting it to the normalization of this work we have
\begin{align}\nonumber
\mathcal{B}(\tau \to \ell \pi^{0})&\approx B_{1}(\tau \to \ell \pi^{0})+B_{2}(\tau \to \ell \pi^{0}) \, ,
\end{align}
with
\begin{align}\nonumber
\mathcal{B}_1(\tau \to \ell \pi^{0})&\approx 1.3 \times 10^8 |\lwc{ ed }{RL}[S][\ell 311]-\lwc{ eu }{RL}[S][\ell 311]-\lwc{ ed }{RR}[S][\ell 311]+\lwc{ eu }{RR}[S][\ell 311]|^2\\&+5.1 \times 10^{7} |\lwc{ ed }{LL}[V][3\ell 11]-\lwc{ eu }{LL}[V][3\ell11]-\lwc{ ed }{LR}[V][3\ell 11]+\lwc{ eu }{LR}[V][3\ell 11]|^2 \, ,
\end{align}
and $\mathcal{B}_2(\tau \to \ell \pi^{0})$ can be obtained from the $B_1(\tau \to \ell \pi^{0})$ expression by making the following transformations
\begin{align}
\lwc{ eq }{LL}[V][3\ell 11] \rightarrow \lwc{ eq }{RR}[V][3\ell 11] ,
\lwc{ eq }{LR}[V][3\ell 11] \rightarrow \lwc{ qe }{LR}[V][11 3 \ell] ,
\lwc{ eq }{RR}[S][\ell 3 qq] \rightarrow \lwc{ eq }{RR^*}[S][3 \ell  qq] ,
\lwc{ eq }{RL}[S][\ell 3 qq] \rightarrow \lwc{ eq }{RL^*}[S][3 \ell qq] .
\end{align}

\item $\tau \to \ell\pi^+\pi^-$

Once again we use the numerical results from Ref.~\cite{Cirigliano:2021img}. We find
\begin{equation}\nonumber
\mathcal{B}(\tau \to \ell \pi \pi)=\mathcal{B}_1(\tau \to \ell \pi \pi)+\mathcal{B}_2(\tau \to \ell \pi \pi) \, .
\end{equation}
\begin{align}\nonumber
\mathcal{B}_1(\tau \to \ell \pi \pi)&\approx 9.2 \times 10^{8}|\lwc{ ed }{RR}[T][3\ell 11]-\lwc{ eu }{RR}[T][3\ell 11]|^2+4.6 \times 10^{8}|\lwc{ ed }{LL}[V][3\ell 11]-\lwc{ eu }{LL}[V][3\ell 11]+\lwc{ ed }{LR}[V][3\ell 11]-\lwc{ eu }{LR}[V][3\ell 11]|^2 \\ \nonumber
&+5.0 \times 10^{8} |\lwc{ e\gamma }{}[][3\ell]|^2+1.6 \times 10^8 |\lwc{ ed }{RL}[S][3\ell 11]+\lwc{ eu }{RL}[S][3\ell 11]+\lwc{ ed }{RR}[S][3\ell 11]+\lwc{ eu }{RR}[S][3\ell 11]|^2\\
&+1.2 \times 10^8 |\lwc{ ed }{RL}[S][3\ell 22]+\lwc{ ed }{RR}[S][3\ell 22]|^2 \, ,
\end{align}
and $\mathcal{B}_2(\tau \to \ell \pi\pi)$ can be obtained from the $B_1(\tau \to \ell  \pi\pi)$ expression by making the following transformations
\begin{align}\nonumber
\lwc{ eq }{RR}[T][3\ell 11] &\rightarrow \lwc{ eq }{RR}[T][\ell 3 11] , 
\lwc{ eq }{LL}[V][3\ell 11] \rightarrow\lwc{ eq }{RR}[V][3 \ell 11] ,
\lwc{ eq }{LR}[V][3\ell 11] \rightarrow \lwc{ qe }{LR}[V][11 3 \ell] ,\\
\lwc{ e\gamma }{}[][3\ell] &\rightarrow\lwc{ e\gamma }{}[][\ell 3] \, ,
\lwc{ eq }{RR}[S][3\ell qq] \rightarrow \lwc{ eq }{RR^*}[S][\ell 3 qq] ,
\lwc{ eq }{RL}[S][3\ell qq] \rightarrow \lwc{ eq }{RL^*}[S][\ell 3 qq] .
\end{align}

\end{itemize}
$\tau \to \ell \eta, \tau \to \ell \eta', \tau \to \ell K, \tau \to \ell \pi K, \tau \to \ell KK$ are also included in our $\chi^2$, adapting the compilation of Ref.~\cite{Cirigliano:2021img}. However they play a marginal phenomenological role in our analysis and thus we refrain from giving the explicit expressions.

\subsection{Matching relations for the Cabibbo angle observables}\label{subapp:cabibbomatch}

The running of the chirality-flipping low-energy couplings to the $M_Z$ mass is large and needs to be taken into account. From Ref.~\cite{Gonzalez-Alonso:2017iyc}, which includes higher-order QCD running effects, one has

\begin{equation}
\begin{pmatrix}
\epsilon_{S} \\
\epsilon_{P} \\
\epsilon_{T}
\end{pmatrix}_{\mu= 2\, \mathrm{GeV}}
=
\begin{pmatrix}
1.72 & 2.46 \times 10^{-6} & -0.0242 \\
2.46 \times 10^{-6} & 1.72 & -0.0242 \\
-2.17 \times 10^{-4} & -2.17 \times 10^{-4} & 0.825
\end{pmatrix}
\begin{pmatrix}
\epsilon_{S} \\
\epsilon_{P} \\
\epsilon_{T}
\end{pmatrix}_{\mu= M_Z} \, .
\end{equation}
At $M_{Z}$ at tree-level one has
\begin{align}
-\frac{2\hat{V}_{ud}}{v^2}\epsilon_{L}^{dse}&=2 \left(\cwc{H\ldoublet}{}[(3)][22]-\cwc{H q}{}[(3)][11]-\cwc{\ldoublet\ldoublet}{}[][1221]+\cwc{\ldoublet q}{}[(3)][1111] \right)\nonumber\\
-\frac{2\hat{V}_{ud}}{v^2}\epsilon_L^{d\ell'/\ell}&=\lwc{\nu edu}{LL}[V][\ell'\ell'11]-\lwc{\nu edu}{LL}[V][\ell \ell 11]=2V_{ud}(\cwc{H\ldoublet}{}[(3)][\ell \ell]-\cwc{H\ldoublet}{}[(3)][\ell'\ell']-\cwc{\ldoublet q}{}[(3)][\ell \ell11]+\cwc{\ldoublet q}{}[(3)][\ell'\ell'11])-2V_{us}(-\cwc{\ldoublet q}{}[(3)][\ell \ell 21]+\cwc{\ldoublet q}{}[(3)][\ell'\ell'21])\, ,\nonumber\\
-\frac{2\hat{V}_{us}}{v^2}\epsilon_L^{s\ell'/\ell}&=\lwc{\nu edu}{LL}[V][\ell'\ell'21]-\lwc{\nu edu}{LL}[V][\ell \ell 21]=2V_{us}(\cwc{H\ldoublet}{}[(3)][\ell \ell]-\cwc{H\ldoublet}{}[(3)][\ell'\ell']-\cwc{\ldoublet q}{}[(3)][\ell \ell11]+\cwc{\ldoublet q}{}[(3)][\ell'\ell'11])+2V_{ud}(-\cwc{\ldoublet q}{}[(3)][\ell \ell 21]+\cwc{\ldoublet q}{}[(3)][\ell'\ell'21])\, ,\nonumber\\
-\frac{2\hat{V}_{uD}}{v^2}\epsilon_R^{D}&=\lwc{\nu edu}{LR}[V][\ell\ell D1]=-\cwc{Hud}{}[][1D]\, ,\nonumber\\
-\frac{2\hat{V}_{ud}}{v^2}\epsilon_S^{d\ell}&=\lwc{\nu edu}{RR}[S][\ell\ell 11]+\lwc{\nu edu}{RL}[S][\ell\ell 11]=\cwc{ledq}{}[][\ell\ell 11]+V_{ud}\cwc{lequ}{}[(1)][\ell\ell 11]-V_{us}\cwc{lequ}{}[(1)][\ell\ell 21]\, ,\nonumber\\
-\frac{2\hat{V}_{us}}{v^2}\epsilon_S^{s\ell}&=\lwc{\nu edu}{RR}[S][\ell\ell 21]+\lwc{\nu edu}{RL}[S][\ell\ell 21]=\cwc{ledq}{}[][\ell\ell 21]+V_{us}\cwc{lequ}{}[(1)][\ell\ell 11]+V_{ud}\cwc{lequ}{}[(1)][\ell\ell 21]\, ,\\
-\frac{2\hat{V}_{ud}}{v^2}\epsilon_P^{d\ell}&=\lwc{\nu edu}{RR}[S][\ell\ell 11]-\lwc{\nu edu}{RL}[S][\ell\ell 11]=-\cwc{ledq}{}[][\ell\ell 11]+V_{ud}\cwc{lequ}{}[(1)][\ell\ell 11]-V_{us}\cwc{lequ}{}[(1)][\ell\ell 21]\, ,\nonumber\\
-\frac{2\hat{V}_{us}}{v^2}\epsilon_P^{s\ell}&=\lwc{\nu edu}{RR}[S][\ell\ell 21]-\lwc{\nu edu}{RL}[S][\ell\ell 21]=-\cwc{ledq}{}[][\ell\ell 21]+V_{us}\cwc{lequ}{}[(1)][\ell\ell 11]+V_{ud}\cwc{lequ}{}[(1)][\ell\ell 21]\, ,\nonumber\\
-\frac{2\hat{V}_{ud}}{v^2}\frac{1}{4}\hat{\epsilon}_T^{d\ell}&=\lwc{\nu edu}{RR}[T][\ell\ell 11]=V_{ud}\cwc{lequ}{}[(3)][\ell\ell 11]-V_{us}\cwc{lequ}{}[(3)][\ell\ell 21]\, , \nonumber\\
-\frac{2\hat{V}_{us}}{v^2}\frac{1}{4}\hat{\epsilon}_T^{s\ell}&=\lwc{\nu edu}{RR}[T][\ell\ell 21]=V_{us}\cwc{lequ}{}[(3)][\ell\ell 11]+V_{ud}\cwc{lequ}{}[(3)][\ell\ell 21]\, . \nonumber
\end{align}

The leading mechanism to generate a top operator is through the large mixing of $\cwc{H\ldoublet}{}[(3)][\alpha\alpha]$ with $\cwc{\ldoublet q}{}[(3)][\alpha\alpha33]$ through a double top-Yukawa insertion \cite{Jenkins:2013wua},\footnote{As in the rest of this work, the DSixTools package has been used to perform the full one-loop running. We have also checked that the loop-induced top operators in the SMEFT-LEFT matching play a marginal role in this sector and, in first approximation, can be neglected.}
\begin{align}\label{eq:leadlogcab}
v^2\cwc{H\ldoublet}{}[(3)][\alpha\alpha](M_Z^2)&\approx  v^2 4 N_c \frac{m_t^2}{(4\pi v)^2}\cwc{\ldoublet q}{}[(3)][\alpha\alpha 33](\Lambda^2_{\mathrm{UV}})\log\frac{\Lambda_{\mathrm{UV}}}{M_Z}=\left[\frac{N_c}{4\pi^2}\frac{m_t^2}{\Lambda^2_{\mathrm{UV}}}\log\frac{\Lambda_{\mathrm{UV}}}{M_Z} \,  \right]\Lambda_{\mathrm{UV}}^2\cwc{\ldoublet q}{}[(3)][\alpha\alpha 33](\Lambda^2_{\mathrm{UV}}) \, .
\end{align}

\subsection{Higgs basis used for EWPT and Higgs fit}
\label{app:HiggsBasis}

The fit performed in Ref.~\cite{Falkowski:2019hvp} employs a specific combinations of Wilson coefficients of the Warsaw basis, called \emph{Higgs basis}. This allows to reduce the correlations between the directions mostly constrained by EW precision data and those constrained by Higgs data, which differ strongly in the precision. The combinations used are the following:
\be\begin{split}
    \label{eq:HiggsBasis}
    \delta g^{W \ell}_L & =   C^{(3)}_{H l} + f(1/2,0) - f(-1/2,-1), \\
    \delta g^{Z\ell}_L & =    - {1 \over 2} C^{(3)}_{H l} - {1\over 2} C_{H l}^{(1)}+   f(-1/2, -1) , \\ 
    \delta g^{Z\ell }_R & =  - {1\over 2} C_{H e}^{(1)}   +  f(0, -1) ,\\ 
    \delta g^{Z u}_L & =   {1 \over 2}  C^{(3)}_{H q} - {1\over 2} C_{H q}^{(1)}   + f(1/2,2/3) ,  \\
    \delta g^{Zd}_L & =   
 - {1 \over 2}  C^{(3)}_{H q} - {1\over 2} C_{H q}^{(1)}   + f(-1/2,-1/3), \\
    \delta g^{Zu}_R & =    - {1\over 2} C_{H u}   +  f(0,2/3), \\
    \delta g^{Zd}_R & =    - {1\over 2} C_{H d}  +  f(0,-1/3), \\
    \delta c_{z} & = 
    C_{H\Box}   - {1 \over 4} C_{H D}
 -{3 \over 2}  \Delta_{G_F}, \\ 
    c_{z\Box} &=    {1 \over 2   g_L^2} \left (
    C_{H D}   
 +2  \Delta_{G_F}   \right ),   \\ 
    c_{gg} &= {4 \over  g_s^2} C_{H G},  \\ 
    c_{\gamma \gamma} &=  4  \left ( {1  \over   g_L^2} C_{H W} + { 1 \over   g_Y^2} C_{H B} - {1 \over   g_L   g_Y}  C_{H WB} \right )  , \\  
    c_{zz} &=    4 \left (  {  g_L^2 C_{H W} +    g_Y^2 C_{H B} +     g_L   g_Y  C_{H WB} \over (  g_L^2 +   g_Y^2)^2} \right ), \\ 
    c_{z\gamma} &=     4 \left ( { C_{H W} -   C_{H B} 
-  {  g_L^2 -    g_Y^2 \over 2   g_L   g_Y} C_{H WB} \over   g_L^2 +   g_Y^2} \right )  ,
\end{split}\ee
where
\be
    f(T^3,Q)  \equiv  \bigg \{  
-   Q  {g_L  g_Y \over  g_L^2 -  g_Y^2} C_{H WB} 
-  {\bf 1}  \left ( {1 \over 4} C_{H D}  
+  {1 \over 2 } \Delta_{G_F}  \right )  \left ( T^3 + Q { g_Y^2 \over  g_L^2 -   g_Y^2} \right )  \bigg \} {\bf 1} ~,
\ee
and $\Delta_{G_F} = [C^{(3)}_{Hl}]_{11} + [C^{(3)}_{Hl}]_{22} -  {1 \over 2}[C_{ll}]_{1221}$. 
To these coefficients, the fit in Ref.~\cite{Falkowski:2019hvp} also adds those for the dimension-6 Yukawa-like operators of top, bottom, muon and tau fermions, as well as the one responsible for modifying the muon decay: $[C_{uH}]_{33}$, $[C_{dH}]_{33}$, $[C_{eH}]_{22}$, $[C_{eH}]_{33}$, $[C_{ll}]_{1221}$.

The set of $\delta g_{V \psi}$ parametrise the corresponding vertex corrections of $Z$ and $W$ bosons, and are mostly constrained by electroweak data from LEP. As mentioned, the coefficient $[C_{ll}]_{1221}$ enters the EW fit by its impact in the muon decay. Finally, the other coefficients do not affect EW observables at tree level and are only constrained by Higgs data.

\subsection{Direct constraints from LHC}
\label{app:LHCconstraints}

Here we provide the 95\% CL bounds on Wilson coefficients, from both individual and global marginalised fits obtained from SMEFiT \cite{Ethier:2021bye}. These are illustrated in Table \ref{tab:SMEFIT}.
Single parameter fits match the findings of Ref.~\cite{Ethier:2021bye} (see Table 5.4 therein).

\begin{table}[t]\renewcommand{\arraystretch}{1.4}
\centering {\small
\begin{tabular}{|c|c|c|c|c|}
\hline
Class & Coefficients & Warsaw basis &  95\% CL Individual &  95\% CL Marginalised\\ \hline
\multirow{3}{*}{Dipoles}
 & $c_{tG}$             & $C_{uG}$          & [0.01,0.11]         & [0.01,0.23]     \\ \cline{2-5}
 & $c_{tW}$             & $C_{uW}$          & [-0.085,0.030]         & [-0.28,0.13]    \\ \cline{2-5}
 & $c_{tZ}$             & $-\mathrm{s}_\theta \, C_{uB}+\mathrm{c}_\theta \, C_{uW}$          & [-0.038,0.090]         & [-0.50,0.14]    \\ \cline{2-5}
 \hline
\multirow{4}{*}{Higgs-Top}
 & $c_{H Q}^{3}$  & $C_{Hq}^{(3)}$        & [-0.39,0.34]        & [-0.42,0.31]     \\ \cline{2-5}
 & $c_{H Q}^{(-)}$& $C_{Hq}^{(1)}-C_{Hq}^{(3)}$  & [-1.1,1.5]     & [-2.7,2.7]     \\ \cline{2-5}
 & $c_{H t}$      & $C_{Hu}$          & [-2.8,2.2]        & [-15,4]   \\ \cline{2-5}
 & $c_{t H}$      & $C_{uH}$          & [-1.3,0.4]        & [-0.5,2.9]     \\ \cline{2-5}
\hline
\multirow{5}{*}{4 quarks}
 & $c_{QQ}^{1}$  & $2C_{qq}^{(1)}-\frac{2}{3}C_{qq}^{(3)}$    & [-2.3,2.0]     & [-3.7,4.4]     \\ \cline{2-5}
 & $c_{QQ}^{8}$         & $8C_{qq}^{(3)}$             & [-6.8,5.9]        & [-13,10]  \\ \cline{2-5}
 & $c_{Qt}^{1}$         & $C_{qu}^{(1)}$             & [-1.8,1.9]        & [-1.5,1.4]    \\ \cline{2-5}
 & $c_{Qt}^{8}$         & $C_{qu}^{(8)}$             & [-4.3,3.3]        & [-3.4,2.5]    \\ \cline{2-5}
 & $c_{tt}^{1}$         & $C_{uu}$            & [-1.1,1.0]        & [-0.88,0.81] \\ \cline{2-5}
\hline
\end{tabular}}
\caption{The 95\% CL bounds from SMEFiT for individual and global fits obtained using quadratic EFT calculations. Top quark, Higgs and diboson data have been used, as discussed in Ref.~\cite{Ethier:2021bye}. Coefficients are expressed in both the SMEFiT and the Warsaw basis conventions.}
\label{tab:SMEFIT}
\end{table}

Even though it is reasonable to expect that individual bounds are comparable or more severe than the global ones, we observe the latter to be stronger in the case of the $C_{qu}^{(1/8)}$ and $C_{uu}$ coefficients. This surprising trend was also observed in Ref.~\cite{Ethier:2021bye}, where the global analysis involved a larger set of operators. 

The direct bounds shown in Table~\ref{tab:SMEFIT} are compared to the indirect limits computed in this work in Section~\ref{sec:globalanalysis}.


\begin{table}[H]
\capstart
\begin{adjustbox}{width=0.85\textwidth,center}
\begin{minipage}[t]{3cm}
\renewcommand{\arraystretch}{1.51}
\small
\begin{align*}
\begin{array}[t]{c|c}
\multicolumn{2}{c}{\boldsymbol{\nu \nu+h.c.}} \\
\hline
\mathcal{O}_{\nu} & (\nu_{Lp}^T C \nu_{Lr})  \\
\end{array}
\end{align*}
\end{minipage}
\begin{minipage}[t]{3cm}
\renewcommand{\arraystretch}{1.51}
\small
\begin{align*}
\begin{array}[t]{c|c}
\multicolumn{2}{c}{\boldsymbol{(\nu \nu) X+h.c.}} \\
\hline
\mathcal{O}_{\nu \gamma} & (\nu_{Lp}^T C   \sigma^{\mu \nu}  \nu_{Lr})  F_{\mu \nu}  \\
\end{array}
\end{align*}
\end{minipage}
\begin{minipage}[t]{3cm}
\renewcommand{\arraystretch}{1.51}
\small
\begin{align*}
\begin{array}[t]{c|c}
\multicolumn{2}{c}{\boldsymbol{(\overline L R ) X+h.c.}} \\
\hline
\mathcal{O}_{e \gamma} & \bar e_{Lp}   \sigma^{\mu \nu} e_{Rr}\, F_{\mu \nu}  \\
\mathcal{O}_{u \gamma} & \bar u_{Lp}   \sigma^{\mu \nu}  u_{Rr}\, F_{\mu \nu}   \\
\mathcal{O}_{d \gamma} & \bar d_{Lp}  \sigma^{\mu \nu} d_{Rr}\, F_{\mu \nu}  \\
\mathcal{O}_{u G} & \bar u_{Lp}   \sigma^{\mu \nu}  T^A u_{Rr}\,  G_{\mu \nu}^A  \\
\mathcal{O}_{d G} & \bar d_{Lp}   \sigma^{\mu \nu} T^A d_{Rr}\,  G_{\mu \nu}^A \\
\end{array}
\end{align*}
\end{minipage}
\begin{minipage}[t]{3cm}
\renewcommand{\arraystretch}{1.51}
\small
\begin{align*}
\begin{array}[t]{c|c}
\multicolumn{2}{c}{\boldsymbol{X^3}} \\
\hline
\mathcal{O}_G     & f^{ABC} G_\mu^{A\nu} G_\nu^{B\rho} G_\rho^{C\mu}  \\
\mathcal{O}_{\widetilde G} & f^{ABC} \widetilde G_\mu^{A\nu} G_\nu^{B\rho} G_\rho^{C\mu}   \\
\end{array}
\end{align*}
\end{minipage}
\end{adjustbox}
\mbox{}\\[-1.25cm]

\begin{adjustbox}{width=1.05\textwidth,center}
\begin{minipage}[t]{3cm}
\renewcommand{\arraystretch}{1.51}
\small
\begin{align*}
\begin{array}[t]{c|c}
\multicolumn{2}{c}{\boldsymbol{(\overline L L)(\overline L  L)}} \\
\hline
\opleft{\nu\nu}{V}{LL} & (\bar \nu_{Lp}  \gamma^\mu \nu_{Lr} )(\bar \nu_{Ls} \gamma_\mu \nu_{Lt})   \\
\opleft{ee}{V}{LL}       & (\bar e_{Lp}  \gamma^\mu e_{Lr})(\bar e_{Ls} \gamma_\mu e_{Lt})   \\
\opleft{\nu e}{V}{LL}       & (\bar \nu_{Lp} \gamma^\mu \nu_{Lr})(\bar e_{Ls}  \gamma_\mu e_{Lt})  \\
\opleft{\nu u}{V}{LL}       & (\bar \nu_{Lp} \gamma^\mu \nu_{Lr}) (\bar u_{Ls}  \gamma_\mu u_{Lt})  \\
\opleft{\nu d}{V}{LL}       & (\bar \nu_{Lp} \gamma^\mu \nu_{Lr})(\bar d_{Ls} \gamma_\mu d_{Lt})     \\
\opleft{eu}{V}{LL}      & (\bar e_{Lp}  \gamma^\mu e_{Lr})(\bar u_{Ls} \gamma_\mu u_{Lt})   \\
\opleft{ed}{V}{LL}       & (\bar e_{Lp}  \gamma^\mu e_{Lr})(\bar d_{Ls} \gamma_\mu d_{Lt})  \\
\opleft{\nu edu}{V}{LL}      & (\bar \nu_{Lp} \gamma^\mu e_{Lr}) (\bar d_{Ls} \gamma_\mu u_{Lt})  + h.c.   \\
\opleft{uu}{V}{LL}        & (\bar u_{Lp} \gamma^\mu u_{Lr})(\bar u_{Ls} \gamma_\mu u_{Lt})    \\
\opleft{dd}{V}{LL}   & (\bar d_{Lp} \gamma^\mu d_{Lr})(\bar d_{Ls} \gamma_\mu d_{Lt})    \\
\opleft{ud}{V1}{LL}     & (\bar u_{Lp} \gamma^\mu u_{Lr}) (\bar d_{Ls} \gamma_\mu d_{Lt})  \\
\opleft{ud}{V8}{LL}     & (\bar u_{Lp} \gamma^\mu T^A u_{Lr}) (\bar d_{Ls} \gamma_\mu T^A d_{Lt})   \\[-0.5cm]
\end{array}
\end{align*}
\renewcommand{\arraystretch}{1.51}
\small
\begin{align*}
\begin{array}[t]{c|c}
\multicolumn{2}{c}{\boldsymbol{(\overline R  R)(\overline R R)}} \\
\hline
\opleft{ee}{V}{RR}     & (\bar e_{Rp} \gamma^\mu e_{Rr})(\bar e_{Rs} \gamma_\mu e_{Rt})  \\
\opleft{eu}{V}{RR}       & (\bar e_{Rp}  \gamma^\mu e_{Rr})(\bar u_{Rs} \gamma_\mu u_{Rt})   \\
\opleft{ed}{V}{RR}     & (\bar e_{Rp} \gamma^\mu e_{Rr})  (\bar d_{Rs} \gamma_\mu d_{Rt})   \\
\opleft{uu}{V}{RR}      & (\bar u_{Rp} \gamma^\mu u_{Rr})(\bar u_{Rs} \gamma_\mu u_{Rt})  \\
\opleft{dd}{V}{RR}      & (\bar d_{Rp} \gamma^\mu d_{Rr})(\bar d_{Rs} \gamma_\mu d_{Rt})    \\
\opleft{ud}{V1}{RR}       & (\bar u_{Rp} \gamma^\mu u_{Rr}) (\bar d_{Rs} \gamma_\mu d_{Rt})  \\
\opleft{ud}{V8}{RR}    & (\bar u_{Rp} \gamma^\mu T^A u_{Rr}) (\bar d_{Rs} \gamma_\mu T^A d_{Rt})  \\
\end{array}
\end{align*}
\end{minipage}
\begin{minipage}[t]{3cm}
\renewcommand{\arraystretch}{1.51}
\small
\begin{align*}
\begin{array}[t]{c|c}
\multicolumn{2}{c}{\boldsymbol{(\overline L  L)(\overline R  R)}} \\
\hline
\opleft{\nu e}{V}{LR}     & (\bar \nu_{Lp} \gamma^\mu \nu_{Lr})(\bar e_{Rs}  \gamma_\mu e_{Rt})  \\
\opleft{ee}{V}{LR}       & (\bar e_{Lp}  \gamma^\mu e_{Lr})(\bar e_{Rs} \gamma_\mu e_{Rt}) \\
\opleft{\nu u}{V}{LR}         & (\bar \nu_{Lp} \gamma^\mu \nu_{Lr})(\bar u_{Rs}  \gamma_\mu u_{Rt})    \\
\opleft{\nu d}{V}{LR}         & (\bar \nu_{Lp} \gamma^\mu \nu_{Lr})(\bar d_{Rs} \gamma_\mu d_{Rt})   \\
\opleft{eu}{V}{LR}        & (\bar e_{Lp}  \gamma^\mu e_{Lr})(\bar u_{Rs} \gamma_\mu u_{Rt})   \\
\opleft{ed}{V}{LR}        & (\bar e_{Lp}  \gamma^\mu e_{Lr})(\bar d_{Rs} \gamma_\mu d_{Rt})   \\
\opleft{ue}{V}{LR}        & (\bar u_{Lp} \gamma^\mu u_{Lr})(\bar e_{Rs}  \gamma_\mu e_{Rt})   \\
\opleft{de}{V}{LR}         & (\bar d_{Lp} \gamma^\mu d_{Lr}) (\bar e_{Rs} \gamma_\mu e_{Rt})   \\
\opleft{\nu edu}{V}{LR}        & (\bar \nu_{Lp} \gamma^\mu e_{Lr})(\bar d_{Rs} \gamma_\mu u_{Rt})  +h.c. \\
\opleft{uu}{V1}{LR}        & (\bar u_{Lp} \gamma^\mu u_{Lr})(\bar u_{Rs} \gamma_\mu u_{Rt})   \\
\opleft{uu}{V8}{LR}       & (\bar u_{Lp} \gamma^\mu T^A u_{Lr})(\bar u_{Rs} \gamma_\mu T^A u_{Rt})    \\ 
\opleft{ud}{V1}{LR}       & (\bar u_{Lp} \gamma^\mu u_{Lr}) (\bar d_{Rs} \gamma_\mu d_{Rt})  \\
\opleft{ud}{V8}{LR}       & (\bar u_{Lp} \gamma^\mu T^A u_{Lr})  (\bar d_{Rs} \gamma_\mu T^A d_{Rt})  \\
\opleft{du}{V1}{LR}       & (\bar d_{Lp} \gamma^\mu d_{Lr})(\bar u_{Rs} \gamma_\mu u_{Rt})   \\
\opleft{du}{V8}{LR}       & (\bar d_{Lp} \gamma^\mu T^A d_{Lr})(\bar u_{Rs} \gamma_\mu T^A u_{Rt}) \\
\opleft{dd}{V1}{LR}      & (\bar d_{Lp} \gamma^\mu d_{Lr})(\bar d_{Rs} \gamma_\mu d_{Rt})  \\
\opleft{dd}{V8}{LR}   & (\bar d_{Lp} \gamma^\mu T^A d_{Lr})(\bar d_{Rs} \gamma_\mu T^A d_{Rt}) \\
\opleft{uddu}{V1}{LR}   & (\bar u_{Lp} \gamma^\mu d_{Lr})(\bar d_{Rs} \gamma_\mu u_{Rt})  + h.c.  \\
\opleft{uddu}{V8}{LR}      & (\bar u_{Lp} \gamma^\mu T^A d_{Lr})(\bar d_{Rs} \gamma_\mu T^A  u_{Rt})  + h.c. \\
\end{array}
\end{align*}
\end{minipage}

\begin{minipage}[t]{3cm}
\renewcommand{\arraystretch}{1.51}
\small
\begin{align*}
\begin{array}[t]{c|c}
\multicolumn{2}{c}{\boldsymbol{(\overline L R)(\overline L R)+h.c.}} \\
\hline
\opleft{ee}{S}{RR} 		& (\bar e_{Lp}   e_{Rr}) (\bar e_{Ls} e_{Rt})   \\
\opleft{eu}{S}{RR}  & (\bar e_{Lp}   e_{Rr}) (\bar u_{Ls} u_{Rt})   \\
\opleft{eu}{T}{RR} & (\bar e_{Lp}   \sigma^{\mu \nu}   e_{Rr}) (\bar u_{Ls}  \sigma_{\mu \nu}  u_{Rt})  \\
\opleft{ed}{S}{RR}  & (\bar e_{Lp} e_{Rr})(\bar d_{Ls} d_{Rt})  \\
\opleft{ed}{T}{RR} & (\bar e_{Lp} \sigma^{\mu \nu} e_{Rr}) (\bar d_{Ls} \sigma_{\mu \nu} d_{Rt})   \\
\opleft{\nu edu}{S}{RR} & (\bar   \nu_{Lp} e_{Rr})  (\bar d_{Ls} u_{Rt} ) \\
\opleft{\nu edu}{T}{RR} &  (\bar  \nu_{Lp}  \sigma^{\mu \nu} e_{Rr} )  (\bar  d_{Ls}  \sigma_{\mu \nu} u_{Rt} )   \\
\opleft{uu}{S1}{RR}  & (\bar u_{Lp}   u_{Rr}) (\bar u_{Ls} u_{Rt})  \\
\opleft{uu}{S8}{RR}   & (\bar u_{Lp}   T^A u_{Rr}) (\bar u_{Ls} T^A u_{Rt})  \\
\opleft{ud}{S1}{RR}   & (\bar u_{Lp} u_{Rr})  (\bar d_{Ls} d_{Rt})   \\
\opleft{ud}{S8}{RR}  & (\bar u_{Lp} T^A u_{Rr})  (\bar d_{Ls} T^A d_{Rt})  \\
\opleft{dd}{S1}{RR}   & (\bar d_{Lp} d_{Rr}) (\bar d_{Ls} d_{Rt}) \\
\opleft{dd}{S8}{RR}  & (\bar d_{Lp} T^A d_{Rr}) (\bar d_{Ls} T^A d_{Rt})  \\
\opleft{uddu}{S1}{RR} &  (\bar u_{Lp} d_{Rr}) (\bar d_{Ls}  u_{Rt})   \\
\opleft{uddu}{S8}{RR}  &  (\bar u_{Lp} T^A d_{Rr}) (\bar d_{Ls}  T^A u_{Rt})  \\[-0.5cm]
\end{array}
\end{align*}
\renewcommand{\arraystretch}{1.51}
\small
\begin{align*}
\begin{array}[t]{c|c}
\multicolumn{2}{c}{\boldsymbol{(\overline L R)(\overline R L) +h.c.}} \\
\hline
\opleft{eu}{S}{RL}  & (\bar e_{Lp} e_{Rr}) (\bar u_{Rs}  u_{Lt})  \\
\opleft{ed}{S}{RL} & (\bar e_{Lp} e_{Rr}) (\bar d_{Rs} d_{Lt}) \\
\opleft{\nu edu}{S}{RL}  & (\bar \nu_{Lp} e_{Rr}) (\bar d_{Rs}  u_{Lt})  \\
\end{array}
\end{align*}
\end{minipage}
\end{adjustbox}
\setlength{\belowcaptionskip}{-3cm}
\caption{LEFT operators of dimension three and five, as well as LEFT operators of dimension six that conserve baryon and lepton number, reproduced from Ref.~\cite{Jenkins:2017jig}.}
\label{tab:oplist1}
\end{table}

\newpage

\bibliographystyle{JHEP}
\bibliography{biblio}

\providecommand{\href}[2]{#2}\begingroup\raggedright\begin{thebibliography}{100}

\bibitem{Glashow:1970gm}
S.~L. Glashow, J.~Iliopoulos and L.~Maiani, \emph{{Weak Interactions with
  Lepton-Hadron Symmetry}},
  \href{https://doi.org/10.1103/PhysRevD.2.1285}{\emph{Phys. Rev. D} {\bfseries
  2} (1970) 1285--1292}.

\bibitem{Franceschini:2023nlp}
R.~Franceschini, \emph{{Beyond-Standard-Model Physics Associated with the Top
  Quark}},  \href{https://arxiv.org/abs/2301.04407}{{\ttfamily 2301.04407}}.

\bibitem{Grzadkowski:2010es}
B.~Grzadkowski, M.~Iskrzynski, M.~Misiak and J.~Rosiek, \emph{{Dimension-Six
  Terms in the Standard Model Lagrangian}},
  \href{https://doi.org/10.1007/JHEP10(2010)085}{\emph{JHEP} {\bfseries 10}
  (2010) 085}, [\href{https://arxiv.org/abs/1008.4884}{{\ttfamily 1008.4884}}].

\bibitem{Barbieri:2011ci}
R.~Barbieri, G.~Isidori, J.~Jones-Perez, P.~Lodone and D.~M. Straub,
  \emph{{$U(2)$ and Minimal Flavour Violation in Supersymmetry}},
  \href{https://doi.org/10.1140/epjc/s10052-011-1725-z}{\emph{Eur. Phys. J. C}
  {\bfseries 71} (2011) 1725},
  [\href{https://arxiv.org/abs/1105.2296}{{\ttfamily 1105.2296}}].

\bibitem{Barbieri:2012uh}
R.~Barbieri, D.~Buttazzo, F.~Sala and D.~M. Straub, \emph{{Flavour physics from
  an approximate $U(2)^3$ symmetry}},
  \href{https://doi.org/10.1007/JHEP07(2012)181}{\emph{JHEP} {\bfseries 07}
  (2012) 181}, [\href{https://arxiv.org/abs/1203.4218}{{\ttfamily 1203.4218}}].

\bibitem{Barbieri:2012tu}
R.~Barbieri, D.~Buttazzo, F.~Sala, D.~M. Straub and A.~Tesi, \emph{{A 125 GeV
  composite Higgs boson versus flavour and electroweak precision tests}},
  \href{https://doi.org/10.1007/JHEP05(2013)069}{\emph{JHEP} {\bfseries 05}
  (2013) 069}, [\href{https://arxiv.org/abs/1211.5085}{{\ttfamily 1211.5085}}].

\bibitem{Faroughy:2020ina}
D.~A. Faroughy, G.~Isidori, F.~Wilsch and K.~Yamamoto, \emph{{Flavour
  symmetries in the SMEFT}},
  \href{https://doi.org/10.1007/JHEP08(2020)166}{\emph{JHEP} {\bfseries 08}
  (2020) 166}, [\href{https://arxiv.org/abs/2005.05366}{{\ttfamily
  2005.05366}}].

\bibitem{Fox:2007in}
P.~J. Fox, Z.~Ligeti, M.~Papucci, G.~Perez and M.~D. Schwartz,
  \emph{{Deciphering top flavor violation at the LHC with $B$ factories}},
  \href{https://doi.org/10.1103/PhysRevD.78.054008}{\emph{Phys. Rev. D}
  {\bfseries 78} (2008) 054008},
  [\href{https://arxiv.org/abs/0704.1482}{{\ttfamily 0704.1482}}].

\bibitem{Grzadkowski:2008mf}
B.~Grzadkowski and M.~Misiak, \emph{{Anomalous Wtb coupling effects in the weak
  radiative B-meson decay}},
  \href{https://doi.org/10.1103/PhysRevD.78.077501}{\emph{Phys. Rev. D}
  {\bfseries 78} (2008) 077501},
  [\href{https://arxiv.org/abs/0802.1413}{{\ttfamily 0802.1413}}]. [Erratum:
  Phys.Rev.D 84, 059903 (2011)].

\bibitem{Drobnak:2011aa}
J.~Drobnak, S.~Fajfer and J.~F. Kamenik, \emph{{Probing anomalous tWb
  interactions with rare B decays}},
  \href{https://doi.org/10.1016/j.nuclphysb.2011.10.004}{\emph{Nucl. Phys. B}
  {\bfseries 855} (2012) 82--99},
  [\href{https://arxiv.org/abs/1109.2357}{{\ttfamily 1109.2357}}].

\bibitem{Brod:2014hsa}
J.~Brod, A.~Greljo, E.~Stamou and P.~Uttayarat, \emph{{Probing anomalous $
  t\overline{t}Z $ interactions with rare meson decays}},
  \href{https://doi.org/10.1007/JHEP02(2015)141}{\emph{JHEP} {\bfseries 02}
  (2015) 141}, [\href{https://arxiv.org/abs/1408.0792}{{\ttfamily 1408.0792}}].

\bibitem{Altmannshofer:2023bfk}
W.~Altmannshofer, S.~Gori, B.~V. Lehmann and J.~Zuo, \emph{{UV physics from IR
  features: New prospects from top flavor violation}},
  \href{https://doi.org/10.1103/PhysRevD.107.095025}{\emph{Phys. Rev. D}
  {\bfseries 107} (2023) 095025},
  [\href{https://arxiv.org/abs/2303.00781}{{\ttfamily 2303.00781}}].

\bibitem{Cirigliano:2016nyn}
V.~Cirigliano, W.~Dekens, J.~de~Vries and E.~Mereghetti, \emph{{Constraining
  the top-Higgs sector of the Standard Model Effective Field Theory}},
  \href{https://doi.org/10.1103/PhysRevD.94.034031}{\emph{Phys. Rev. D}
  {\bfseries 94} (2016) 034031},
  [\href{https://arxiv.org/abs/1605.04311}{{\ttfamily 1605.04311}}].

\bibitem{Bissmann:2019gfc}
S.~Bi\ss{}mann, J.~Erdmann, C.~Grunwald, G.~Hiller and K.~Kr\"oninger,
  \emph{{Constraining top-quark couplings combining top-quark and
  $\boldsymbol{B}$ decay observables}},
  \href{https://doi.org/10.1140/epjc/s10052-020-7680-9}{\emph{Eur. Phys. J. C}
  {\bfseries 80} (2020) 136},
  [\href{https://arxiv.org/abs/1909.13632}{{\ttfamily 1909.13632}}].

\bibitem{Bissmann:2020mfi}
S.~Bi\ss{}mann, C.~Grunwald, G.~Hiller and K.~Kr\"oninger, \emph{{Top and
  Beauty synergies in SMEFT-fits at present and future colliders}},
  \href{https://doi.org/10.1007/JHEP06(2021)010}{\emph{JHEP} {\bfseries 06}
  (2021) 010}, [\href{https://arxiv.org/abs/2012.10456}{{\ttfamily
  2012.10456}}].

\bibitem{Jenkins:2017jig}
E.~E. Jenkins, A.~V. Manohar and P.~Stoffer, \emph{{Low-Energy Effective Field
  Theory below the Electroweak Scale: Operators and Matching}},
  \href{https://doi.org/10.1007/JHEP03(2018)016}{\emph{JHEP} {\bfseries 03}
  (2018) 016}, [\href{https://arxiv.org/abs/1709.04486}{{\ttfamily
  1709.04486}}].

\bibitem{Celis:2017hod}
A.~Celis, J.~Fuentes-Martin, A.~Vicente and J.~Virto, \emph{{DsixTools: The
  Standard Model Effective Field Theory Toolkit}},
  \href{https://doi.org/10.1140/epjc/s10052-017-4967-6}{\emph{Eur. Phys. J. C}
  {\bfseries 77} (2017) 405},
  [\href{https://arxiv.org/abs/1704.04504}{{\ttfamily 1704.04504}}].

\bibitem{Fuentes-Martin:2020zaz}
J.~Fuentes-Martin, P.~Ruiz-Femenia, A.~Vicente and J.~Virto, \emph{{DsixTools
  2.0: The Effective Field Theory Toolkit}},
  \href{https://doi.org/10.1140/epjc/s10052-020-08778-y}{\emph{Eur. Phys. J. C}
  {\bfseries 81} (2021) 167},
  [\href{https://arxiv.org/abs/2010.16341}{{\ttfamily 2010.16341}}].

\bibitem{Jenkins:2013zja}
E.~E. Jenkins, A.~V. Manohar and M.~Trott, \emph{{Renormalization Group
  Evolution of the Standard Model Dimension Six Operators I: Formalism and
  lambda Dependence}},
  \href{https://doi.org/10.1007/JHEP10(2013)087}{\emph{JHEP} {\bfseries 10}
  (2013) 087}, [\href{https://arxiv.org/abs/1308.2627}{{\ttfamily 1308.2627}}].

\bibitem{Jenkins:2013wua}
E.~E. Jenkins, A.~V. Manohar and M.~Trott, \emph{{Renormalization Group
  Evolution of the Standard Model Dimension Six Operators II: Yukawa
  Dependence}}, \href{https://doi.org/10.1007/JHEP01(2014)035}{\emph{JHEP}
  {\bfseries 01} (2014) 035},
  [\href{https://arxiv.org/abs/1310.4838}{{\ttfamily 1310.4838}}].

\bibitem{Alonso:2013hga}
R.~Alonso, E.~E. Jenkins, A.~V. Manohar and M.~Trott, \emph{{Renormalization
  Group Evolution of the Standard Model Dimension Six Operators III: Gauge
  Coupling Dependence and Phenomenology}},
  \href{https://doi.org/10.1007/JHEP04(2014)159}{\emph{JHEP} {\bfseries 04}
  (2014) 159}, [\href{https://arxiv.org/abs/1312.2014}{{\ttfamily 1312.2014}}].

\bibitem{Braathen:2017jvs}
J.~Braathen, M.~D. Goodsell, M.~E. Krauss, T.~Opferkuch and F.~Staub,
  \emph{{$N$-loop running should be combined with $N$-loop matching}},
  \href{https://doi.org/10.1103/PhysRevD.97.015011}{\emph{Phys. Rev. D}
  {\bfseries 97} (2018) 015011},
  [\href{https://arxiv.org/abs/1711.08460}{{\ttfamily 1711.08460}}].

\bibitem{Dekens:2019ept}
W.~Dekens and P.~Stoffer, \emph{{Low-energy effective field theory below the
  electroweak scale: matching at one loop}},
  \href{https://doi.org/10.1007/JHEP10(2019)197}{\emph{JHEP} {\bfseries 10}
  (2019) 197}, [\href{https://arxiv.org/abs/1908.05295}{{\ttfamily
  1908.05295}}]. [Erratum: JHEP 11, 148 (2022)].

\bibitem{Aebischer:2015fzz}
J.~Aebischer, A.~Crivellin, M.~Fael and C.~Greub, \emph{{Matching of gauge
  invariant dimension-six operators for $b\to s$ and $b\to c$ transitions}},
  \href{https://doi.org/10.1007/JHEP05(2016)037}{\emph{JHEP} {\bfseries 05}
  (2016) 037}, [\href{https://arxiv.org/abs/1512.02830}{{\ttfamily
  1512.02830}}].

\bibitem{Jenkins:2017dyc}
E.~E. Jenkins, A.~V. Manohar and P.~Stoffer, \emph{{Low-Energy Effective Field
  Theory below the Electroweak Scale: Anomalous Dimensions}},
  \href{https://doi.org/10.1007/JHEP01(2018)084}{\emph{JHEP} {\bfseries 01}
  (2018) 084}, [\href{https://arxiv.org/abs/1711.05270}{{\ttfamily
  1711.05270}}].

\bibitem{vanRitbergen:1997va}
T.~van Ritbergen, J.~A.~M. Vermaseren and S.~A. Larin, \emph{{The Four loop
  beta function in quantum chromodynamics}},
  \href{https://doi.org/10.1016/S0370-2693(97)00370-5}{\emph{Phys. Lett. B}
  {\bfseries 400} (1997) 379--384},
  [\href{https://arxiv.org/abs/hep-ph/9701390}{{\ttfamily hep-ph/9701390}}].

\bibitem{Vermaseren:1997fq}
J.~A.~M. Vermaseren, S.~A. Larin and T.~van Ritbergen, \emph{{The four loop
  quark mass anomalous dimension and the invariant quark mass}},
  \href{https://doi.org/10.1016/S0370-2693(97)00660-6}{\emph{Phys. Lett. B}
  {\bfseries 405} (1997) 327--333},
  [\href{https://arxiv.org/abs/hep-ph/9703284}{{\ttfamily hep-ph/9703284}}].

\bibitem{Baikov:2017ujl}
P.~A. Baikov, K.~G. Chetyrkin and J.~H. K\"uhn, \emph{{Five-loop fermion
  anomalous dimension for a general gauge group from four-loop massless
  propagators}}, \href{https://doi.org/10.1007/JHEP04(2017)119}{\emph{JHEP}
  {\bfseries 04} (2017) 119},
  [\href{https://arxiv.org/abs/1702.01458}{{\ttfamily 1702.01458}}].

\bibitem{Chetyrkin:2000yt}
K.~G. Chetyrkin, J.~H. Kuhn and M.~Steinhauser, \emph{{RunDec: A Mathematica
  package for running and decoupling of the strong coupling and quark masses}},
  \href{https://doi.org/10.1016/S0010-4655(00)00155-7}{\emph{Comput. Phys.
  Commun.} {\bfseries 133} (2000) 43--65},
  [\href{https://arxiv.org/abs/hep-ph/0004189}{{\ttfamily hep-ph/0004189}}].

\bibitem{Bednyakov:2012rb}
A.~V. Bednyakov, A.~F. Pikelner and V.~N. Velizhanin, \emph{{Anomalous
  dimensions of gauge fields and gauge coupling beta-functions in the Standard
  Model at three loops}},
  \href{https://doi.org/10.1007/JHEP01(2013)017}{\emph{JHEP} {\bfseries 01}
  (2013) 017}, [\href{https://arxiv.org/abs/1210.6873}{{\ttfamily 1210.6873}}].

\bibitem{Bednyakov:2012en}
A.~V. Bednyakov, A.~F. Pikelner and V.~N. Velizhanin, \emph{{Yukawa coupling
  beta-functions in the Standard Model at three loops}},
  \href{https://doi.org/10.1016/j.physletb.2013.04.038}{\emph{Phys. Lett. B}
  {\bfseries 722} (2013) 336--340},
  [\href{https://arxiv.org/abs/1212.6829}{{\ttfamily 1212.6829}}].

\bibitem{Bednyakov:2013eba}
A.~V. Bednyakov, A.~F. Pikelner and V.~N. Velizhanin, \emph{{Higgs
  self-coupling beta-function in the Standard Model at three loops}},
  \href{https://doi.org/10.1016/j.nuclphysb.2013.07.015}{\emph{Nucl. Phys. B}
  {\bfseries 875} (2013) 552--565},
  [\href{https://arxiv.org/abs/1303.4364}{{\ttfamily 1303.4364}}].

\bibitem{Bednyakov:2014pia}
A.~V. Bednyakov, A.~F. Pikelner and V.~N. Velizhanin, \emph{{Three-loop SM
  beta-functions for matrix Yukawa couplings}},
  \href{https://doi.org/10.1016/j.physletb.2014.08.049}{\emph{Phys. Lett. B}
  {\bfseries 737} (2014) 129--134},
  [\href{https://arxiv.org/abs/1406.7171}{{\ttfamily 1406.7171}}].

\bibitem{LHCb:2022qnv}
{\scshape LHCb} collaboration, R.~Aaij et~al., \emph{{Test of lepton
  universality in $b \rightarrow s \ell^+ \ell^-$ decays}},
  \href{https://doi.org/10.1103/PhysRevLett.131.051803}{\emph{Phys. Rev. Lett.}
  {\bfseries 131} (2023) 051803},
  [\href{https://arxiv.org/abs/2212.09152}{{\ttfamily 2212.09152}}].

\bibitem{Becirevic:2023aov}
D.~Be\v{c}irevi\'c, G.~Piazza and O.~Sumensari, \emph{{Revisiting $B\rightarrow
  K^{(*)} \nu {\bar{\nu }}$ decays in the Standard Model and beyond}},
  \href{https://doi.org/10.1140/epjc/s10052-023-11388-z}{\emph{Eur. Phys. J. C}
  {\bfseries 83} (2023) 252},
  [\href{https://arxiv.org/abs/2301.06990}{{\ttfamily 2301.06990}}].

\bibitem{Belle-II:2023esi}
{\scshape Belle-II} collaboration, I.~Adachi et~al., \emph{{Evidence for
  $B^{+}\to K^{+}\nu\bar{\nu}$ Decays}},
  \href{https://arxiv.org/abs/2311.14647}{{\ttfamily 2311.14647}}.

\bibitem{Belle:2017oht}
{\scshape Belle} collaboration, J.~Grygier et~al., \emph{{Search for
  $\boldsymbol{B\to h\nu\bar{\nu}}$ decays with semileptonic tagging at
  Belle}}, \href{https://doi.org/10.1103/PhysRevD.96.091101}{\emph{Phys. Rev.
  D} {\bfseries 96} (2017) 091101},
  [\href{https://arxiv.org/abs/1702.03224}{{\ttfamily 1702.03224}}]. [Addendum:
  Phys.Rev.D 97, 099902 (2018)].

\bibitem{Workman:2022ynf}
{\scshape Particle Data Group} collaboration, R.~L. Workman et~al.,
  \emph{{Review of Particle Physics}},
  \href{https://doi.org/10.1093/ptep/ptac097}{\emph{PTEP} {\bfseries 2022}
  (2022) 083C01}.

\bibitem{BELLE:2019xld}
{\scshape BELLE} collaboration, S.~Choudhury et~al., \emph{{Test of lepton
  flavor universality and search for lepton flavor violation in $B \rightarrow
  K\ell \ell$ decays}},
  \href{https://doi.org/10.1007/JHEP03(2021)105}{\emph{JHEP} {\bfseries 03}
  (2021) 105}, [\href{https://arxiv.org/abs/1908.01848}{{\ttfamily
  1908.01848}}].

\bibitem{BaBar:2012azg}
{\scshape BaBar} collaboration, J.~P. Lees et~al., \emph{{A search for the
  decay modes $B^{+-} \to h^{+-} \tau^{+-}l$}},
  \href{https://doi.org/10.1103/PhysRevD.86.012004}{\emph{Phys. Rev. D}
  {\bfseries 86} (2012) 012004},
  [\href{https://arxiv.org/abs/1204.2852}{{\ttfamily 1204.2852}}].

\bibitem{LHCb:2020khb}
{\scshape LHCb} collaboration, R.~Aaij et~al., \emph{{Search for the lepton
  flavour violating decay $B^+ \rightarrow K^+ \mu^- \tau^+$ using
  $B_{s2}^{*0}$ decays}},
  \href{https://doi.org/10.1007/JHEP06(2020)129}{\emph{JHEP} {\bfseries 06}
  (2020) 129}, [\href{https://arxiv.org/abs/2003.04352}{{\ttfamily
  2003.04352}}].

\bibitem{LHCb:2020pcv}
{\scshape LHCb} collaboration, R.~Aaij et~al., \emph{{Search for the Rare
  Decays $B^0_s\to e^+e^-$ and $B^0\to e^+e^-$}},
  \href{https://doi.org/10.1103/PhysRevLett.124.211802}{\emph{Phys. Rev. Lett.}
  {\bfseries 124} (2020) 211802},
  [\href{https://arxiv.org/abs/2003.03999}{{\ttfamily 2003.03999}}].

\bibitem{LHCb:2021awg}
{\scshape LHCb} collaboration, R.~Aaij et~al., \emph{{Measurement of the
  $B^0_s\to\mu^+\mu^-$ decay properties and search for the $B^0\to\mu^+\mu^-$
  and $B^0_s\to\mu^+\mu^-\gamma$ decays}},
  \href{https://doi.org/10.1103/PhysRevD.105.012010}{\emph{Phys. Rev. D}
  {\bfseries 105} (2022) 012010},
  [\href{https://arxiv.org/abs/2108.09283}{{\ttfamily 2108.09283}}].

\bibitem{LHCb:2017myy}
{\scshape LHCb} collaboration, R.~Aaij et~al., \emph{{Search for the decays
  $B_s^0\to\tau^+\tau^-$ and $B^0\to\tau^+\tau^-$}},
  \href{https://doi.org/10.1103/PhysRevLett.118.251802}{\emph{Phys. Rev. Lett.}
  {\bfseries 118} (2017) 251802},
  [\href{https://arxiv.org/abs/1703.02508}{{\ttfamily 1703.02508}}].

\bibitem{LHCb:2017hag}
{\scshape LHCb} collaboration, R.~Aaij et~al., \emph{{Search for the
  lepton-flavour violating decays B$_{(s)}^{0} \to e^{\pm}\mu^{\mp}$}},
  \href{https://doi.org/10.1007/JHEP03(2018)078}{\emph{JHEP} {\bfseries 03}
  (2018) 078}, [\href{https://arxiv.org/abs/1710.04111}{{\ttfamily
  1710.04111}}].

\bibitem{LHCb:2019ujz}
{\scshape LHCb} collaboration, R.~Aaij et~al., \emph{{Search for the
  lepton-flavour-violating decays $B^{0}_{s}\to\tau^{\pm}\mu^{\mp}$ and
  $B^{0}\to\tau^{\pm}\mu^{\mp}$}},
  \href{https://doi.org/10.1103/PhysRevLett.123.211801}{\emph{Phys. Rev. Lett.}
  {\bfseries 123} (2019) 211801},
  [\href{https://arxiv.org/abs/1905.06614}{{\ttfamily 1905.06614}}].

\bibitem{NA62:2021zjw}
{\scshape NA62} collaboration, E.~Cortina~Gil et~al., \emph{{Measurement of the
  very rare K$^{+}$\textrightarrow{}$ {\pi}^{+}\nu \overline{\nu} $ decay}},
  \href{https://doi.org/10.1007/JHEP06(2021)093}{\emph{JHEP} {\bfseries 06}
  (2021) 093}, [\href{https://arxiv.org/abs/2103.15389}{{\ttfamily
  2103.15389}}].

\bibitem{E949:2008btt}
{\scshape E949} collaboration, A.~V. Artamonov et~al., \emph{{New measurement
  of the $K^{+} \to \pi^{+} \nu \bar{\nu}$ branching ratio}},
  \href{https://doi.org/10.1103/PhysRevLett.101.191802}{\emph{Phys. Rev. Lett.}
  {\bfseries 101} (2008) 191802},
  [\href{https://arxiv.org/abs/0808.2459}{{\ttfamily 0808.2459}}].

\bibitem{KOTO:2018dsc}
{\scshape KOTO} collaboration, J.~K. Ahn et~al., \emph{{Search for the $K_L
  \!\to\! \pi^0 \nu \overline{\nu}$ and $K_L \!\to\! \pi^0 X^0$ decays at the
  J-PARC KOTO experiment}},
  \href{https://doi.org/10.1103/PhysRevLett.122.021802}{\emph{Phys. Rev. Lett.}
  {\bfseries 122} (2019) 021802},
  [\href{https://arxiv.org/abs/1810.09655}{{\ttfamily 1810.09655}}].

\bibitem{LHCb:2020ycd}
{\scshape LHCb} collaboration, R.~Aaij et~al., \emph{{Constraints on the $K^0_S
  \rightarrow \mu^+ \mu^-$ Branching Fraction}},
  \href{https://doi.org/10.1103/PhysRevLett.125.231801}{\emph{Phys. Rev. Lett.}
  {\bfseries 125} (2020) 231801},
  [\href{https://arxiv.org/abs/2001.10354}{{\ttfamily 2001.10354}}].

\bibitem{Isidori:2003ts}
G.~Isidori and R.~Unterdorfer, \emph{{On the short distance constraints from
  K(L,S) ---\ensuremath{>} mu+ mu-}},
  \href{https://doi.org/10.1088/1126-6708/2004/01/009}{\emph{JHEP} {\bfseries
  01} (2004) 009}, [\href{https://arxiv.org/abs/hep-ph/0311084}{{\ttfamily
  hep-ph/0311084}}].

\bibitem{BNL:1998apv}
{\scshape BNL} collaboration, D.~Ambrose et~al., \emph{{New limit on muon and
  electron lepton number violation from K0(L) ---\ensuremath{>} mu+- e-+
  decay}}, \href{https://doi.org/10.1103/PhysRevLett.81.5734}{\emph{Phys. Rev.
  Lett.} {\bfseries 81} (1998) 5734--5737},
  [\href{https://arxiv.org/abs/hep-ex/9811038}{{\ttfamily hep-ex/9811038}}].

\bibitem{KTEV:2000ngj}
{\scshape KTEV} collaboration, A.~Alavi-Harati et~al., \emph{{Search for the
  Decay $K_L \to \pi^0 \mu^+ \mu^-$}},
  \href{https://doi.org/10.1103/PhysRevLett.84.5279}{\emph{Phys. Rev. Lett.}
  {\bfseries 84} (2000) 5279--5282},
  [\href{https://arxiv.org/abs/hep-ex/0001006}{{\ttfamily hep-ex/0001006}}].

\bibitem{KTeV:2003sls}
{\scshape KTeV} collaboration, A.~Alavi-Harati et~al., \emph{{Search for the
  rare decay K(L) ---\ensuremath{>} pi0 e+ e-}},
  \href{https://doi.org/10.1103/PhysRevLett.93.021805}{\emph{Phys. Rev. Lett.}
  {\bfseries 93} (2004) 021805},
  [\href{https://arxiv.org/abs/hep-ex/0309072}{{\ttfamily hep-ex/0309072}}].

\bibitem{KTeV:2007cvy}
{\scshape KTeV} collaboration, E.~Abouzaid et~al., \emph{{Search for lepton
  flavor violating decays of the neutral kaon}},
  \href{https://doi.org/10.1103/PhysRevLett.100.131803}{\emph{Phys. Rev. Lett.}
  {\bfseries 100} (2008) 131803},
  [\href{https://arxiv.org/abs/0711.3472}{{\ttfamily 0711.3472}}].

\bibitem{NA62:2021zxl}
{\scshape NA62} collaboration, E.~Cortina~Gil et~al., \emph{{Search for Lepton
  Number and Flavor Violation in $K^+$ and $\pi^0$ Decays}},
  \href{https://doi.org/10.1103/PhysRevLett.127.131802}{\emph{Phys. Rev. Lett.}
  {\bfseries 127} (2021) 131802},
  [\href{https://arxiv.org/abs/2105.06759}{{\ttfamily 2105.06759}}].

\bibitem{Buras:2022wpw}
A.~J. Buras and E.~Venturini, \emph{{The exclusive vision of rare K and B
  decays and of the quark mixing in the standard model}},
  \href{https://doi.org/10.1140/epjc/s10052-022-10583-8}{\emph{Eur. Phys. J. C}
  {\bfseries 82} (2022) 615},
  [\href{https://arxiv.org/abs/2203.11960}{{\ttfamily 2203.11960}}].

\bibitem{Aebischer:2020dsw}
J.~Aebischer, C.~Bobeth, A.~J. Buras and J.~Kumar, \emph{{SMEFT ATLAS of
  $\Delta$F = 2 transitions}},
  \href{https://doi.org/10.1007/JHEP12(2020)187}{\emph{JHEP} {\bfseries 12}
  (2020) 187}, [\href{https://arxiv.org/abs/2009.07276}{{\ttfamily
  2009.07276}}].

\bibitem{Cirigliano:2021yto}
V.~Cirigliano, D.~D\'\i{}az-Calder\'on, A.~Falkowski, M.~Gonz\'alez-Alonso and
  A.~Rodr\'\i{}guez-S\'anchez, \emph{{Semileptonic tau decays beyond the
  Standard Model}}, \href{https://doi.org/10.1007/JHEP04(2022)152}{\emph{JHEP}
  {\bfseries 04} (2022) 152},
  [\href{https://arxiv.org/abs/2112.02087}{{\ttfamily 2112.02087}}].

\bibitem{Gonzalez-Alonso:2016etj}
M.~Gonz\'alez-Alonso and J.~Martin~Camalich, \emph{{Global
  Effective-Field-Theory analysis of New-Physics effects in (semi)leptonic kaon
  decays}}, \href{https://doi.org/10.1007/JHEP12(2016)052}{\emph{JHEP}
  {\bfseries 12} (2016) 052},
  [\href{https://arxiv.org/abs/1605.07114}{{\ttfamily 1605.07114}}].

\bibitem{Falkowski:2020pma}
A.~Falkowski, M.~Gonz\'alez-Alonso and O.~Naviliat-Cuncic, \emph{{Comprehensive
  analysis of beta decays within and beyond the Standard Model}},
  \href{https://doi.org/10.1007/JHEP04(2021)126}{\emph{JHEP} {\bfseries 04}
  (2021) 126}, [\href{https://arxiv.org/abs/2010.13797}{{\ttfamily
  2010.13797}}].

\bibitem{HFLAV:2022pwe}
{\scshape Heavy Flavor Averaging Group, HFLAV} collaboration, Y.~S. Amhis
  et~al., \emph{{Averages of b-hadron, c-hadron, and \ensuremath{\tau}-lepton
  properties as of 2021}},
  \href{https://doi.org/10.1103/PhysRevD.107.052008}{\emph{Phys. Rev. D}
  {\bfseries 107} (2023) 052008},
  [\href{https://arxiv.org/abs/2206.07501}{{\ttfamily 2206.07501}}].

\bibitem{Aebischer:2021uvt}
J.~Aebischer, W.~Dekens, E.~E. Jenkins, A.~V. Manohar, D.~Sengupta and
  P.~Stoffer, \emph{{Effective field theory interpretation of lepton magnetic
  and electric dipole moments}},
  \href{https://doi.org/10.1007/JHEP07(2021)107}{\emph{JHEP} {\bfseries 07}
  (2021) 107}, [\href{https://arxiv.org/abs/2102.08954}{{\ttfamily
  2102.08954}}].

\bibitem{Morel:2020dww}
L.~Morel, Z.~Yao, P.~Clad\'e and S.~Guellati-Kh\'elifa, \emph{{Determination of
  the fine-structure constant with an accuracy of 81 parts per trillion}},
  \href{https://doi.org/10.1038/s41586-020-2964-7}{\emph{Nature} {\bfseries
  588} (2020) 61--65}.

\bibitem{Parker:2018vye}
R.~H. Parker, C.~Yu, W.~Zhong, B.~Estey and H.~M\"uller, \emph{{Measurement of
  the fine-structure constant as a test of the Standard Model}},
  \href{https://doi.org/10.1126/science.aap7706}{\emph{Science} {\bfseries 360}
  (2018) 191}, [\href{https://arxiv.org/abs/1812.04130}{{\ttfamily
  1812.04130}}].

\bibitem{Aoyama:2020ynm}
T.~Aoyama et~al., \emph{{The anomalous magnetic moment of the muon in the
  Standard Model}},
  \href{https://doi.org/10.1016/j.physrep.2020.07.006}{\emph{Phys. Rept.}
  {\bfseries 887} (2020) 1--166},
  [\href{https://arxiv.org/abs/2006.04822}{{\ttfamily 2006.04822}}].

\bibitem{Aoyama:2012wk}
T.~Aoyama, M.~Hayakawa, T.~Kinoshita and M.~Nio, \emph{{Complete Tenth-Order
  QED Contribution to the Muon g-2}},
  \href{https://doi.org/10.1103/PhysRevLett.109.111808}{\emph{Phys. Rev. Lett.}
  {\bfseries 109} (2012) 111808},
  [\href{https://arxiv.org/abs/1205.5370}{{\ttfamily 1205.5370}}].

\bibitem{Aoyama:2019ryr}
T.~Aoyama, T.~Kinoshita and M.~Nio, \emph{{Theory of the Anomalous Magnetic
  Moment of the Electron}},
  \href{https://doi.org/10.3390/atoms7010028}{\emph{Atoms} {\bfseries 7} (2019)
  28}.

\bibitem{Czarnecki:2002nt}
A.~Czarnecki, W.~J. Marciano and A.~Vainshtein, \emph{{Refinements in
  electroweak contributions to the muon anomalous magnetic moment}},
  \href{https://doi.org/10.1103/PhysRevD.67.073006}{\emph{Phys. Rev. D}
  {\bfseries 67} (2003) 073006},
  [\href{https://arxiv.org/abs/hep-ph/0212229}{{\ttfamily hep-ph/0212229}}].
  [Erratum: Phys.Rev.D 73, 119901 (2006)].

\bibitem{Gnendiger:2013pva}
C.~Gnendiger, D.~St\"ockinger and H.~St\"ockinger-Kim, \emph{{The electroweak
  contributions to $(g-2)_\mu$ after the Higgs boson mass measurement}},
  \href{https://doi.org/10.1103/PhysRevD.88.053005}{\emph{Phys. Rev. D}
  {\bfseries 88} (2013) 053005},
  [\href{https://arxiv.org/abs/1306.5546}{{\ttfamily 1306.5546}}].

\bibitem{Davier:2017zfy}
M.~Davier, A.~Hoecker, B.~Malaescu and Z.~Zhang, \emph{{Reevaluation of the
  hadronic vacuum polarisation contributions to the Standard Model predictions
  of the muon $g-2$ and ${\alpha (m_Z^2)}$ using newest hadronic cross-section
  data}}, \href{https://doi.org/10.1140/epjc/s10052-017-5161-6}{\emph{Eur.
  Phys. J. C} {\bfseries 77} (2017) 827},
  [\href{https://arxiv.org/abs/1706.09436}{{\ttfamily 1706.09436}}].

\bibitem{Keshavarzi:2018mgv}
A.~Keshavarzi, D.~Nomura and T.~Teubner, \emph{{Muon $g-2$ and $\alpha(M_Z^2)$:
  a new data-based analysis}},
  \href{https://doi.org/10.1103/PhysRevD.97.114025}{\emph{Phys. Rev. D}
  {\bfseries 97} (2018) 114025},
  [\href{https://arxiv.org/abs/1802.02995}{{\ttfamily 1802.02995}}].

\bibitem{Colangelo:2018mtw}
G.~Colangelo, M.~Hoferichter and P.~Stoffer, \emph{{Two-pion contribution to
  hadronic vacuum polarization}},
  \href{https://doi.org/10.1007/JHEP02(2019)006}{\emph{JHEP} {\bfseries 02}
  (2019) 006}, [\href{https://arxiv.org/abs/1810.00007}{{\ttfamily
  1810.00007}}].

\bibitem{Hoferichter:2019mqg}
M.~Hoferichter, B.-L. Hoid and B.~Kubis, \emph{{Three-pion contribution to
  hadronic vacuum polarization}},
  \href{https://doi.org/10.1007/JHEP08(2019)137}{\emph{JHEP} {\bfseries 08}
  (2019) 137}, [\href{https://arxiv.org/abs/1907.01556}{{\ttfamily
  1907.01556}}].

\bibitem{Davier:2019can}
M.~Davier, A.~Hoecker, B.~Malaescu and Z.~Zhang, \emph{{A new evaluation of the
  hadronic vacuum polarisation contributions to the muon anomalous magnetic
  moment and to $\mathbf{\boldsymbol\alpha(m_Z^2)}$}},
  \href{https://doi.org/10.1140/epjc/s10052-020-7792-2}{\emph{Eur. Phys. J. C}
  {\bfseries 80} (2020) 241},
  [\href{https://arxiv.org/abs/1908.00921}{{\ttfamily 1908.00921}}]. [Erratum:
  Eur.Phys.J.C 80, 410 (2020)].

\bibitem{Keshavarzi:2019abf}
A.~Keshavarzi, D.~Nomura and T.~Teubner, \emph{{$g-2$ of charged leptons,
  $\alpha (M^2_Z)$ , and the hyperfine splitting of muonium}},
  \href{https://doi.org/10.1103/PhysRevD.101.014029}{\emph{Phys. Rev. D}
  {\bfseries 101} (2020) 014029},
  [\href{https://arxiv.org/abs/1911.00367}{{\ttfamily 1911.00367}}].

\bibitem{Kurz:2014wya}
A.~Kurz, T.~Liu, P.~Marquard and M.~Steinhauser, \emph{{Hadronic contribution
  to the muon anomalous magnetic moment to next-to-next-to-leading order}},
  \href{https://doi.org/10.1016/j.physletb.2014.05.043}{\emph{Phys. Lett. B}
  {\bfseries 734} (2014) 144--147},
  [\href{https://arxiv.org/abs/1403.6400}{{\ttfamily 1403.6400}}].

\bibitem{Melnikov:2003xd}
K.~Melnikov and A.~Vainshtein, \emph{{Hadronic light-by-light scattering
  contribution to the muon anomalous magnetic moment revisited}},
  \href{https://doi.org/10.1103/PhysRevD.70.113006}{\emph{Phys. Rev. D}
  {\bfseries 70} (2004) 113006},
  [\href{https://arxiv.org/abs/hep-ph/0312226}{{\ttfamily hep-ph/0312226}}].

\bibitem{Masjuan:2017tvw}
P.~Masjuan and P.~Sanchez-Puertas, \emph{{Pseudoscalar-pole contribution to the
  $(g_{\mu}-2)$: a rational approach}},
  \href{https://doi.org/10.1103/PhysRevD.95.054026}{\emph{Phys. Rev. D}
  {\bfseries 95} (2017) 054026},
  [\href{https://arxiv.org/abs/1701.05829}{{\ttfamily 1701.05829}}].

\bibitem{Colangelo:2017fiz}
G.~Colangelo, M.~Hoferichter, M.~Procura and P.~Stoffer, \emph{{Dispersion
  relation for hadronic light-by-light scattering: two-pion contributions}},
  \href{https://doi.org/10.1007/JHEP04(2017)161}{\emph{JHEP} {\bfseries 04}
  (2017) 161}, [\href{https://arxiv.org/abs/1702.07347}{{\ttfamily
  1702.07347}}].

\bibitem{Hoferichter:2018kwz}
M.~Hoferichter, B.-L. Hoid, B.~Kubis, S.~Leupold and S.~P. Schneider,
  \emph{{Dispersion relation for hadronic light-by-light scattering: pion
  pole}}, \href{https://doi.org/10.1007/JHEP10(2018)141}{\emph{JHEP} {\bfseries
  10} (2018) 141}, [\href{https://arxiv.org/abs/1808.04823}{{\ttfamily
  1808.04823}}].

\bibitem{Gerardin:2019vio}
A.~G\'erardin, H.~B. Meyer and A.~Nyffeler, \emph{{Lattice calculation of the
  pion transition form factor with $N_f=2+1$ Wilson quarks}},
  \href{https://doi.org/10.1103/PhysRevD.100.034520}{\emph{Phys. Rev. D}
  {\bfseries 100} (2019) 034520},
  [\href{https://arxiv.org/abs/1903.09471}{{\ttfamily 1903.09471}}].

\bibitem{Bijnens:2019ghy}
J.~Bijnens, N.~Hermansson-Truedsson and A.~Rodr\'\i{}guez-S\'anchez,
  \emph{{Short-distance constraints for the HLbL contribution to the muon
  anomalous magnetic moment}},
  \href{https://doi.org/10.1016/j.physletb.2019.134994}{\emph{Phys. Lett. B}
  {\bfseries 798} (2019) 134994},
  [\href{https://arxiv.org/abs/1908.03331}{{\ttfamily 1908.03331}}].

\bibitem{Colangelo:2019uex}
G.~Colangelo, F.~Hagelstein, M.~Hoferichter, L.~Laub and P.~Stoffer,
  \emph{{Longitudinal short-distance constraints for the hadronic
  light-by-light contribution to $(g-2)_\mu$ with large-$N_c$ Regge models}},
  \href{https://doi.org/10.1007/JHEP03(2020)101}{\emph{JHEP} {\bfseries 03}
  (2020) 101}, [\href{https://arxiv.org/abs/1910.13432}{{\ttfamily
  1910.13432}}].

\bibitem{Blum:2019ugy}
T.~Blum, N.~Christ, M.~Hayakawa, T.~Izubuchi, L.~Jin, C.~Jung et~al.,
  \emph{{Hadronic Light-by-Light Scattering Contribution to the Muon Anomalous
  Magnetic Moment from Lattice QCD}},
  \href{https://doi.org/10.1103/PhysRevLett.124.132002}{\emph{Phys. Rev. Lett.}
  {\bfseries 124} (2020) 132002},
  [\href{https://arxiv.org/abs/1911.08123}{{\ttfamily 1911.08123}}].

\bibitem{Colangelo:2014qya}
G.~Colangelo, M.~Hoferichter, A.~Nyffeler, M.~Passera and P.~Stoffer,
  \emph{{Remarks on higher-order hadronic corrections to the muon
  g\ensuremath{-}2}},
  \href{https://doi.org/10.1016/j.physletb.2014.06.012}{\emph{Phys. Lett. B}
  {\bfseries 735} (2014) 90--91},
  [\href{https://arxiv.org/abs/1403.7512}{{\ttfamily 1403.7512}}].

\bibitem{Borsanyi:2020mff}
S.~Borsanyi et~al., \emph{{Leading hadronic contribution to the muon magnetic
  moment from lattice QCD}},
  \href{https://doi.org/10.1038/s41586-021-03418-1}{\emph{Nature} {\bfseries
  593} (2021) 51--55}, [\href{https://arxiv.org/abs/2002.12347}{{\ttfamily
  2002.12347}}].

\bibitem{Lehner:2020crt}
C.~Lehner and A.~S. Meyer, \emph{{Consistency of hadronic vacuum polarization
  between lattice QCD and the R-ratio}},
  \href{https://doi.org/10.1103/PhysRevD.101.074515}{\emph{Phys. Rev. D}
  {\bfseries 101} (2020) 074515},
  [\href{https://arxiv.org/abs/2003.04177}{{\ttfamily 2003.04177}}].

\bibitem{Ce:2022eix}
M.~C\`e, A.~G\'erardin, G.~von Hippel, H.~B. Meyer, K.~Miura, K.~Ottnad et~al.,
  \emph{{The hadronic running of the electromagnetic coupling and the
  electroweak mixing angle from lattice QCD}},
  \href{https://doi.org/10.1007/JHEP08(2022)220}{\emph{JHEP} {\bfseries 08}
  (2022) 220}, [\href{https://arxiv.org/abs/2203.08676}{{\ttfamily
  2203.08676}}].

\bibitem{Ce:2022kxy}
M.~C\`e et~al., \emph{{Window observable for the hadronic vacuum polarization
  contribution to the muon g-2 from lattice QCD}},
  \href{https://doi.org/10.1103/PhysRevD.106.114502}{\emph{Phys. Rev. D}
  {\bfseries 106} (2022) 114502},
  [\href{https://arxiv.org/abs/2206.06582}{{\ttfamily 2206.06582}}].

\bibitem{Wang:2022lkq}
{\scshape chiQCD} collaboration, G.~Wang, T.~Draper, K.-F. Liu and Y.-B. Yang,
  \emph{{Muon g-2 with overlap valence fermions}},
  \href{https://doi.org/10.1103/PhysRevD.107.034513}{\emph{Phys. Rev. D}
  {\bfseries 107} (2023) 034513},
  [\href{https://arxiv.org/abs/2204.01280}{{\ttfamily 2204.01280}}].

\bibitem{Alexandrou:2022amy}
{\scshape Extended Twisted Mass} collaboration, C.~Alexandrou et~al.,
  \emph{{Lattice calculation of the short and intermediate time-distance
  hadronic vacuum polarization contributions to the muon magnetic moment using
  twisted-mass fermions}},
  \href{https://doi.org/10.1103/PhysRevD.107.074506}{\emph{Phys. Rev. D}
  {\bfseries 107} (2023) 074506},
  [\href{https://arxiv.org/abs/2206.15084}{{\ttfamily 2206.15084}}].

\bibitem{Colangelo:2022vok}
G.~Colangelo, A.~X. El-Khadra, M.~Hoferichter, A.~Keshavarzi, C.~Lehner,
  P.~Stoffer et~al., \emph{{Data-driven evaluations of Euclidean windows to
  scrutinize hadronic vacuum polarization}},
  \href{https://doi.org/10.1016/j.physletb.2022.137313}{\emph{Phys. Lett. B}
  {\bfseries 833} (2022) 137313},
  [\href{https://arxiv.org/abs/2205.12963}{{\ttfamily 2205.12963}}].

\bibitem{Aubin:2022hgm}
C.~Aubin, T.~Blum, M.~Golterman and S.~Peris, \emph{{Muon anomalous magnetic
  moment with staggered fermions: Is the lattice spacing small enough?}},
  \href{https://doi.org/10.1103/PhysRevD.106.054503}{\emph{Phys. Rev. D}
  {\bfseries 106} (2022) 054503},
  [\href{https://arxiv.org/abs/2204.12256}{{\ttfamily 2204.12256}}].

\bibitem{Blum:2023qou}
T.~Blum et~al., \emph{{An update of Euclidean windows of the hadronic vacuum
  polarization}},  \href{https://arxiv.org/abs/2301.08696}{{\ttfamily
  2301.08696}}.

\bibitem{Bazavov:2023has}
{\scshape Fermilab Lattice, HPQCD,, MILC} collaboration, A.~Bazavov et~al.,
  \emph{{Light-quark connected intermediate-window contributions to the muon
  g-2 hadronic vacuum polarization from lattice QCD}},
  \href{https://doi.org/10.1103/PhysRevD.107.114514}{\emph{Phys. Rev. D}
  {\bfseries 107} (2023) 114514},
  [\href{https://arxiv.org/abs/2301.08274}{{\ttfamily 2301.08274}}].

\bibitem{Davier:2023hhn}
M.~Davier, D.~D\'\i{}az-Calder\'on, B.~Malaescu, A.~Pich,
  A.~Rodr\'\i{}guez-S\'anchez and Z.~Zhang, \emph{{The Euclidean Adler function
  and its interplay with $ \Delta {\alpha}_{\textrm{QED}}^{\textrm{had}} $ and
  \ensuremath{\alpha}$_{s}$}},
  \href{https://doi.org/10.1007/JHEP04(2023)067}{\emph{JHEP} {\bfseries 04}
  (2023) 067}, [\href{https://arxiv.org/abs/2302.01359}{{\ttfamily
  2302.01359}}].

\bibitem{CMD-3:2023alj}
{\scshape CMD-3} collaboration, F.~V. Ignatov et~al., \emph{{Measurement of the
  $e^+e^-\to\pi^+\pi^-$ cross section from threshold to 1.2 GeV with the CMD-3
  detector}},  \href{https://arxiv.org/abs/2302.08834}{{\ttfamily 2302.08834}}.

\bibitem{Benton:2023dci}
G.~Benton, D.~Boito, M.~Golterman, A.~Keshavarzi, K.~Maltman and S.~Peris,
  \emph{{Data-driven determination of the light-quark connected component of
  the intermediate-window contribution to the muon $g-2$}},
  \href{https://arxiv.org/abs/2306.16808}{{\ttfamily 2306.16808}}.

\bibitem{Davier:2023cyp}
M.~Davier, Z.~Fodor, A.~Gerardin, L.~Lellouch, B.~Malaescu, F.~M. Stokes
  et~al., \emph{{Hadronic vacuum polarization: comparing lattice QCD and
  data-driven results in systematically improvable ways}},
  \href{https://arxiv.org/abs/2308.04221}{{\ttfamily 2308.04221}}.

\bibitem{Pich:2013lsa}
A.~Pich, \emph{{Precision Tau Physics}},
  \href{https://doi.org/10.1016/j.ppnp.2013.11.002}{\emph{Prog. Part. Nucl.
  Phys.} {\bfseries 75} (2014) 41--85},
  [\href{https://arxiv.org/abs/1310.7922}{{\ttfamily 1310.7922}}].

\bibitem{MEG:2016leq}
{\scshape MEG} collaboration, A.~M. Baldini et~al., \emph{{Search for the
  lepton flavour violating decay $\mu ^+ \rightarrow \mathrm {e}^+ \gamma $
  with the full dataset of the MEG experiment}},
  \href{https://doi.org/10.1140/epjc/s10052-016-4271-x}{\emph{Eur. Phys. J. C}
  {\bfseries 76} (2016) 434},
  [\href{https://arxiv.org/abs/1605.05081}{{\ttfamily 1605.05081}}].

\bibitem{SINDRUM:1987nra}
{\scshape SINDRUM} collaboration, U.~Bellgardt et~al., \emph{{Search for the
  Decay mu+ ---\ensuremath{>} e+ e+ e-}},
  \href{https://doi.org/10.1016/0550-3213(88)90462-2}{\emph{Nucl. Phys. B}
  {\bfseries 299} (1988) 1--6}.

\bibitem{SINDRUMII:2006dvw}
{\scshape SINDRUM II} collaboration, W.~H. Bertl et~al., \emph{{A Search for
  muon to electron conversion in muonic gold}},
  \href{https://doi.org/10.1140/epjc/s2006-02582-x}{\emph{Eur. Phys. J. C}
  {\bfseries 47} (2006) 337--346}.

\bibitem{BaBar:2009hkt}
{\scshape BaBar} collaboration, B.~Aubert et~al., \emph{{Searches for Lepton
  Flavor Violation in the Decays tau+- ---\ensuremath{>} e+- gamma and tau+-
  ---\ensuremath{>} mu+- gamma}},
  \href{https://doi.org/10.1103/PhysRevLett.104.021802}{\emph{Phys. Rev. Lett.}
  {\bfseries 104} (2010) 021802},
  [\href{https://arxiv.org/abs/0908.2381}{{\ttfamily 0908.2381}}].

\bibitem{Hayasaka:2010np}
K.~Hayasaka et~al., \emph{{Search for Lepton Flavor Violating Tau Decays into
  Three Leptons with 719 Million Produced Tau+Tau- Pairs}},
  \href{https://doi.org/10.1016/j.physletb.2010.03.037}{\emph{Phys. Lett. B}
  {\bfseries 687} (2010) 139--143},
  [\href{https://arxiv.org/abs/1001.3221}{{\ttfamily 1001.3221}}].

\bibitem{Belle:2007cio}
{\scshape Belle} collaboration, Y.~Miyazaki et~al., \emph{{Search for lepton
  flavor violating tau- decays into l- eta, l- eta-prime and l- pi0}},
  \href{https://doi.org/10.1016/j.physletb.2007.03.027}{\emph{Phys. Lett. B}
  {\bfseries 648} (2007) 341--350},
  [\href{https://arxiv.org/abs/hep-ex/0703009}{{\ttfamily hep-ex/0703009}}].

\bibitem{Belle:2012unr}
{\scshape Belle} collaboration, Y.~Miyazaki et~al., \emph{{Search for
  Lepton-Flavor-Violating and Lepton-Number-Violating $\tau \to \ell h
  h^\prime$ Decay Modes}},
  \href{https://doi.org/10.1016/j.physletb.2013.01.032}{\emph{Phys. Lett. B}
  {\bfseries 719} (2013) 346--353},
  [\href{https://arxiv.org/abs/1206.5595}{{\ttfamily 1206.5595}}].

\bibitem{Belle:2021ysv}
{\scshape Belle} collaboration, A.~Abdesselam et~al., \emph{{Search for
  lepton-flavor-violating tau-lepton decays to $\ell\gamma$ at Belle}},
  \href{https://doi.org/10.1007/JHEP10(2021)019}{\emph{JHEP} {\bfseries 10}
  (2021) 19}, [\href{https://arxiv.org/abs/2103.12994}{{\ttfamily
  2103.12994}}].

\bibitem{BaBar:2006jhm}
{\scshape BaBar} collaboration, B.~Aubert et~al., \emph{{Search for Lepton
  Flavor Violating Decays $\tau^\pm \to \ell^\pm \pi^0$, $\ell^\pm \eta$,
  $\ell^\pm \eta^\prime$}},
  \href{https://doi.org/10.1103/PhysRevLett.98.061803}{\emph{Phys. Rev. Lett.}
  {\bfseries 98} (2007) 061803},
  [\href{https://arxiv.org/abs/hep-ex/0610067}{{\ttfamily hep-ex/0610067}}].

\bibitem{Davidson:2022jai}
S.~Davidson, B.~Echenard, R.~H. Bernstein, J.~Heeck and D.~G. Hitlin,
  \emph{{Charged Lepton Flavor Violation}},
  \href{https://arxiv.org/abs/2209.00142}{{\ttfamily 2209.00142}}.

\bibitem{Crivellin:2017rmk}
A.~Crivellin, S.~Davidson, G.~M. Pruna and A.~Signer,
  \emph{{Renormalisation-group improved analysis of $\mu\to e$ processes in a
  systematic effective-field-theory approach}},
  \href{https://doi.org/10.1007/JHEP05(2017)117}{\emph{JHEP} {\bfseries 05}
  (2017) 117}, [\href{https://arxiv.org/abs/1702.03020}{{\ttfamily
  1702.03020}}].

\bibitem{MEGII:2018kmf}
{\scshape MEG II} collaboration, A.~M. Baldini et~al., \emph{{The design of the
  MEG II experiment}},
  \href{https://doi.org/10.1140/epjc/s10052-018-5845-6}{\emph{Eur. Phys. J. C}
  {\bfseries 78} (2018) 380},
  [\href{https://arxiv.org/abs/1801.04688}{{\ttfamily 1801.04688}}].

\bibitem{Mu2e:2014fns}
{\scshape Mu2e} collaboration, L.~Bartoszek et~al., \emph{{Mu2e Technical
  Design Report}},  \href{https://arxiv.org/abs/1501.05241}{{\ttfamily
  1501.05241}}.

\bibitem{Mu3e:2020gyw}
{\scshape Mu3e} collaboration, K.~Arndt et~al., \emph{{Technical design of the
  phase I Mu3e experiment}},
  \href{https://doi.org/10.1016/j.nima.2021.165679}{\emph{Nucl. Instrum. Meth.
  A} {\bfseries 1014} (2021) 165679},
  [\href{https://arxiv.org/abs/2009.11690}{{\ttfamily 2009.11690}}].

\bibitem{Celis:2014asa}
A.~Celis, V.~Cirigliano and E.~Passemar, \emph{{Model-discriminating power of
  lepton flavor violating $\tau$ decays}},
  \href{https://doi.org/10.1103/PhysRevD.89.095014}{\emph{Phys. Rev. D}
  {\bfseries 89} (2014) 095014},
  [\href{https://arxiv.org/abs/1403.5781}{{\ttfamily 1403.5781}}].

\bibitem{Cirigliano:2021img}
V.~Cirigliano, K.~Fuyuto, C.~Lee, E.~Mereghetti and B.~Yan, \emph{{Charged
  Lepton Flavor Violation at the EIC}},
  \href{https://doi.org/10.1007/JHEP03(2021)256}{\emph{JHEP} {\bfseries 03}
  (2021) 256}, [\href{https://arxiv.org/abs/2102.06176}{{\ttfamily
  2102.06176}}].

\bibitem{Husek:2020fru}
T.~Husek, K.~Monsalvez-Pozo and J.~Portoles, \emph{{Lepton-flavour violation in
  hadronic tau decays and $\mu-\tau$ conversion in nuclei}},
  \href{https://doi.org/10.1007/JHEP01(2021)059}{\emph{JHEP} {\bfseries 01}
  (2021) 059}, [\href{https://arxiv.org/abs/2009.10428}{{\ttfamily
  2009.10428}}].

\bibitem{Banerjee:2022xuw}
S.~Banerjee et~al., \emph{{Snowmass 2021 White Paper: Charged lepton flavor
  violation in the tau sector}},
  \href{https://arxiv.org/abs/2203.14919}{{\ttfamily 2203.14919}}.

\bibitem{Belle-II:2022cgf}
{\scshape Belle-II} collaboration, L.~Aggarwal et~al., \emph{{Snowmass White
  Paper: Belle II physics reach and plans for the next decade and beyond}},
  \href{https://arxiv.org/abs/2207.06307}{{\ttfamily 2207.06307}}.

\bibitem{Falkowski:2019hvp}
A.~Falkowski and D.~Straub, \emph{{Flavourful SMEFT likelihood for Higgs and
  electroweak data}},
  \href{https://doi.org/10.1007/JHEP04(2020)066}{\emph{JHEP} {\bfseries 04}
  (2020) 066}, [\href{https://arxiv.org/abs/1911.07866}{{\ttfamily
  1911.07866}}].

\bibitem{Efrati:2015eaa}
A.~Efrati, A.~Falkowski and Y.~Soreq, \emph{{Electroweak constraints on
  flavorful effective theories}},
  \href{https://doi.org/10.1007/JHEP07(2015)018}{\emph{JHEP} {\bfseries 07}
  (2015) 018}, [\href{https://arxiv.org/abs/1503.07872}{{\ttfamily
  1503.07872}}].

\bibitem{Janot:2019oyi}
P.~Janot and S.~Jadach, \emph{{Improved Bhabha cross section at LEP and the
  number of light neutrino species}},
  \href{https://doi.org/10.1016/j.physletb.2020.135319}{\emph{Phys. Lett. B}
  {\bfseries 803} (2020) 135319},
  [\href{https://arxiv.org/abs/1912.02067}{{\ttfamily 1912.02067}}].

\bibitem{Aguilar-Saavedra:2018ksv}
D.~Barducci et~al., \emph{{Interpreting top-quark LHC measurements in the
  standard-model effective field theory}},
  \href{https://arxiv.org/abs/1802.07237}{{\ttfamily 1802.07237}}.

\bibitem{Maltoni:2019aot}
F.~Maltoni, L.~Mantani and K.~Mimasu, \emph{{Top-quark electroweak interactions
  at high energy}}, \href{https://doi.org/10.1007/JHEP10(2019)004}{\emph{JHEP}
  {\bfseries 10} (2019) 004},
  [\href{https://arxiv.org/abs/1904.05637}{{\ttfamily 1904.05637}}].

\bibitem{Brivio:2019ius}
I.~Brivio, S.~Bruggisser, F.~Maltoni, R.~Moutafis, T.~Plehn, E.~Vryonidou
  et~al., \emph{{O new physics, where art thou? A global search in the top
  sector}}, \href{https://doi.org/10.1007/JHEP02(2020)131}{\emph{JHEP}
  {\bfseries 02} (2020) 131},
  [\href{https://arxiv.org/abs/1910.03606}{{\ttfamily 1910.03606}}].

\bibitem{Durieux:2019rbz}
G.~Durieux, A.~Irles, V.~Miralles, A.~Pe\~nuelas, R.~P\"oschl, M.~Perell\'o
  et~al., \emph{{The electro-weak couplings of the top and bottom quarks
  \textemdash{} Global fit and future prospects}},
  \href{https://doi.org/10.1007/JHEP12(2019)098}{\emph{JHEP} {\bfseries 12}
  (2019) 98}, [\href{https://arxiv.org/abs/1907.10619}{{\ttfamily
  1907.10619}}]. [Erratum: JHEP 01, 195 (2021)].

\bibitem{Hartland:2019bjb}
N.~P. Hartland, F.~Maltoni, E.~R. Nocera, J.~Rojo, E.~Slade, E.~Vryonidou
  et~al., \emph{{A Monte Carlo global analysis of the Standard Model Effective
  Field Theory: the top quark sector}},
  \href{https://doi.org/10.1007/JHEP04(2019)100}{\emph{JHEP} {\bfseries 04}
  (2019) 100}, [\href{https://arxiv.org/abs/1901.05965}{{\ttfamily
  1901.05965}}].

\bibitem{Bruggisser:2021duo}
S.~Bruggisser, R.~Sch\"afer, D.~van Dyk and S.~Westhoff, \emph{{The Flavor of
  UV Physics}}, \href{https://doi.org/10.1007/JHEP05(2021)257}{\emph{JHEP}
  {\bfseries 05} (2021) 257},
  [\href{https://arxiv.org/abs/2101.07273}{{\ttfamily 2101.07273}}].

\bibitem{Ethier:2021bye}
{\scshape SMEFiT} collaboration, J.~J. Ethier, G.~Magni, F.~Maltoni,
  L.~Mantani, E.~R. Nocera, J.~Rojo et~al., \emph{{Combined SMEFT
  interpretation of Higgs, diboson, and top quark data from the LHC}},
  \href{https://doi.org/10.1007/JHEP11(2021)089}{\emph{JHEP} {\bfseries 11}
  (2021) 089}, [\href{https://arxiv.org/abs/2105.00006}{{\ttfamily
  2105.00006}}].

\bibitem{Miralles:2021dyw}
V.~Miralles, M.~M. L\'opez, M.~M. Ll\'acer, A.~Pe\~nuelas, M.~Perell\'o and
  M.~Vos, \emph{{The top quark electro-weak couplings after LHC Run 2}},
  \href{https://doi.org/10.1007/JHEP02(2022)032}{\emph{JHEP} {\bfseries 02}
  (2022) 032}, [\href{https://arxiv.org/abs/2107.13917}{{\ttfamily
  2107.13917}}].

\bibitem{Durieux:2022cvf}
G.~Durieux, A.~G. Camacho, L.~Mantani, V.~Miralles, M.~M. L\'opez,
  M.~Ll\'acer~Moreno et~al., \emph{{Snowmass White Paper: prospects for the
  measurement of top-quark couplings}},  in \emph{{Snowmass 2021}}, 5, 2022,
  \href{https://arxiv.org/abs/2205.02140}{{\ttfamily 2205.02140}}.

\bibitem{deBlas:2022ofj}
J.~de~Blas, Y.~Du, C.~Grojean, J.~Gu, V.~Miralles, M.~E. Peskin et~al.,
  \emph{{Global SMEFT Fits at Future Colliders}},  in \emph{{Snowmass 2021}},
  6, 2022, \href{https://arxiv.org/abs/2206.08326}{{\ttfamily 2206.08326}}.

\bibitem{Bruggisser:2022rhb}
S.~Bruggisser, D.~van Dyk and S.~Westhoff, \emph{{Resolving the flavor
  structure in the MFV-SMEFT}},
  \href{https://doi.org/10.1007/JHEP02(2023)225}{\emph{JHEP} {\bfseries 02}
  (2023) 225}, [\href{https://arxiv.org/abs/2212.02532}{{\ttfamily
  2212.02532}}].

\bibitem{Giani:2023gfq}
T.~Giani, G.~Magni and J.~Rojo, \emph{{SMEFiT: a flexible toolbox for global
  interpretations of particle physics data with effective field theories}},
  \href{https://doi.org/10.1140/epjc/s10052-023-11534-7}{\emph{Eur. Phys. J. C}
  {\bfseries 83} (2023) 393},
  [\href{https://arxiv.org/abs/2302.06660}{{\ttfamily 2302.06660}}].

\bibitem{Kassabov:2023hbm}
Z.~Kassabov, M.~Madigan, L.~Mantani, J.~Moore, M.~Morales~Alvarado, J.~Rojo
  et~al., \emph{{The top quark legacy of the LHC Run II for PDF and SMEFT
  analyses}}, \href{https://doi.org/10.1007/JHEP05(2023)205}{\emph{JHEP}
  {\bfseries 05} (2023) 205},
  [\href{https://arxiv.org/abs/2303.06159}{{\ttfamily 2303.06159}}].

\bibitem{Grunwald:2023nli}
C.~Grunwald, G.~Hiller, K.~Kr\"oninger and L.~Nollen, \emph{{More Synergies
  from Beauty, Top, $Z$ and Drell-Yan Measurements in SMEFT}},
  \href{https://arxiv.org/abs/2304.12837}{{\ttfamily 2304.12837}}.

\bibitem{Gavela:2016bzc}
B.~M. Gavela, E.~E. Jenkins, A.~V. Manohar and L.~Merlo, \emph{{Analysis of
  General Power Counting Rules in Effective Field Theory}},
  \href{https://doi.org/10.1140/epjc/s10052-016-4332-1}{\emph{Eur. Phys. J. C}
  {\bfseries 76} (2016) 485},
  [\href{https://arxiv.org/abs/1601.07551}{{\ttfamily 1601.07551}}].

\bibitem{Dorsner:2016wpm}
I.~Dor\v{s}ner, S.~Fajfer, A.~Greljo, J.~F. Kamenik and N.~Ko\v{s}nik,
  \emph{{Physics of leptoquarks in precision experiments and at particle
  colliders}}, \href{https://doi.org/10.1016/j.physrep.2016.06.001}{\emph{Phys.
  Rept.} {\bfseries 641} (2016) 1--68},
  [\href{https://arxiv.org/abs/1603.04993}{{\ttfamily 1603.04993}}].

\bibitem{deBlas:2017xtg}
J.~de~Blas, J.~C. Criado, M.~Perez-Victoria and J.~Santiago, \emph{{Effective
  description of general extensions of the Standard Model: the complete
  tree-level dictionary}},
  \href{https://doi.org/10.1007/JHEP03(2018)109}{\emph{JHEP} {\bfseries 03}
  (2018) 109}, [\href{https://arxiv.org/abs/1711.10391}{{\ttfamily
  1711.10391}}].

\bibitem{Arnan:2019olv}
P.~Arnan, D.~Becirevic, F.~Mescia and O.~Sumensari, \emph{{Probing low energy
  scalar leptoquarks by the leptonic $W$ and $Z$ couplings}},
  \href{https://doi.org/10.1007/JHEP02(2019)109}{\emph{JHEP} {\bfseries 02}
  (2019) 109}, [\href{https://arxiv.org/abs/1901.06315}{{\ttfamily
  1901.06315}}].

\bibitem{Crivellin:2019dwb}
A.~Crivellin, D.~M\"uller and F.~Saturnino, \emph{{Flavor Phenomenology of the
  Leptoquark Singlet-Triplet Model}},
  \href{https://doi.org/10.1007/JHEP06(2020)020}{\emph{JHEP} {\bfseries 06}
  (2020) 020}, [\href{https://arxiv.org/abs/1912.04224}{{\ttfamily
  1912.04224}}].

\bibitem{Saad:2020ihm}
S.~Saad, \emph{{Combined explanations of $(g-2)_{\mu}$, $R_{D^{(*)}}$,
  $R_{K^{(*)}}$ anomalies in a two-loop radiative neutrino mass model}},
  \href{https://doi.org/10.1103/PhysRevD.102.015019}{\emph{Phys. Rev. D}
  {\bfseries 102} (2020) 015019},
  [\href{https://arxiv.org/abs/2005.04352}{{\ttfamily 2005.04352}}].

\bibitem{Crivellin:2020ukd}
A.~Crivellin, D.~M\"uller and F.~Saturnino, \emph{{Leptoquarks in oblique
  corrections and Higgs signal strength: status and prospects}},
  \href{https://doi.org/10.1007/JHEP11(2020)094}{\emph{JHEP} {\bfseries 11}
  (2020) 094}, [\href{https://arxiv.org/abs/2006.10758}{{\ttfamily
  2006.10758}}].

\bibitem{Gherardi:2020qhc}
V.~Gherardi, D.~Marzocca and E.~Venturini, \emph{{Low-energy phenomenology of
  scalar leptoquarks at one-loop accuracy}},
  \href{https://doi.org/10.1007/JHEP01(2021)138}{\emph{JHEP} {\bfseries 01}
  (2021) 138}, [\href{https://arxiv.org/abs/2008.09548}{{\ttfamily
  2008.09548}}].

\bibitem{Marzocca:2021miv}
D.~Marzocca, S.~Trifinopoulos and E.~Venturini, \emph{{From B-meson anomalies
  to Kaon physics with scalar leptoquarks}},
  \href{https://doi.org/10.1140/epjc/s10052-022-10271-7}{\emph{Eur. Phys. J. C}
  {\bfseries 82} (2022) 320},
  [\href{https://arxiv.org/abs/2106.15630}{{\ttfamily 2106.15630}}].

\bibitem{Gherardi:2020det}
V.~Gherardi, D.~Marzocca and E.~Venturini, \emph{{Matching scalar leptoquarks
  to the SMEFT at one loop}},
  \href{https://doi.org/10.1007/JHEP07(2020)225}{\emph{JHEP} {\bfseries 07}
  (2020) 225}, [\href{https://arxiv.org/abs/2003.12525}{{\ttfamily
  2003.12525}}]. [Erratum: JHEP 01, 006 (2021)].

\bibitem{ATLAS:2021jyv}
{\scshape ATLAS} collaboration, G.~Aad et~al., \emph{{Search for new phenomena
  in $pp$ collisions in final states with tau leptons, b-jets, and missing
  transverse momentum with the ATLAS detector}},
  \href{https://doi.org/10.1103/PhysRevD.104.112005}{\emph{Phys. Rev. D}
  {\bfseries 104} (2021) 112005},
  [\href{https://arxiv.org/abs/2108.07665}{{\ttfamily 2108.07665}}].

\bibitem{Belfatto:2019swo}
B.~Belfatto, R.~Beradze and Z.~Berezhiani, \emph{{The CKM unitarity problem: A
  trace of new physics at the TeV scale?}},
  \href{https://doi.org/10.1140/epjc/s10052-020-7691-6}{\emph{Eur. Phys. J. C}
  {\bfseries 80} (2020) 149},
  [\href{https://arxiv.org/abs/1906.02714}{{\ttfamily 1906.02714}}].

\bibitem{Cirigliano:2022qdm}
V.~Cirigliano, W.~Dekens, J.~de~Vries, E.~Mereghetti and T.~Tong,
  \emph{{Beta-decay implications for the W-boson mass anomaly}},
  \href{https://doi.org/10.1103/PhysRevD.106.075001}{\emph{Phys. Rev. D}
  {\bfseries 106} (2022) 075001},
  [\href{https://arxiv.org/abs/2204.08440}{{\ttfamily 2204.08440}}].

\bibitem{Crivellin:2022ctt}
A.~Crivellin, \emph{{Explaining the Cabibbo Angle Anomaly}},  7, 2022,
  \href{https://arxiv.org/abs/2207.02507}{{\ttfamily 2207.02507}}.

\bibitem{Buras:2014fpa}
A.~J. Buras, J.~Girrbach-Noe, C.~Niehoff and D.~M. Straub, \emph{{$ B\to
  {K}^{\left(\ast \right)}\nu \overline{\nu} $ decays in the Standard Model and
  beyond}}, \href{https://doi.org/10.1007/JHEP02(2015)184}{\emph{JHEP}
  {\bfseries 02} (2015) 184},
  [\href{https://arxiv.org/abs/1409.4557}{{\ttfamily 1409.4557}}].

\bibitem{Becirevic:2016zri}
D.~Be\v{c}irevi\'c, O.~Sumensari and R.~Zukanovich~Funchal, \emph{{Lepton
  flavor violation in exclusive $b\rightarrow s$ decays}},
  \href{https://doi.org/10.1140/epjc/s10052-016-3985-0}{\emph{Eur. Phys. J. C}
  {\bfseries 76} (2016) 134},
  [\href{https://arxiv.org/abs/1602.00881}{{\ttfamily 1602.00881}}].

\bibitem{DeBruyn:2012wj}
K.~De~Bruyn, R.~Fleischer, R.~Knegjens, P.~Koppenburg, M.~Merk and N.~Tuning,
  \emph{{Branching Ratio Measurements of $B_s$ Decays}},
  \href{https://doi.org/10.1103/PhysRevD.86.014027}{\emph{Phys. Rev. D}
  {\bfseries 86} (2012) 014027},
  [\href{https://arxiv.org/abs/1204.1735}{{\ttfamily 1204.1735}}].

\bibitem{DeBruyn:2012wk}
K.~De~Bruyn, R.~Fleischer, R.~Knegjens, P.~Koppenburg, M.~Merk, A.~Pellegrino
  et~al., \emph{{Probing New Physics via the $B^0_s\to \mu^+\mu^-$ Effective
  Lifetime}}, \href{https://doi.org/10.1103/PhysRevLett.109.041801}{\emph{Phys.
  Rev. Lett.} {\bfseries 109} (2012) 041801},
  [\href{https://arxiv.org/abs/1204.1737}{{\ttfamily 1204.1737}}].

\bibitem{Aoki:2016frl}
S.~Aoki et~al., \emph{{Review of lattice results concerning low-energy particle
  physics}}, \href{https://doi.org/10.1140/epjc/s10052-016-4509-7}{\emph{Eur.
  Phys. J. C} {\bfseries 77} (2017) 112},
  [\href{https://arxiv.org/abs/1607.00299}{{\ttfamily 1607.00299}}].

\bibitem{Hiller:2014ula}
G.~Hiller and M.~Schmaltz, \emph{{Diagnosing lepton-nonuniversality in $b \to s
  \ell \ell$}}, \href{https://doi.org/10.1007/JHEP02(2015)055}{\emph{JHEP}
  {\bfseries 02} (2015) 055},
  [\href{https://arxiv.org/abs/1411.4773}{{\ttfamily 1411.4773}}].

\bibitem{Ciuchini:2019usw}
M.~Ciuchini, A.~M. Coutinho, M.~Fedele, E.~Franco, A.~Paul, L.~Silvestrini
  et~al., \emph{{New Physics in $b \to s \ell^+ \ell^-$ confronts new data on
  Lepton Universality}},
  \href{https://doi.org/10.1140/epjc/s10052-019-7210-9}{\emph{Eur. Phys. J. C}
  {\bfseries 79} (2019) 719},
  [\href{https://arxiv.org/abs/1903.09632}{{\ttfamily 1903.09632}}].

\bibitem{Hiller:2017bzc}
G.~Hiller and I.~Nisandzic, \emph{{$R_K$ and $R_{K^{\ast}}$ beyond the standard
  model}}, \href{https://doi.org/10.1103/PhysRevD.96.035003}{\emph{Phys. Rev.
  D} {\bfseries 96} (2017) 035003},
  [\href{https://arxiv.org/abs/1704.05444}{{\ttfamily 1704.05444}}].

\bibitem{Geng:2017svp}
L.-S. Geng, B.~Grinstein, S.~J\"ager, J.~Martin~Camalich, X.-L. Ren and R.-X.
  Shi, \emph{{Towards the discovery of new physics with lepton-universality
  ratios of $b\to s\ell\ell$ decays}},
  \href{https://doi.org/10.1103/PhysRevD.96.093006}{\emph{Phys. Rev. D}
  {\bfseries 96} (2017) 093006},
  [\href{https://arxiv.org/abs/1704.05446}{{\ttfamily 1704.05446}}].

\bibitem{Altmannshofer:2008dz}
W.~Altmannshofer, P.~Ball, A.~Bharucha, A.~J. Buras, D.~M. Straub and M.~Wick,
  \emph{{Symmetries and Asymmetries of $B \to K^{*} \mu^{+} \mu^{-}$ Decays in
  the Standard Model and Beyond}},
  \href{https://doi.org/10.1088/1126-6708/2009/01/019}{\emph{JHEP} {\bfseries
  01} (2009) 019}, [\href{https://arxiv.org/abs/0811.1214}{{\ttfamily
  0811.1214}}].

\bibitem{Bharucha:2015bzk}
A.~Bharucha, D.~M. Straub and R.~Zwicky, \emph{{$B\to V\ell^+\ell^-$ in the
  Standard Model from light-cone sum rules}},
  \href{https://doi.org/10.1007/JHEP08(2016)098}{\emph{JHEP} {\bfseries 08}
  (2016) 098}, [\href{https://arxiv.org/abs/1503.05534}{{\ttfamily
  1503.05534}}].

\bibitem{Aebischer:2022vky}
J.~Aebischer, A.~J. Buras and J.~Kumar, \emph{{On the Importance of Rare Kaon
  Decays: A Snowmass 2021 White Paper}},  in \emph{{Snowmass 2021}}, 3, 2022,
  \href{https://arxiv.org/abs/2203.09524}{{\ttfamily 2203.09524}}.

\bibitem{Angelescu:2020uug}
A.~Angelescu, D.~A. Faroughy and O.~Sumensari, \emph{{Lepton Flavor Violation
  and Dilepton Tails at the LHC}},
  \href{https://doi.org/10.1140/epjc/s10052-020-8210-5}{\emph{Eur. Phys. J. C}
  {\bfseries 80} (2020) 641},
  [\href{https://arxiv.org/abs/2002.05684}{{\ttfamily 2002.05684}}].

\bibitem{Misiak:2015xwa}
M.~Misiak et~al., \emph{{Updated NNLO QCD predictions for the weak radiative
  B-meson decays}},
  \href{https://doi.org/10.1103/PhysRevLett.114.221801}{\emph{Phys. Rev. Lett.}
  {\bfseries 114} (2015) 221801},
  [\href{https://arxiv.org/abs/1503.01789}{{\ttfamily 1503.01789}}].

\bibitem{Czakon:2015exa}
M.~Czakon, P.~Fiedler, T.~Huber, M.~Misiak, T.~Schutzmeier and M.~Steinhauser,
  \emph{{The $(Q_{7}, Q_{1,2})$ contribution to $ \overline{B}\to {X}_s\gamma $
  at $ \mathcal{O}\left({\alpha}_{\mathrm{s}}^2\right) $}},
  \href{https://doi.org/10.1007/JHEP04(2015)168}{\emph{JHEP} {\bfseries 04}
  (2015) 168}, [\href{https://arxiv.org/abs/1503.01791}{{\ttfamily
  1503.01791}}].

\bibitem{Dekens:2018pbu}
W.~Dekens, E.~E. Jenkins, A.~V. Manohar and P.~Stoffer, \emph{{Non-perturbative
  effects in $\mu \to e \gamma$}},
  \href{https://doi.org/10.1007/JHEP01(2019)088}{\emph{JHEP} {\bfseries 01}
  (2019) 088}, [\href{https://arxiv.org/abs/1810.05675}{{\ttfamily
  1810.05675}}].

\bibitem{Gonzalez-Alonso:2017iyc}
M.~Gonz\'alez-Alonso, J.~Martin~Camalich and K.~Mimouni,
  \emph{{Renormalization-group evolution of new physics contributions to
  (semi)leptonic meson decays}},
  \href{https://doi.org/10.1016/j.physletb.2017.07.003}{\emph{Phys. Lett. B}
  {\bfseries 772} (2017) 777--785},
  [\href{https://arxiv.org/abs/1706.00410}{{\ttfamily 1706.00410}}].

\end{thebibliography}\endgroup

\end{document}